\documentclass[11pt,a4paper]{article}
\usepackage{jheppub}
\usepackage{tikz}
\usepackage{mathtools}
\usepackage{subcaption}
\usepackage[shortlabels]{enumitem}

\usepackage{footmisc}

\allowdisplaybreaks

\title{\boldmath The Geometry of Decoupling Fields}

\author[a,b]{Ibrahima Bah,}
\author[c]{Federico Bonetti,}
\author[a]{Enoch Leung}
\author[a]{and Peter Weck} 
\affiliation[a]{Department of Physics and Astronomy, Johns Hopkins University,\\
3400 North Charles Street, Baltimore, MD 21218, USA}
\affiliation[b]{Institute for Advanced Study,\\
1 Einstein Drive, Princeton, NJ 08540, USA}
\affiliation[c]{Mathematical Institute, University of Oxford,\\
Woodstock Road, Oxford, OX2 6GG, UK}

\emailAdd{iboubah@jhu.edu}
\emailAdd{federico.bonetti@maths.ox.ac.uk}
\emailAdd{yleung5@jhu.edu}
\emailAdd{pweck1@jhu.edu}

\abstract{
We consider 4d field theories obtained by reducing     
the 6d (1,0) SCFT
of $N$ M5-branes probing a $\mathbb C^2/\mathbb Z_k$ singularity
 on 
a Riemann surface with fluxes.
We follow two different routes.
On the one hand, we consider the integration of
the anomaly polynomial of the parent
6d SCFT on the Riemann surface. On the other hand, we perform an anomaly inflow analysis directly
from eleven dimensions, from a setup with M5-branes probing a resolved 
 $\mathbb C^2/\mathbb Z_k$ singularity fibered over the Riemann surface.
By comparing the 4d anomaly polynomials,
 we provide a characterization of a class of modes that decouple
along the RG flow from six to four dimensions, for
 generic $N$, $k$, and genus.
These modes are identified with the flip fields encountered in
the Lagrangian descriptions of these 4d models, when they are available.
We show that such fields couple to operators originating from  M2-branes
wrapping the resolution cycles.
This provides a geometric origin of flip fields. They interpolate between the 6d theory in the UV,
where  the M2-brane operators are projected out, and the 4d theory in the IR, 
where these M2-brane operators
are part of the spectrum.

}


\newcommand{\cN}{\mathcal{N}}
\newcommand{\cS}{\mathcal{S}}

\newcommand{\cO}{\mathcal{O}}

\newcommand{\cZ}{\mathcal{Z}}

\newcommand{\beq}{\begin{equation}}
\newcommand{\eeq}{\end{equation}}

\newcommand{\nn}{\nonumber}

\newcommand{\U}{\mathrm U}

\newcommand{\SU}{\mathrm {SU}}

\begin{document}


\maketitle
\flushbottom

\addtocontents{toc}{\protect\setcounter{tocdepth}{2}}

 
\section{Introduction and summary}

The reduction of higher-dimensional superconformal field theories (SCFTs)
to lower dimensions has proven to be a powerful framework to construct non-trivial field theories,
study their properties, and, more broadly, organize the space of quantum field theories (QFTs)
using topological and geometric tools. One of the most prominent realizations
of this paradigm is provided by class $\mathcal S$ constructions \cite{Gaiotto:2009we,Gaiotto:2009hg} and their generalizations
\cite{Maruyoshi:2009uk,Benini:2009mz,Bah:2011je,Bah:2011vv,Bah:2012dg,Gaiotto:2015usa,Franco:2015jna,DelZotto:2015rca,Razamat:2016dpl,Bah:2017gph,Kim:2017toz}. The central idea is to start with a 6d SCFT and reduce it on a Riemann surface, triggering a renormalization group (RG) flow
that can yield a non-trivial 4d SCFT in the~IR.


In most instances, the parent 6d SCFT 
admits a realization in string/M/F-theory.
Indeed, an atomic classification of 6d (1,0) SCFTs 
has been proposed in F-theory \cite{Heckman:2015bfa}.
When an explicit string theoretic construction of the parent SCFT is available,
we have two ways of thinking about the
resulting 4d SCFT: it is the IR fixed point of a purely field-theoretic RG flow
from six dimensions, and it is also the QFT capturing the low-energy dynamics of a string theory setup
with four non-compact dimensions of spacetime.
As a simple example, we may consider the 6d (2,0) SCFT of type $A_{N-1}$, which
can be realized by a stack of $N$ M5-branes.
This 6d SCFT can be reduced on a smooth Riemann surface
preserving 4d $\mathcal N = 2$ or $\mathcal N =1$ supersymmetry,
depending on the choice of topological twist \cite{Gaiotto:2009we,Bah:2012dg}.
In M-theory, the M5-brane stack is wrapped on the Riemann surface,
and the choice of topological twist is mapped to the topology of the normal bundle
to the M5-branes.

In this work, we explore a generalization of this circle of ideas
to 4d $\cN = 1$ SCFTs of class $\mathcal S_k$ \cite{Gaiotto:2015usa}.
In the class $\mathcal S_k$ program, the parent 6d SCFT is the 6d (1,0) theory
realized by a stack of $N$ M5-branes probing a $\mathbb C^2/\mathbb Z_k$ singularity. 
The global symmetries of this SCFT for generic $N$, $k$ include
the $\SU(2)_R$ R-symmetry and a $\U(1) \times \SU(k) \times \SU(k)$ flavor
symmetry.\footnote{For $k=2$, the  $\U(1)$ factor enhances to $\SU(2)$.
For $N = 2$ the flavor symmetry enhances to $\SU(2k)$.
For $k=2$, $N=2$, it enhances to $\mathrm{SO}(7)$. For the remainder of this work,
we focus on the $\U(1) \times \SU(k) \times \SU(k)$ symmetry of the general case.
Moreover, we are cavalier about the global form of the symmetry group,
since it does not play a role in our discussion.
}
This 6d (1,0) SCFT
is reduced on a Riemann surface
with a topological twist that preserves 4d $\mathcal N =1$ supersymmetry.
The data that specify the construction include the topology of the Riemann surface, 
possible defects (punctures) that can decorate the Riemann surface,
and a choice of fluxes for global symmetries of the 6d (1,0) SCFT.

For simplicity, in this paper we restrict our attention to
the case of a smooth Riemann surface without punctures.
Even in this simpler class of setups,
the RG flow from six to four dimensions can exhibit
subtle features, such as a non-trivial pattern
of modes that decouple along the flow,
as demonstrated in \cite{Bah:2017gph} for reductions on tori with flux.

In the construction of Lagrangian models for 4d SCFTs originating
from reductions of a 6d SCFT, it is not unusual to encounter flip fields,
i.e.~gauge singlets $\phi$ that participate in a superpotential coupling
$W_{\rm flip}= \phi \, \mathcal O$, where $\mathcal O$ is a gauge invariant operator
(for example, a baryonic operator constructed from a bifundamental  field in a quiver
gauge theory). 
For a gauge theory of sufficiently large rank,
the superpotential coupling $W_{\rm flip}= \phi \, \mathcal O$
is irrelevant: in the deep IR, the flip field $\phi$ is expected to behave as a free field,
and therefore decouple from 
the interacting SCFT.
Flip fields 
 are ubiquitous in the literature on class $\cS$ and its generalizations \cite{Agarwal:2014rua,Gaiotto:2015usa,Razamat:2016dpl,Maruyoshi:2016tqk, Maruyoshi:2016aim,Nardoni:2016ffl,Kim:2017toz,Agarwal:2017roi,Benvenuti:2017bpg,
 Giacomelli:2017ckh,Zafrir:2018hkr,Chen:2019njf,Hwang:2021xyw},
 and appear in particular in many models studied in 
\cite{Bah:2017gph}. One of the aims of this work is to revisit this class of models,
with the aim of shedding light on the origin of flip fields
from a geometric M-theory perspective.

\begin{figure}
\centering
\includegraphics[width = 11.5 cm]{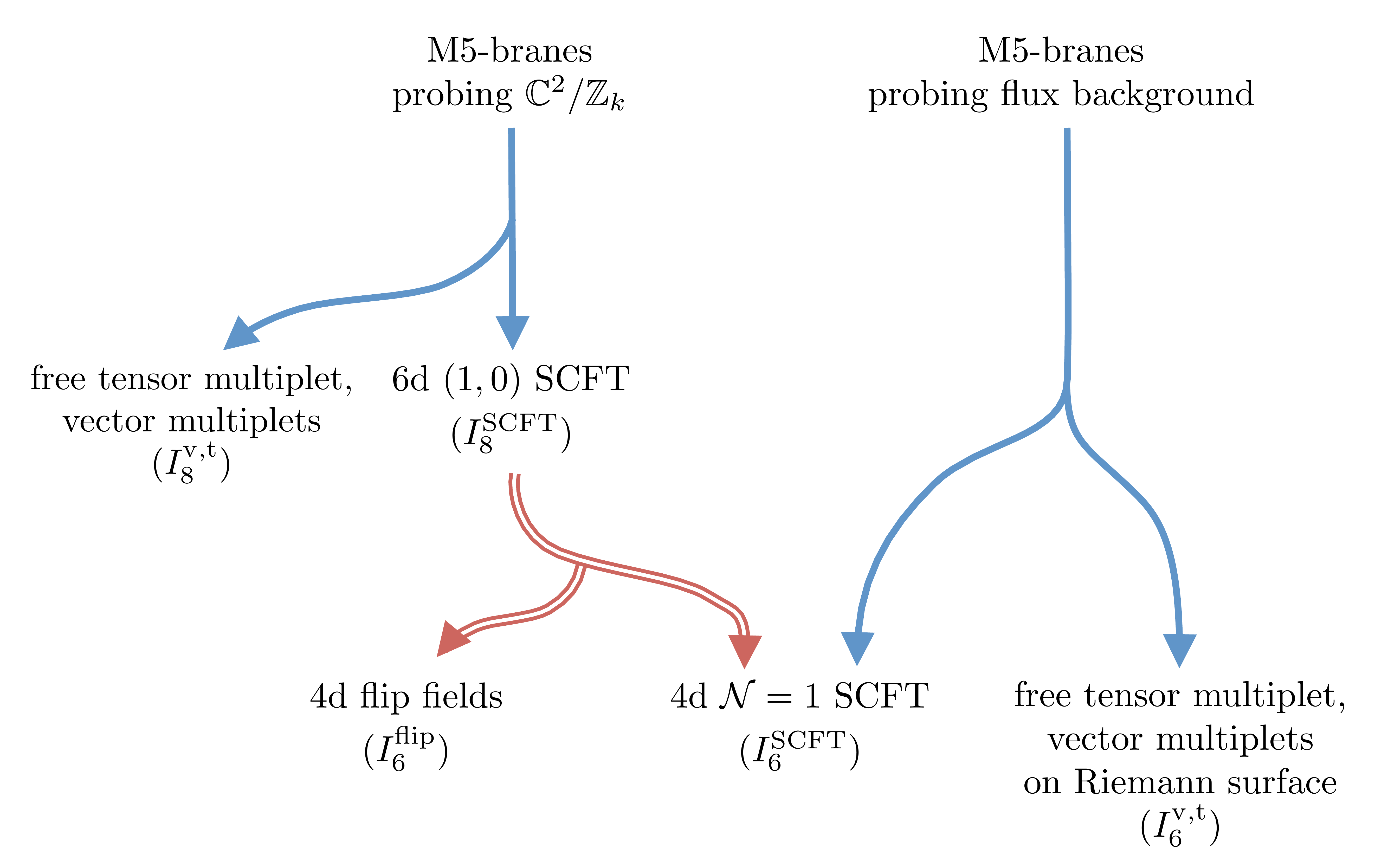}
\caption{
Schematic depiction of the flows across dimensions considered in this work.
On the left: a stack of M5-branes 
with flat 6d worldvolume
probes a $\mathbb C^2/\mathbb Z_k$ singularity, yielding 
a 6d (1,0) SCFT, plus 6d modes associated to free tensor, vector multiplets.
The interacting 6d (1,0) SCFT
is reduced on a Riemann surface with fluxes. The outcome is organized into
an interacting 4d SCFT, and a collection of 4d free fields, interpreted as flip fields.
On the right: a stack of M5-branes probes a resolved $\mathbb C^2/\mathbb Z_k$ singularity
and is wrapped on a Riemann surface. At low energies, this M-theory setup gives
the same
4d SCFT, plus other 4d modes coming from free tensor, vector multiplets on
the Riemann surface. 
The blue, solid arrows denote anomaly inflow
from the 11d bulk onto the M5-branes. The red, hollow arrows
denote the purely field-theoretical reduction of the 6d (1,0) SCFT
on the Riemann surface.
}
\label{fig_flows}
\end{figure}

The main goal of this paper is to contrast
the field-theoretic point of view on $\mathcal S_k$ constructions
with a point of view based on a direct construction from M-theory,
as illustrated in figure \ref{fig_flows}.
Our objective is two-fold:
\begin{enumerate}[(i)]
\item Identify the  M-theory setups that correspond
to class $\mathcal S_k$ reductions on a smooth Riemann surface
with non-zero fluxes for the global $\SU(k)^2$ flavor symmetry of the parent 6d SCFT.
\item Exploit these M-theory setups
to gain insights on    the 
field theory flow from 6d to 4d.
\end{enumerate}
Let us now proceed to summarize the main results of this paper.

\subsection*{Inflow for wrapped M5-branes probing flux backgrounds.}

As far as objective (i)~is concerned,
a natural proposal is as follows: in M-theory, we should consider
a stack of M5-branes that probes a \emph{resolved}
$\mathbb C^2/\mathbb Z_k$ singularity, further wrapped on a Riemann surface.
The 11d background probed by the M5-branes is expected to be a flux background:
a non-zero $G_4$-flux threads non-trivial 4-cycles, obtained 
combining the Riemann surface with the  2-cycles
originating from the resolution of the 
$\mathbb C^2/\mathbb Z_k$ singularity.

The above discussion can be made more precise for
$k=2$. Indeed, in this  case
we can establish a connection to a class of 
$AdS_5$ solutions in 11d supergravity, first discussed by
Gauntlett, Martelli, Sparks, and Waldram (GMSW) \cite{Gauntlett:2004zh}.
These solutions take the form of a warped product $AdS_5 \times _w M_6$,
in which the internal space $M_6$ is a fibration of a 4-manifold $M_4$
over a Riemann surface $\Sigma_g$. The space $M_4$ 
can be regarded as a resolution of the orbifold $S^4/\mathbb Z_2$:
the north and south pole of the $S^4$ are fixed points of the $\mathbb Z_2$ action,
yielding locally $\mathbb R^4/\mathbb Z_2$ singularities; the singularity at each pole
is resolved introducing a 2-cycle. 
 The internal
$G_4$-flux configuration is   specified by three positive integer flux quanta,
which we denote by $N$, $N_\mathrm {N_1}$, $N_{\mathrm S_1}$.
The flux quantum $N$ measures $G_4$-flux on the 4-cycle 
given by the fiber $M_4$ at a generic point of the base $\Sigma_g$.
The integer $N_{\mathrm N_1}$ quantifies instead
the $G_4$-flux on the 4-cycle obtained by combining
the Riemann surface with the resolution 2-cycle at the north pole of $S^4$.
Similar remarks apply to $N_{\mathrm S_1}$.
We propose the following interpretation of this class of solutions:
they describe the near-horizon geometry of a stack of $N$ wrapped M5-branes probing
a resolved $\mathbb C^2/\mathbb Z_2$ singularity \cite{Bah:2019vmq}.

As discussed in \cite{Bah:2021brs}, the topology and $G_4$-flux configuration of 
the internal space
$M_6$ for $k=2$ admit a natural generalization
for higher $k$. In this case, $M_6$ is still taken to be a fibration of a 4-manifold
$M_4$ over a Riemann surface $\Sigma$, but $M_4$ is identified with the resolution
of the orbifold $S^4/\mathbb Z_k$.
The fixed points of the orbifold action are locally $\mathbb R^4/\mathbb Z_k$
and can be resolved by blow-up, introducing a collection of $k-1$ resolution 2-cycles
near each pole of the $S^4$.
The $G_4$-flux configuration is described by a total of $2k-1$ flux quanta, 
$N$, $\{ N_{\mathrm N_i} \}_{i=1}^{k-1}$, $\{ N_{\mathrm S_i} \}_{i=1}^{k-1}$,
in direct analogy to the GMSW solution.

Explicit $AdS_5$ solutions in which the internal space $M_6$ has the topology
described in the previous paragraph are not known for $k>2$.
While the relevant BPS system is well-studied \cite{Gauntlett:2004zh,Bah:2015fwa},
the search for such solutions proves to be a challenging task.
This is certainly an important problem, but one which we choose to set aside
for the purposes of this paper. Our working assumption  is that
the topology  and flux configuration for $k>2$ can be realized 
in the near horizon limit of a well-defined M-theory setup.

Crucially, in order to extract the physical consequences 
of our working assumption  
we do not need an explicit holographic solution.
Building on \cite{Freed:1998tg,Harvey:1998bx}, systematic methods
have been developed \cite{Bah:2019rgq} (see also \cite{Hosseini:2020vgl})
to compute 
the inflow anomaly polynomial $I_6^{\rm inflow}$
for the wrapped M5-branes probing the resolved singularity,
using as input 
  the topology and flux configuration of $M_6$.
The quantity $I_6^{\rm inflow}$ is a   6-form characteristic class
that captures the anomalous variation of the bulk 11d supergravity action
in the presence of the wrapped M5-branes.
According to the standard anomaly inflow paradigm,
$I_6^{\rm inflow}$ is expected to be canceled by the
't Hooft anomalies of the 4d degrees of freedom 
living along the non-compact directions of the M5-brane stack.
The computation of $I_6^{\rm inflow}$ was performed in \cite{Bah:2019vmq}
for $k=2$ and in \cite{Bah:2021brs} for general $k$, at cubic order
in the flux quanta (which originate from the 2-derivative $C_3 G_4 G_4$ coupling
in the M-theory effective action). In this work we complete the computation of 
$I_6^{\rm inflow}$ by deriving the terms linear in the flux quanta
(which originate from the higher-derivative coupling $C_3 X_8$).

\subsection*{Comparison with the integrated anomaly polynomial in field theory.}

In order to contrast the field-theoretical and M-theory perspectives
on class $\mathcal S_k$ theories,
it is natural to compare the quantity $I_6^{\rm inflow}$ on the M-theory side
with the quantity $\int_{\Sigma_g} I_8^{\rm SCFT}$ on the field theory side.
Here, $I_8^{\rm SCFT}$ denotes the anomaly polynomial of the parent interacting
6d (1,0) SCFT \cite{Ohmori:2014kda}, and $\int_{\Sigma_g} I_8^{\rm SCFT}$ denotes the 6-form
characteristic class obtained upon integrating $I_8^{\rm SCFT}$
on the Riemann surface, taking into account both R-symmetry and flavor fluxes
\cite{Bah:2017gph}.
In particular, the quantity $\int_{\Sigma_g} I_8^{\rm SCFT}$
depends not only on $N$, $k$, and the Euler characteristic
$\chi$ of the Riemann surface, but also on the flavor fluxes
for the Cartan subgroups of the $SU(k)^2$ flavor symmetry,
denoted by $N_{b_i}$, $N_{c_i}$, $i=1,\dots,k$
(subject to the constraints $\sum_{i=1}^k N_{b_i} = 0$,
$\sum_{i=1}^k N_{c_i} = 0$).

One of the main results of this paper
is a precise match between $\int_{\Sigma_g} I_8^{\rm SCFT}$ and 
$I_6^{\rm inflow}$, including terms of order 1 in the flux
parameters. 
The match takes the following   form,
\beq \label{main_equation}
 -I_6^{\rm inflow} - I_6^\text{v,t} = \int_{\Sigma_g} I_8^{\rm SCFT}  - I_6^{\rm flip} =I_6^{\rm SCFT}   \ .
\eeq
The 6-form $I_6^{\rm SCFT}$ is the anomaly polynomial of the
\emph{interacting} 4d SCFT of class $\mathcal S_k$ that we want to study.\footnote{While
we expect that the flows from six to four dimensions studied in this work
yield non-trivial SCFTs in the IR, we do not have a general proof. Our analysis
of the case of genus one in section \ref{sec_genus_one} and of the central charges in section \ref{sec_central_charges} 
supports our expectations.
}
The quantity $I_6^\text{v,t}$ is the anomaly polynomial of a collection of free 4d fields,
which we interpret as the reduction on $\Sigma_g$ of a free 6d tensor multiplet and  free 6d vector
multiplets.
The quantity $I_6^\text{flip}$ is also the anomaly polynomial of a collection of free 4d fields,
but with a different interpretation: they are flip chiral multiplets, i.e.~gauge singlets coupled
to 4d operators of the interacting 4d SCFT via irrelevant interactions.
They are free fields, but due to their superpotential couplings
to non-trivial operators in the interacting SCFT, they give large contributions
to the integrated anomaly polynomial $\int_\Sigma I_8^{\rm SCFT}$
(here ``large'' refers to the fact that these contributions scale 
with $N$ and the flux quanta, and are not order one fixed numbers).
These contributions have to be suitably accounted for
in order to extract the anomaly polynomial $I_6^{\rm SCFT}$
of the interacting 4d SCFT from the integrated  polynomial $\int_\Sigma I_8^{\rm SCFT}$.
The role of flip fields in the field-theory flow from 6d to 4d 
is studied in detail in \cite{Bah:2017gph} with several examples where the Riemann
surface is a torus. 

Our analysis furnishes a general expression for the contribution
$I_6^{\rm flip}$ of flip fields, for a Riemann surface of arbitrary genus 
and for arbitrary values of $k$ and the flux parameters. Our findings
match the results of \cite{Bah:2017gph} in the case of genus one.

\subsection*{\boldmath The case $k=2$ at genus one: D3-branes at the tip of ${\rm Cone}(Y^{p,q})$.}

In the special case $k=2$ and genus one, the GMSW solution
is related by a chain of string theory dualities to the $AdS_5 \times Y^{p,q}$ solutions
in Type IIB string theory \cite{Gauntlett:2004zh}, which are holographically dual
to the SCFTs engineered by a stack of D3-branes at the tip of the Calabi-Yau cone over
$Y^{p,q}$.
By analyzing this  special case, we find further evidence in favor
of   \eqref{main_equation}, by verifying the relation
\beq \label{match_with_Ypq}
\text{$k=2$, genus $g=1$:} \qquad - I_6^{\rm inflow} = I_{6}^{\rm SCFT}(Y^{p,q}) \ .
\eeq
The above equality holds exactly in $N$, and not only at large $N$.
We confirm the map between the $p$, $q$ integers of $Y^{p,q}$
and the flux quanta on the M-theory side, established in \cite{Gauntlett:2006ai}.
Notice that the term $I_6^{\rm v,t}$ in the general relation \eqref{main_equation}
is absent in \eqref{match_with_Ypq}, because 
it is proportional to the Euler characteristic of the Riemann surface.

\subsection*{Flip fields and M2-brane operators.}

The anomaly polynomial $I_6^{\rm flip}$ encodes the anomalies
of the flip fields we encounter upon reducing the 6d (1,0) SCFT on the Riemann surface.
From the data in $I_6^{\rm flip}$ it is straightforward to 
extract the charges and multiplicities of the operators that get flipped.
Such charges and multiplicities can be matched precisely with those of wrapped M2-brane states
in the M-theory setup.

In the case $k=2$, one can resort to the explicit $AdS_5$ GMSW solution
to study supersymmetric M2-brane probes, by identifying the calibrated
2-cycles in the internal space $M_6$ \cite{Gauntlett:2006ai}. Moreover,
the charges of the operators originating from these wrapped M2-brane probes
 can be extracted systematically from the 
terms in the uplift ansatz for $G_4$ that are linear in the external   gauge fields.
These are in turn conveniently extracted from  
the same 4-form $E_4$
that we utilize in the anomaly inflow computation.

For $k >2$ we do not have an explicit holographic solution,
nor do we have a solution describing the flux background probed by the M5-brane stack.
For these reasons, a direct analysis of 
the calibration conditions for 
wrapped M2-brane probes is challenging.
Nonetheless, we can identify non-trivial 2-cycles in the internal space $M_6$.
Motivated by the analogy with the $k=2$ case, we make the working assumption
that the relevant non-trivial 2-homology classes in $M_6$ admit a calibrated representative,
so that the associated wrapped M2-brane operators are BPS.
We can then proceed to compute their charges
from the 4-form $E_4$ utilized in the anomaly inflow analysis.
We obtain a perfect match with the charges of the operators that are flipped
by the fields in $I_6^{\rm flip}$. Moreover,
we can also reproduce their multiplicities: they are simply given
by the units of $G_4$ flux threading the relevant 2-cycle (combined with the Riemann surface),
by virtue of a standard Landau-level degeneracy argument \cite{Gaiotto:2009gz},
which we review in section \ref{sec_wrappedM2}.

The identification of charges and multiplicities of  flipped operators, and wrapped M2-brane operators,
suggests the following physical picture.
The wrapped M2-branes operators are associated to  
blow-up modes for the $\mathbb C^2/\mathbb Z_k$
singularity. In the 6d (1,0) SCFT, however, such  modes are not present 
\cite{Witten:1995zh,Ganor:1996pc,Hanany:1996ie,Brunner:1997gk,Blum:1997fw,Blum:1997mm,Intriligator:1997dh,Brunner:1997gf,Hanany:1997gh}. In contrast, 
we expect the M2-brane operators to be part of the spectrum of the 4d theory
obtained by reduction on the Riemann surface.
Indeed, in Lagrangian models, they   are baryonic operators.
As a result, a mechanism is needed to interpolate between
six and four dimensions; this mechanism is precisely given by the flip fields from
the term $I_6^{\rm flip}$. They act as Lagrange multipliers
that project away the wrapped M2-brane operators in the integration of the anomaly
polynomial on $\Sigma_g$. In the 4d theory, they are expected to be free fields and decouple,
thus effectively reintroducing the M2-brane operators.

\subsection*{Central charges.}

We study $a$-maximization \cite{Intriligator:2003jj}  on the combination
$- I_6^{\rm inflow} - I_6^{\rm v,t}$, in order to support the interpretation
of this quantity as the anomaly polynomial of the interacting SCFT of class $\mathcal S_k$.
By a combination of analytic and numeric methods, we explore vast regions
of the flux parameter space. Our findings provide evidence for 
the existence of a unique local maximum of the trial $a$ central charge.
The resulting $a$ and $c$   are
  compatible with the 
  Hofman-Maldacena bounds \cite{Hofman:2008ar}.

\subsection*{Organization of the paper.}
The rest of this paper is organized as follows. In section \ref{sec_review_flux}
we review the M-theory flux setups studied in \cite{Bah:2021brs}, 
giving a brief account of the isometries and topology of the relevant internal space $M_6$.
In section \ref{sec_comparison} we 
discuss the anomaly inflow computation (including the $E_4 X_8$ contribution) and
we present in detail the main relation \eqref{main_equation},
giving the explicit expressions for $I_6^{\rm v,t}$ and $I_6^{\rm flip}$. 
Section \ref{sec_wrappedM2} is devoted to the match between the charges and
multiplicities of the flip fields entering $I_6^{\rm flip}$, and those of 
operators originating from M2-branes wrapping resolution 2-cycles in $M_6$.
In section \ref{sec_genus_one} we focus on the case in which the Riemann
surface is a torus, establishing a precise correspondence with D3-brane theories
dual to $AdS_5 \times Y^{p,q}$, for $k=2$. We also consider some explicit Lagrangian
examples with $k=2,3$. 
In section \ref{sec_central_charges} we use $a$-maximization to 
compute conformal and flavor central charges from $- I_6^{\rm inflow} - I_6^{\rm v,t}$
and establish various properties of these quantities.
We conclude with an outlook in section \ref{outlook}.
Several appendices collect useful material and detailed derivations.

\section{Review of the eleven dimensional flux setups} \label{sec_review_flux}

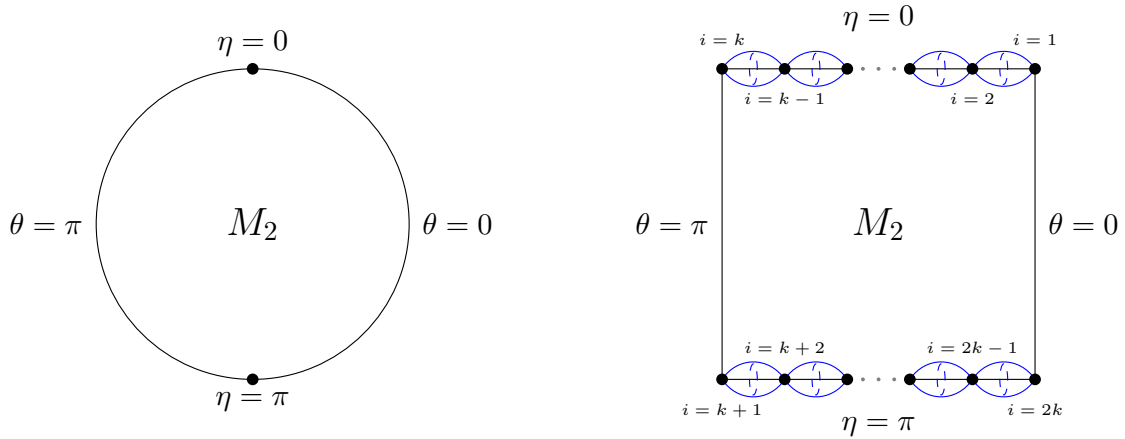
\begin{figure}[t!]
	\centering
		\resizebox{\textwidth}{!}{
		\begin{tikzpicture}
			\draw (2,2) circle (2);
			\fill (2,0) circle (0.075) node[below=1] {$\eta = \pi$};
			\fill (2,4) circle (0.075) node[above=1] {$\eta = 0$};
			\fill (4,2) circle (0) node[right=1] {$\theta = 0$};
			\fill (0,2) circle (0) node[left=1] {$\theta = \pi$};
			\node (S4Zk) at (2,2) {\Large $M_2$};
			\draw (8,0) -- (9.6,0);
			\fill[black!50] (9.8,0) circle (0.025);
			\fill[black!50] (10,0) circle (0.025) node[color=black,below=10] {$\eta = \pi$};
			\fill[black!50] (10.2,0) circle (0.025);
			\draw (10.4,0) -- (12,0);
			\draw (8,4) -- (9.6,4);
			\fill[black!50] (9.8,4) circle (0.025);
			\fill[black!50] (10,4) circle (0.025) node[color=black,above=10] {$\eta = 0$};
			\fill[black!50] (10.2,4) circle (0.025);
			\draw (10.4,4) -- (12,4);
			\draw (8,0) -- node[left=1] {$\theta = \pi$} (8,4);
			\draw (12,0) -- node[right=1] {$\theta = 0$} (12,4);
			\node (M4) at (10,2) {\Large $M_2$};
			\draw[color=blue] (8,0) .. controls (8.2,0.3) and (8.6,0.3) .. (8.8,0);
			\draw[color=blue] (8,0) .. controls (8.2,-0.3) and (8.6,-0.3) .. (8.8,0);
			\draw[dashed,color=blue] (8.4,0) ellipse (0.05 and 0.2);
			\draw[color=blue] (8.8,0) .. controls (9,0.3) and (9.4,0.3) .. (9.6,0);
			\draw[color=blue] (8.8,0) .. controls (9,-0.3) and (9.4,-0.3) .. (9.6,0);
			\draw[dashed,color=blue] (9.2,0) ellipse (0.05 and 0.2);
			\draw[color=blue] (10.4,0) .. controls (10.6,0.3) and (11,0.3) .. (11.2,0);
			\draw[color=blue] (10.4,0) .. controls (10.6,-0.3) and (11,-0.3) .. (11.2,0);
			\draw[dashed,color=blue] (10.8,0) ellipse (0.05 and 0.2);
			\draw[color=blue] (11.2,0) .. controls (11.4,0.3) and (11.8,0.3) .. (12,0);
			\draw[color=blue] (11.2,0) .. controls (11.4,-0.3) and (11.8,-0.3) .. (12,0);
			\draw[dashed,color=blue] (11.6,0) ellipse (0.05 and 0.2);
			\draw[color=blue] (8,4) .. controls (8.2,4.3) and (8.6,4.3) .. (8.8,4);
			\draw[color=blue] (8,4) .. controls (8.2,3.7) and (8.6,3.7) .. (8.8,4);
			\draw[dashed,color=blue] (8.4,4) ellipse (0.05 and 0.2);
			\draw[color=blue] (8.8,4) .. controls (9,4.3) and (9.4,4.3) .. (9.6,4);
			\draw[color=blue] (8.8,4) .. controls (9,3.7) and (9.4,3.7) .. (9.6,4);
			\draw[dashed,color=blue] (9.2,4) ellipse (0.05 and 0.2);
			\draw[color=blue] (10.4,4) .. controls (10.6,4.3) and (11,4.3) .. (11.2,4);
			\draw[color=blue] (10.4,4) .. controls (10.6,3.7) and (11,3.7) .. (11.2,4);
			\draw[dashed,color=blue] (10.8,4) ellipse (0.05 and 0.2);
			\draw[color=blue] (11.2,4) .. controls (11.4,4.3) and (11.8,4.3) .. (12,4);
			\draw[color=blue] (11.2,4) .. controls (11.4,3.7) and (11.8,3.7) .. (12,4);
			\draw[dashed,color=blue] (11.6,4) ellipse (0.05 and 0.2);
			\fill (8,0) circle (0.075) node[below=5] {\tiny $i=k+1$};
			\fill (8.8,0) circle (0.075) node[above=5] {\tiny $i=k+2$};
			\fill (9.6,0) circle (0.075);
			\fill (10.4,0) circle (0.075);
			\fill (11.2,0) circle (0.075) node[above=5] {\tiny $i=2k-1$};
			\fill (12,0) circle (0.075) node[below=5] {\tiny $i=2k$};
			\fill (8,4) circle (0.075) node[above=5] {\tiny $i=k$};
			\fill (8.8,4) circle (0.075) node[below=5] {\tiny $i=k-1$};
			\fill (9.6,4) circle (0.075);
			\fill (10.4,4) circle (0.075);
			\fill (11.2,4) circle (0.075) node[below=5] {\tiny $i=2$};
			\fill (12,4) circle (0.075) node[above=5] {\tiny $i=1$};
		\end{tikzpicture}
		}
	\caption{Illustration of the topology of the unresolved $S^4/ \mathbb{Z}_k$ space ({\it left}) and resolved $M_4$ space ({\it right}), with the $\psi$ and $\varphi$ angles suppressed, taken from \cite{Bah:2021brs}. The circle $S^1_\psi$ vanishes along the entire boundary $\partial M_2$, while $S^1_\varphi$ vanishes only at the $2k$  monopoles, labeled by the index $i$. The blue bubbles represent the resolution 2-cycles connected by unit-charge Kaluza-Klein monopoles.}
	\label{base_space_illustration}
\end{figure}

In this section we summarize the basic features of the internal space $M_6$ in the putative 11d flux backgrounds of relevance for the 4d field theories of $\mathcal{S}_k$. For a more detailed account of the geometry and homology of $M_6$, we refer the reader to \cite{Bah:2021brs}. $M_6$ is characterized by the fibration
\begin{equation}\label{M6_fibration}
	M_4 \hookrightarrow M_6 \rightarrow \Sigma_g \, ,
\end{equation}
where $\Sigma_g$ is a Riemann surface of genus $g$, and $M_4$ is the manifold obtained by resolving the fixed points of the orbifold $S^4/\mathbb{Z}_k$ via a blow-up procedure,
\begin{equation}
	M_4 = [S^4/\mathbb{Z}_k]_\mathrm{resolved} \, .
\end{equation}
This $M_4$ is locally a multi-center Gibbons-Hawking space, with $k-1$ 2-cycles separated by $k$ (unit-charge) Kaluza-Klein monopoles aligned along a common axis. It admits two $\mathrm{U}(1)$ isometries, and can be expressed in turn as a fibration
\begin{equation}
	S^1_\varphi \hookrightarrow M_4 \rightarrow S^1_\psi \times M_2 \, ,
\end{equation}
where $M_2$ is a compact 2d space. Schematically, the metric on $M_4$ can be cast in the form
\begin{equation}
	\begin{gathered}
		ds^2(M_4) = ds^2(M_2) + R_\psi^2(\eta,\theta) d\psi^2 + R_\varphi^2(\eta,\theta) D\varphi^2 \, ,\label{M4_metric}\\
		D\varphi = d\varphi - L(\eta,\theta) d\psi \, ,
	\end{gathered}
\end{equation}
where the angular coordinates $\eta$ and $\theta$ span the 2d space $M_2$. The $S^1_\psi$ circle shrinks everywhere on the boundary $\partial M_2$, at $\eta,\theta = 0,\pi$. Before the blow-up, the orbifold fixed points are labeled by $\eta=0, \pi$, which we refer to here as the north and south poles, respectively. After the blow-up, each pole is replaced by a chain of $k-1$ resolution 2-cycles. The $k$ monopoles in the north carry Kaluza-Klein charge $+1$, while the $k$ monopoles in the south carry charge $-1$, with the relative sign accounting for the opposite orientations relative to $M_2$ at $\eta = 0,\pi$.  There is a $\mathrm{U}(1)$ gauge symmetry associated with each resolution 2-cycle, so there is an overall $\mathrm{U}(1)^{k-1}$ symmetry in both the north and the south. The topology of the $S^4 /\mathbb{Z}_k$ and its resolved counterpart $M_4$ are illustrated in figure \ref{base_space_illustration}. 

The function $L(\eta, \theta)$ is piecewise constant on $\partial M_2$, with its difference across a given monopole measuring that monopole's Kaluza-Klein charge.
Labeling the resolution 2-cycles in the north by $i=1, \dots, k-1$ and those in the south by $i=k+1, \dots, 2k-1$, we have explicitly
\begin{equation}
	\ell_i \equiv L(t_i < t < t_{i+1}) = \begin{cases} \displaystyle i - \frac{k}{2} & \mathrm{for} \quad i=1, \dots, k ,\\[2ex] \displaystyle \frac{3k}{2} - i & \mathrm{for} \quad i= k + 1, \dots, 2k \, ,\end{cases}\label{l_i}
\end{equation}
where $t$ is a periodic coordinate parameterizing the boundary $\partial M_2$.

In the full internal space $M_6$, twisting of $M_4$ over the Riemann surface introduces a $\mathrm{U}(1)$ connection over $\Sigma_g$ to the form $d\psi$. $\mathcal{N} = 1$ supersymmetry is preserved specifically by a topological twist \cite{Bah:2011vv,Bah:2012dg}
\begin{equation}\label{top_twist}
	d\psi \to D\psi = d\psi - 2 \pi \chi A_\Sigma \, .
\end{equation}
The quantity $\chi= 2(1-g)$ is the Euler characteristic of the genus-$g$ Riemann surface, with volume form $V_\Sigma$ normalized as $\int_{\Sigma_g} V_\Sigma= 1$, and $A_\Sigma$ is the local antiderivative of $V_\Sigma$. This topological twist leads to nontrivial relations in the homology of $M_6$. Consider first the 2-cycles in $M_6$. There are two types: the boundary 2-cycles in $M_4$, and the Riemann surface $\Sigma_g$ itself, at the position of each of the monopoles. We thus have $4k$ total 2-cycles,
\begin{equation}\label{two_cycles}
	\mathcal{C}_2^{\Sigma,i} = \Sigma_g|_{t=t_i} \, , \qquad \mathcal{C}_2^i = [t_i,t_{i+1}] \times S^1_\varphi \, ,
\end{equation}
one of each type for every $i=1 , \dots, 2k$. However, as described in \cite{Bah:2021brs}, only $2k-1$ of these 2-cycles are independent. The situation is analogous for the 4-cycles. The region $M_4$ constitutes one 4-cycle in the full $M_6$ space, as do the $2k$ pairings of the boundary 2-cycles with $\Sigma_g$,
\begin{equation}
	\mathcal{C}_{4}^\mathrm{C} = M_4 \, , \qquad\mathcal{C}_{4}^{i} = \mathcal{C}_2^i \times \Sigma_g = [t_i,t_{i+1}] \times S^1_\varphi \times \Sigma_g \, .\label{four_cycles}
\end{equation}
The topological twist \eqref{top_twist} trivializes certain linear combinations of these 4-cycles, however,
\begin{equation}
	\sum_{i=1}^{2k} \mathcal{C}_{4}^i = \chi \mathcal{C}_{4}^{\mathrm{C}} \, , \qquad \sum_{i=1}^{2k} \ell_i \, \mathcal{C}_{4}^i = 0 \, ,\label{four_cycles_sum_rules}
\end{equation}
leaving $2k-1$ independent 4-cycles. It is convenient to adopt a complete basis of 4-cycles $\mathcal{C}_{4,\alpha}$ consisting of the $M_4$ bulk 4-cycle, the $k-1$ 4-cycles in the north, and the $k-1$ 4-cycles in the south,
\begin{equation}\label{natural_four_cycles}
\mathcal{C}_{4,\alpha=C}=\mathcal C_{4}^\mathrm{C} \, , \quad \mathcal{C}_{4, \alpha=\mathrm{N}_{i=1, \dots, k-1}}=\mathcal{C}_{4}^{i=1, \dots, k-1}, \quad  \mathcal{C}_{4, \alpha=\mathrm{S}_{i=1, \dots, k-1}}=\mathcal{C}_{4}^{i=2k-1, \dots, k+1} \, .
\end{equation}
The corresponding basis of $2k-1$ 2-cycles $\mathcal{C}_2^\alpha$ can be taken to be Poincar\'{e}-dual to these 4-cycles. Quantization of the M-theory 4-form flux $G_4$ associates each 4-cycle with an integer, 
\begin{equation}
	N = \int_{\mathcal{C}_{4}^{\mathrm{C}}} \frac{G_4}{2\pi} \, , \qquad N_i = \int_{\mathcal{C}_{4}^{i}} \frac{G_4}{2\pi} \, ,
\end{equation}
subject to linear constraints inherited from \eqref{four_cycles_sum_rules}, namely,
\begin{equation}
	\sum_{i=1}^{2k} N_i = \chi N \, , \qquad \sum_{i=1}^{2k} \ell_i N_i = 0\, .\label{flux_quanta_sum_rules}
\end{equation}
In the basis \eqref{natural_four_cycles}, we have $2k-1$ independent flux quanta,
\begin{equation}
N_{\mathrm{N}_{i=1, \dots,k-1}}=N_{i=1, \dots, k-1} \, , \quad N_{\mathrm{S}_{i=1, \dots,k-1}}=N_{i=2k-1, \dots, k+1} \, , \quad N_\mathrm{C}=N \, .
\end{equation}
All told, the space $M_6$ is characterized by the integer parameters $k$, $\chi$, $N$, $N_{\mathrm{N}_i}$, $N_{\mathrm{S}_i}$.

 
\section{\boldmath Anomaly polynomials in class $\mathcal{S}_k$ from inflow} \label{sec_comparison}

In this section we argue that the inflow anomaly polynomial
 for wrapped M5-branes probing
a resolved $\mathbb C^2/\mathbb Z_k$ singularity is to be identified
with the anomaly polynomial of a class $\mathcal S_k$ theory,
obtained from reduction of the parent 6d (1,0) SCFT on a smooth Riemann
surface with non-trivial $\SU(k)^2$ flavor fluxes.
The identification holds up to the contribution of a suitable collection of free
fields, which we discuss in detail.

\subsection{Integrated anomaly polynomial from six dimensions} \label{sec_integrated}

Here we review briefly the
integration of the 6d 8-form anomaly polynomial $I_8^{\rm SCFT}$
on a smooth genus-$g$ Riemann surface, with a non-trivial topological twist and flavor fluxes \cite{Bah:2017gph}.
Let us stress that $I_8^{\rm SCFT}$ denotes the anomaly polynonomial
of the \emph{interacting} 6d (1,0) SCFT realized by a stack of M5-branes
probing a $\mathbb C^2/\mathbb Z_k$ singularity \cite{Ohmori:2014kda},
without the inclusion of decoupled sectors,
such as the center-of-mass mode of the stack.

The R-symmetry   of the parent 6d (1,0) theory is
$\SU(2)_R$. In the reduction to four dimensions,
the Chern root of the $\SU(2)_R$ bundle
is shifted to implement the topological twist that preserves 4d $\mathcal N =1$
supersymmetry,
\beq 
c_2(\SU(2)_R) = - \Big[ c_1(R') - \frac \chi 2 \, V_\Sigma  \Big]^2 \ .
\eeq
The label $R'$ on the 4d background Chern class $c_1(R')$
is a reminder that this is a reference R-symmetry, which does not generically 
coincide with the 4d $\cN = 1$ superconformal R-symmetry $R_{\cN = 1}$.

For generic $N$, $k$, the 6d SCFT admits a $\U(1)_s \times \SU(k)_b \times \SU(k)_c$
flavor symmetry. The Chern roots of the $\SU(k)_b$, $\SU(k)_c$ bundles are denoted by
$b_i$, $c_i$ ($i=1,\dots, k$), respectively, and they are subject to the constraints
$\sum_{i=1}^k b_i = 0 =\sum_{i=1}^k c_i$. 
The reduction to four dimensions is performed with the following
Chern root shifts,
\beq
c_1(\U(1)_s) = c_1(t) \ , \qquad b_i= c_1(\beta_i)  + N_{\beta_i} \, V_\Sigma
\ , \qquad c_i= c_1(\gamma_i)  + N_{\gamma_i} \, V_\Sigma \ .
\eeq
Notice in particular that, for simplicity, in this work we do not turn
on a flavor flux for the $\U(1)_s$ symmetry.
The quantities $c_1(t)$, $c_1(\beta_i)$, $c_1(\gamma_i)$
are the first Chern classes of background fields for 4d $\U(1)$ global symmetries.
We observe that $c_1(\beta_i)$, $c_1(\gamma_i)$, as well as the flavor fluxes
$N_{\beta_i}$, $N_{\gamma_i}$, are constrained quantities,
\beq
\sum_{i=1}^k c_1(\beta_i) = \sum_{i=1}^k c_1(\gamma_i) 
= 0\, , \qquad \sum_{i=1}^k N_{\beta_i} = \sum_{i=1}^k N_{\gamma_i}  = 0 \ .
\eeq
We find it convenient to adopt the following parametrizations of the above
constraints,
\begin{align} \label{constraint_param}
N_{\beta_i} &= N_{\widetilde \beta_i} - N_{\widetilde \beta_{i-1}} \ , &
c_1(\beta_i) &= c_1(\widetilde \beta_i) - c_1(\widetilde \beta_{i-1}) \ , \nn \\
N_{\gamma_i} &= N_{\widetilde \gamma_i} - N_{\widetilde \gamma_{i-1}} \ , &
c_1(\gamma_i) &= c_1(\widetilde \gamma_i) - c_1(\widetilde \gamma_{i-1}) \ , & i = 1,\dots,k \ ,
\end{align}
with the conventions $N_{\widetilde \beta_0} \coloneqq0$, $N_{\widetilde \beta_k} \coloneqq0$,
$c_1(\widetilde \beta_0)\coloneqq 0$,
$c_1(\widetilde \beta_k)\coloneqq 0$,
$N_{\widetilde \gamma_0} \coloneqq0$, $N_{\widetilde \gamma_k} \coloneqq0$,
$c_1(\widetilde \gamma_0)\coloneqq 0$,
$c_1(\widetilde \gamma_k)\coloneqq 0$.

We are now in a position to quote the result of the integration
of the anomaly polynomial $I_8^{\rm SCFT}$ 
of the parent 6d (1,0) SCFT
on the Riemann surface $\Sigma_g$,
\begin{align}
	\int_{\Sigma_g} I_8^{\rm SCFT} & = -\frac{\big[k^2 (N^2 + N - 1) + 2\big] \chi (N-1)}{12} \, c_1^3(R') + \frac{k^2 \chi N (N^2 - 1)}{12} \, c_1(R') \, c_1^2(t)\nonumber\\
	& \phantom{=\ } - \frac{1}{2} \big[k N (N-1) c_1^2(R') - k N^2 c_1^2(t)\big] \sum_{i=1}^k \big[N_{\beta_i} c_1(\beta_i) + N_{\gamma_i} c_1(\gamma_i)\big]\nonumber\\
	& \phantom{=\ } + \frac{k \chi N^2 (N-1)}{4} \, c_1(R') \sum_{i=1}^k \big[c_1^2(\beta_i) + c_1^2(\gamma_i)\big]\nonumber\\
	& \phantom{=\ } + \frac{k N^2}{2} \sum_{i=1}^k \big[N_{\beta_i} c_1(t) \, c_1^2(\beta_i) - N_{\gamma_i} c_1(t) \, c_1^2(\gamma_i)\big]  \\
	& \phantom{=\ } + \frac{k N^3}{6} \sum_{i=1}^k \big[N_{\beta_i} c_1^3(\beta_i) + N_{\gamma_i} c_1^3(\gamma_i)\big] + \frac{N^2 (N-1)}{2} \sum_{i=1}^k c_1^2(\beta_i) \sum_{j=1}^k N_{\beta_j} c_1(\beta_j)\nonumber\\
	& \phantom{=\ } + \frac{N^2 (N-1)}{2} \sum_{i=1}^k c_1^2(\gamma_i) \sum_{j=1}^k N_{\gamma_j} c_1(\gamma_j)\nonumber\\
	& \phantom{=\ } + \frac{N^2}{2} \Bigg[\sum_{i=1}^k c_1^2(\beta_i) \sum_{j=1}^k N_{\gamma_j} c_1(\gamma_j) + \sum_i c_1^2(\gamma_i) \sum_{j=1}^k N_{\beta_j} c_1(\beta_j)\Bigg]\nonumber\\
	& \phantom{=\ } - \frac{(k^2-2) \chi (N-1)}{48} \, c_1(R') \, p_1(T_4) - \frac{k N}{24} \sum_{i=1}^k \big[N_{\beta_i} c_1(\beta_i) + N_{\gamma_i} c_1(\gamma_i)\big] \, p_1(T_4) \, \nn .\label{BHMRTZ_I6_expression}
\end{align}

\subsection{Anomaly inflow  from eleven dimensions}

The inflow anomaly polynomial for a stack of M5-branes probing a background
associated with the internal geometry $M_6$ 
and background flux configuration $\overline G_4$
is given by \cite{Bah:2019rgq}
\beq
- I_6^{\rm inflow} = \int_{M_6} \bigg[  \frac 16 \, E_4^3 + E_4 \wedge X_8 \bigg] \ , 
\qquad
X_8 = \frac{1}{192} \, \bigg[ p_1(TM_{11})^2 - 4 \, p_2(TM_{11}) \bigg]  \ .
\eeq
The closed and gauge-invariant 4-form $E_4$ is constructed from $\overline G_4$
by including the external gauge fields associated with the isometries  
and the non-trivial cohomology classes of $M_6$. The 8-form characteristic class $X_8$
is built with the Pontryagin classes $p_i(TM_{11})$ of the   tangent bundle of the 11d spacetime.

\begin{figure}[ht!]
\centering
\includegraphics[width = 14.5 cm]{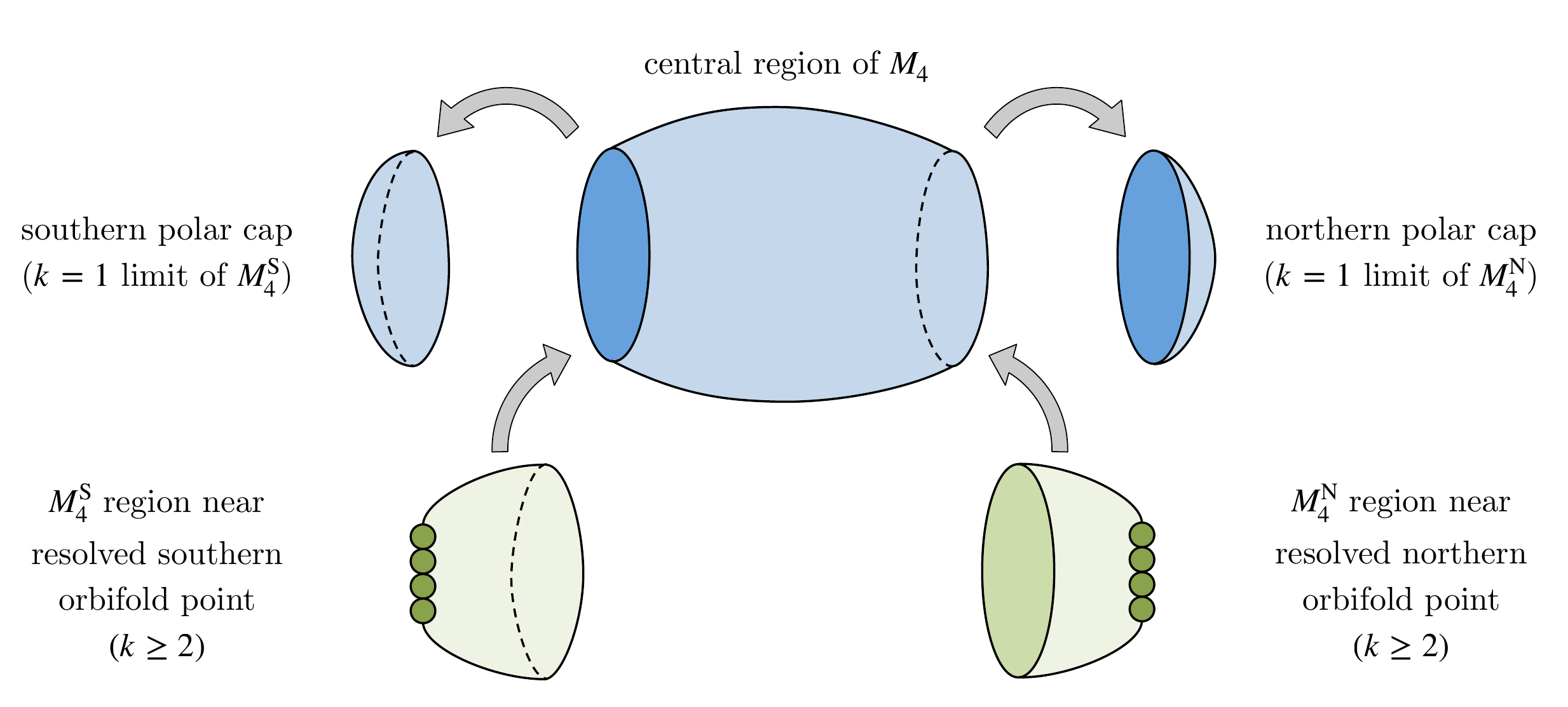}
\caption{
Pictorial depiction of the strategy used to evaluate the $E_4 X_8$ contribution to the inflow
anomaly polynomial. In order to obtain the resolved orbifold $M_4$, we consider the central
``cylindrical'' region of $S^4$ away from the poles, and we glue in
the resolved orbifolds at each pole. These include resolution 2-cycles, depicted as 
small green circles in the figure. The central contribution is obtained by taking a full 
$S^4$  and removing small polar caps. The contribution of the latter is computed
by taking the limit $k=1$ in the contribution of the resolved orbifolds.
}
\label{fig_collage}
\end{figure}

The computation of the inflow anomaly polynomial
for the flux setups reviewed in section \ref{sec_review_flux}
is discussed in detail in \cite{Bah:2021brs} for general $k$,
where the
contribution of the $E_4^3$ term was studied. The contribution of the $E_4 \, X_8$
term for general $k$ is derived in appendix \ref{app_E4X8}.
We have a completely explicit expression for 
$- I_6^{\rm inflow}$ in terms of $k$, $N$, $\chi$, the flux parameters $N_{\mathrm N_i}$,
$N_{\mathrm S_i}$, the external field strengths   $F_2^\psi$,  $F_2^\varphi$,
 $F_2^{\mathrm N_i}$,  $F_2^{\mathrm S_i}$, and the first Pontryagin 
 class $p_1(TW_4)$ of the external spacetime.
Since this expression is quite complicated, however, we refrain
from reproducing it in the main text. The $E_4^3$ contribution to  
$- I_6^{\rm inflow}$
is given in appendix \ref{app_E4cube}, while the result for the $E_4 X_8$ part
can be found in appendix \ref{app_E4X8}.

Without going into the technical details of the $E_4 X_8$ computation,
it is interesting to comment on the general strategy we follow,
which is depicted pictorially in figure \ref{fig_collage}. 
The main idea is to organize the contributions to $E_4 X_8$ into
a bulk part, originating from the central region of $S^4$ away from the poles,
and the parts associated to the resolved orbifold points at the north and south poles of the $S^4$. 
Operationally, we obtain an expression for the resolved orbifolds at the poles
for generic $k$. If we specialize to $k=1$, i.e.~no orbifolding, this contribution is identified
with the contribution of the polar caps of $S^4$, i.e.~small neighborhoods
of the two poles. If we take the full $S^4$ contribution, and we subtract these polar caps for $k=1$,
we get the contribution of the central ``cylindrical'' region. Finally, we glue back in
the resolved orbifolds with the appropriate value $k >1$, in order to get the final
desired result.

In concluding, we observe that the $E_4^3$ contribution was computed
in \cite{Bah:2021brs} using a different strategy, but is nonetheless compatible
with this cut-and-glue picture.
Both for $E_4 X_8$, and for $E_4^3$, the contribution associated to  the central region of $M_4$ 
is equal to the inflow anomaly polynomial for a 4d $\mathcal N = 1$ SCFT
originating from the 6d $(2,0)$ SCFT of type $A_{N-1}$ with twist parameters $p$, $q$
satisfying $p=q$ \cite{Bah:2012dg}.

\subsection{Matching the two sides:   decoupled modes and flip fields}

Let us now discuss the relation between
$\int_{\Sigma_g} I_8^{\rm SCFT}$ and $-I_6^{\rm inflow}$.
To this end, it is convenient to introduce the following notation,
\begin{align} \label{free_fermion}
I_6^{\rm free}(q_{R'} ; q_t ;  q_{\widetilde \beta_i}  ; q_{\widetilde \gamma_i}) &= \frac 16\,  \Big[
q_{R'} \, c_1(R') + q_t \, c_1(t)
+ \sum_{i=1}^{k-1} q_{\widetilde \beta_i} \, c_1(\widetilde \beta_i)
+ \sum_{i=1}^{k-1} q_{\widetilde \gamma_i} \, c_1(\widetilde \gamma_i)
\Big]^3
  \\
& \phantom{=\ } - \frac{1}{24} \, p_1(TW_4) \, \Big[
q_{R'} \, c_1(R') + q_t \, c_1(t)
+ \sum_{i=1}^{k-1} q_{\widetilde \beta_i} \, c_1(\widetilde \beta_i)
+ \sum_{i=1}^{k-1} q_{\widetilde \gamma_i} \, c_1(\widetilde \gamma_i)
\Big] \ . \nonumber
\end{align}
This quantity is the anomaly polynomial of a free, positive-chirality Weyl fermion
in four dimensions, with prescribed charges $q_{R'}$, $q_t$, $q_{\widetilde \beta_i}$,
$q_{\widetilde \gamma_i}$ under the 4d  $\U(1)$ symmetries
$\U(1)_{R'}$, $\U(1)_t$, $\U(1)_{\widetilde \beta_i}$, $\U(1)_{\widetilde \gamma_i}$,
in the notation of section \ref{sec_integrated}.
The quantities $c_1(\widetilde \beta_i)$, $c_1(\widetilde \gamma_i)$ are the
unconstrained Chern roots defined by \eqref{constraint_param}.

We can write the difference between
$\int_{\Sigma_g} I_8^{\rm SCFT}$ and $-I_6^{\rm inflow}$ in 
terms of a collection of free fermions, with anomaly given by \eqref{free_fermion}
for appropriate charges. More precisely, we find
\beq  \label{comparison_eq}  
-I_6^{\rm inflow} - I_6^\text{v,t} = \int_{\Sigma_g} I_8^{\rm SCFT}  - I_6^{\rm flip} \ ,
\eeq
where $I_6^\text{v,t}$ and $I_6^{\rm flip}$ are given by
\begin{align}
I_6^\text{v,t} & = \frac \chi 2 \, (k-2) \, I_6^{\rm free} \big (q_{R'} = 1; \, q_t = 0; \, q_{\widetilde \beta_i} = 0;
\, q_{\widetilde \gamma_i}  = 0 \big)
\nn \\
& \phantom{=\ }+ \frac \chi 2 \, \sum_{1 \le a < b \le k} 
I_6^{\rm free} \big (q_{R'} = 1; \, q_t = 0;  \, q_{\widetilde \beta_i} = N \, L_{a,b}^i;  \, 
q_{\widetilde \gamma_i} = 0 \big)
\nn \\
& \phantom{=\ } +  \frac \chi 2 \, \sum_{1 \le a < b \le k}  I_6^{\rm free} \big( q_{R'} = 1; \, q_t  = 0;  \, 
q_{\widetilde \beta_i} =  0;  \, 
q_{\widetilde \gamma_i} = 
N \,  L_{a,b}^i  \big)
  \ , \label{dec_from6d} \\
I_6^{\rm flip} & =  \sum_{1\le a < b \le k}   
m_{a,b}^{(\beta)} \, I_6^{\rm free}  \big(q_{R'} = 1; \, q_t = 0; \, q_{\widetilde \beta_i} = N \, L_{a,b}^i; \, 
q_{\widetilde \gamma_i} =  0 \big)
\nn \\
& \phantom{=\ } +   \sum_{1\le a < b \le k}    m_{a,b}^{(\gamma)} \, I_6^{\rm free} \big( q_{R'} = 1; \, 
q_t = 0; \,
q_{\widetilde \beta_i} =  0;  \,
q_{\widetilde \gamma_i} =   N \, L_{a,b}^i  \big ) \ .  \label{dec_and_flip}
\end{align}
Some remarks on our notation are in order.
The quantities $L_{a,b}^i$ can be identified with the components
of     the positive roots of the Lie algebra $\mathfrak{su}(k)$,
\beq
L_{a,b}^i = L_a^i - L_b^i \ , \quad L_a^i = \delta_{a,i} - \delta_{a,i+1} \ , \quad 
1 \le a < b \le k \ , \quad i =1,\dots,k-1 \ .
\eeq
The multiplicity factors $m_{a,b}^{(\beta)}$, $m_{a,b}^{(\gamma)}$
are given 
in terms of the unconstrained flavor fluxes $N_{\widetilde \beta_i}$,
$N_{\widetilde \gamma_i}$ introduced in \eqref{constraint_param} by the following
expressions,
\beq \label{multiplicity}
m^{(\beta)}_{a,b} =  N_{\widetilde \beta_a} 
- N_{\widetilde \beta_{a-1}}
- N_{\widetilde \beta_b}
+ N_{\widetilde \beta_{b-1}} \ , \quad
m^{(\gamma)}_{a,b} =  N_{\widetilde \gamma_a} 
- N_{\widetilde \gamma_{a-1}}
- N_{\widetilde \gamma_b}
+ N_{\widetilde \gamma_{b-1}} \ .
\eeq
The relation \eqref{comparison_eq} holds
provided that we make the following
identifications among quantities 
related to  anomaly inflow from eleven dimensions, and quantities
related to integration from six dimensions,
\begin{align} \label{dictionary}
\frac{F_2^\psi}{2\pi} & = 2 \, c_1(R') \ , &
\frac{F_2^\varphi}{2\pi} & = - k \, c_1(t) \ , \nn \\
\frac{F_2^{\mathrm N_i}}{  2\pi } & = c_1(\widetilde \beta_i) + \frac{2}{N \, \chi} \, N_{\widetilde \beta_i} \, c_1(R') \ ,
&
\frac{F_2^{\mathrm S_{k-i}}}{  2\pi } & = - c_1(\widetilde \gamma_i) - \frac{2}{N \, \chi} \, N_{\widetilde \gamma_i} \, c_1(R') \ , \nn \\
N_{\mathrm N_i} & = \sum_{j=1}^{k-1} (\mathsf A_{k-1})_{ij} \, N_{\widetilde \beta_j} \ , &
N_{\mathrm S_{k-i}} & =  \sum_{j=1}^{k-1} \, (\mathsf A_{k-1})_{ij} \, N_{\widetilde \gamma_j} \ .
\end{align}
The quantities $(\mathsf A_{k-1})_{ij}$ are the entries of the Cartan
matrix of $\mathfrak{su}(k)$,
\beq
(\mathsf A_{k-1})_{ij} = 2 \, \delta_{i,j} - \delta_{i,j+1} - \delta_{i+1,j} \ , \qquad i,j =1,\dots,k-1 \ .
\eeq
Notice the $k-i$ label on southern quantities, as opposed to the $i$
label on northern quantities.

\paragraph{\boldmath Interpretation of $I_6^\text{v,t}$}
The free-field contributions in $I_6^\text{v,t}$ are interpreted as originating from the 
reduction on $\Sigma_g$ of free fields in six dimensions,
as suggested by the fact that they are proportional to the Euler
characteristic $\chi$. 
The 8-form anomaly polynomials of a free 6d (1,0) tensor multiplet,
and a free 6d (1,0) vector multiplet, are readily computed and integrated on the Riemann surface,
with the result
\begin{align}
I_8^\text{(1,0) tens}  = - I_8^\text{(1,0) vec} =  - \frac \chi 2 \,  I_6^{\rm free} \big (q_{R'} = 1; \, q_t = 0; \, q_{\widetilde \beta_i} = 0;
\, q_{\widetilde \gamma_i}  = 0 \big) \ .
\end{align}
This observation suggests us to interpret 
 the first line of
$I_6^\text{v,t}$ in \eqref{dec_from6d} as a contribution of
one     tensor and $k-1$ vectors.
The former is associated with the center of mass
of the M5-brane stack, while the latter is thought of
as the Cartan generators of $SU(k)$.
By a similar token, we interpret the other terms in \eqref{dec_from6d}
as coming from the reduction of 6d W-bosons of $\SU(k)$,
whose charges are indeed given by the roots of $\mathfrak{su}(k)$.

\paragraph{Interpretation of $I_6^{\rm flip}$.}

While the free-field contributions in $I_6^\text{v,t}$ have a 6d interpretation,
those in  
$I_6^{\rm flip}$
are interpreted in  four-dimensional terms. More precisely, we identify  $I_6^{\rm flip}$
with the anomaly polynomial of a collection of flip fields.
Here, by flip field we mean a 4d $\cN  =1$ chiral multiplet $\phi$ that
couples to an operator $\cO$ of the interacting 4d SCFT with a superpotential coupling of the form
\beq \label{W_flip}
W_{\rm flip} = \phi \, \cO \ .
\eeq
The field $\phi$ has canonical kinetic terms and, if it were not for \eqref{W_flip},
would be completely decoupled from the 4d SCFT. 
Since the superpotential has R-charge 2 and is neutral under other global symmetries,
the coupling \eqref{W_flip} implies
\beq
q_{R'}[\Psi_\phi] =   (2 - q_{R'}[\cO]   ) -1 \ , \qquad 
q_{X}[\Psi_ \phi]  = -q_X[\cO] \ ,
\eeq
where 
$\Psi_\phi$ denotes the fermion in the chiral multiplet $\phi$, and
$q_X$ is a shorthand notation for $q_t$, $q_{\widetilde \beta_i}$, $q_{\widetilde \gamma_i}$.
The charges $q_{R'}[\Psi_\phi]$,  $q_X[\Psi_ \phi]$
are the charges given  $I_6^{\rm flip}$ in \eqref{dec_and_flip}. They are readily translated 
into charges of the operators $\cO$ that get flipped,
\begin{align} \label{flipped_charges}
&\text{flipped op.'s $\cO^{(\beta)}_{a,b}$:} & 
q_{R'}[\cO^{(\beta)}_{a,b}] & = 0 \ , &
q_{\widetilde \beta_i}[\cO^{(\beta)}_{a,b}] & = -N \, L^i_{a,b} \ , &
q_{\widetilde \gamma_i}[\cO^{(\beta)}_{a,b}]& = 0 \ ,   \\
&\text{flipped op.'s $\cO^{(\gamma)}_{a,b}$:} & 
q_{R'}[\cO^{(\gamma)}_{a,b}] & = 0 \ , &
q_{\widetilde \beta_i} [\cO^{(\gamma)}_{a,b}]  & = 0 \ , &
q_{\widetilde \gamma_i} [\cO^{(\gamma)}_{a,b}]  & = -N \, L^i_{a,b} \ . \nn
\end{align}
Notice how all flipped operators have zero charge under $\U(1)_{R'}$.

In section \ref{sec_wrappedM2} we match the charges of the flipped operators 
\eqref{flipped_charges} with charges of wrapped M2-brane states, and we comment further
on the physical mechanism underlying the flipping of these operators.

\paragraph{Flavor fluxes versus resolution fluxes: a geometric picture.}\label{basis_interp}

Let us motivate the identification \eqref{dictionary} by considering the geometric interpretation of the flux quanta $N_{\tilde{\beta}_i}, N_{\tilde{\gamma}_i}$ 
in the 6d setup. Prior to compactification over the Riemann surface, the internal space of the 6d theories is $S^4/\mathbb{Z}_k$, and has two orbifold fixed points which can be resolved by the set of 2-cycles $\mathcal{C}_2^i$ defined in \eqref{two_cycles}. Associated with each 2-cycle is a $\mathrm{U}(1)$ flavor symmetry. The only 4-cycle in such a setup is the (unresolved) $S^4/\mathbb{Z}_k$ that is analogous to $\mathcal{C}_4^\mathrm{C}$; there is no analogue of the 4-cycles $\mathcal{C}_{4}^{i}$ defined in \eqref{four_cycles}. As a consequence, the flux quanta $N_{\tilde{\beta}_i}, N_{\tilde{\gamma}_i}$ assigned to the $\mathrm{U}(1)_{\tilde{\beta}_i}, \mathrm{U}(1)_{\tilde{\gamma}_i}$ flavor symmetries in the compactification are naturally associated with $\mathcal{C}_2^i$ but not $\mathcal{C}_{4}^{i}$. From the perspective of the expansion 
\begin{equation}
	E_4 \supset N_\alpha (\Omega_4^\alpha)^\mathrm{eq} + N \, \frac{F_2^\alpha}{2\pi} \wedge (\omega_{2,\alpha})^\mathrm{eq} \, ,
\end{equation}
flux quanta are intrinsically paired to 4-cycles \cite{Bah:2021brs}. So we conjecture that $N_{\tilde{\beta}_i}, N_{\tilde{\gamma}_i}$ are really flux quanta with respect to the 4-cycles Poincar\'{e}-dual to the resolution 2-cycles $\mathcal{C}_2^i$ (after reduction on the Riemann surface). In contrast, the flux quanta $N_{\mathrm{N}_i}, N_{\mathrm{S}_i}$ were defined with respect to the 4-cycles $\mathcal{C}_{4,\mathrm{N_i}},\mathcal{C}_{4,\mathrm{S_i}}$ described in \eqref{natural_four_cycles}. Direct comparison 
between $\int_{\Sigma_g} I_8^{\rm SCFT}$ and $-I_6^{\rm inflow}$ therefore requires that we find the transformation matrices relating these two distinct bases of homology classes. As worked out in appendix \ref{app_change_of_basis}, these transformation matrices turn out to be block diagonal, with each of the two nontrivial blocks proportional to $\mathsf{A}_{k-1}$, the Cartan matrix of $\mathfrak{su}(k)$. One may heuristically interpret this as a remnant of the enhanced $\mathrm{SU}(k)$ symmetry present at each orbifold fixed point before being resolved into $k-1$ 2-cycles with $\mathrm{U}(1)$ symmetries. Indeed, the identification in \eqref{dictionary} is precisely the change of basis we have described, with an additional sign change for the southern flux quanta to preserve their positivity.

Next, we argue that the factor of $1/k$ in the identification \eqref{dictionary} of the field  strength $F_2^\varphi$ with the Chern root $c_1(t)$ can be attributed to the different periodicities of $t$ and $\varphi$. More specifically, the periodicity of the angular variable $t$ in the 6d theories is reduced from $2\pi$ to $2\pi/k$ by the $\mathbb{Z}_k$ quotient, but the periodicity of $\varphi$ in the inflow computation directly from 11d is simply $2\pi$ by construction. In the former case, we gauge the $\mathrm{U}(1)_t$ isometry as $(Dt)^\mathrm{g} = Dt - A_1^t$, while in the latter case we gauge the $\mathrm{U}(1)_\varphi$ isometry as $(D\varphi)^\mathrm{g} = D\varphi + A_1^\varphi$. This motivates the identifications,
\begin{equation}
	t = \frac{1}{k} \, \varphi \, , \quad A_1^t = -\frac{1}{k} \, A_1^\varphi \, , \quad F_2^t = -\frac{1}{k} \, F_2^\varphi \, .
\end{equation}
Lastly, the factor of 2 appearing in the identification \eqref{dictionary} between $F_2^\psi/2\pi$ and $c_1(R')$ is needed to ensure an appropriate R-charge normalization \cite{Gauntlett:2006ai}. 


 
\section{Wrapped M2-branes and flip fields}\label{sec_wrappedM2}

\paragraph{Operators from M2-branes wrapping calibrated 2-cycles.}

The calibration conditions for probe M2-branes wrapping 2-cycles in the internal space
$M_6$ were derived in \cite{Gauntlett:2006ai}, where they were also analyzed for the $k=2$ GMSW solutions.
The homology classes of the 
resolution 2-cycles of $M_4$ (at a generic point on the Riemann surface)
admit a calibrated representative. 
In the notation of section \ref{sec_review_flux}, these homology classes are the  $\mathcal C_2^i$ in \eqref{two_cycles}
with labels $i=1$ and $i=3$.
Wrapping M2-brane probes on such 2-cycles
yields BPS particle states in the external spacetime.

A direct analysis of the calibration conditions for $k \ge 3$ is challenging.
Indeed, we do not have an explicit $AdS_5$ solution in which the internal space
has the topology and flux quanta of $M_6$. Solutions describing the flux background probed by the M5-brane stack are not known, either. For these reasons, we refrain from studying the calibration conditions
for M2-brane probes for $k \ge 3$, and we make the following working assumption:
the homology classes of the resolution 2-cycles of $M_4$ admit calibrated representative 2-cycles.
We wrap probe M2-branes on such cycles, getting BPS states in the external spacetime.

In what follows, we study for generic $k \ge 2$ the charges and multiplicities of such states.
Crucially, these data do not depend on the (putative, for $k >2$) explicit calibrated representative, but only on its homology class.

\paragraph{Charges of  M2-branes operators.}

The method for the computation of the charges of wrapped M2-brane operators
is explained in \cite{Gauntlett:2006ai}. The key point is to use the standard coupling of the 11d
3-form $C_3$ to the worldvolume of the M2-brane probe. The desired charges are extracted
by integrating this 3d coupling along the compact directions of the 2-cycle wrapped by the M2-brane,
thus extracting terms linear in the external gauge fields. 

In order to determine how the external gauge fields enter the 11d 3-form $C_3$,
we resort to the 4-form $E_4$ used in the anomaly inflow computation.
More precisely,
we     proceed as follows:
extract the terms in $E_4$ that are linear in the external 2-form field strengths
$F_2^\psi$, $F_2^\varphi$, $F_2^{\mathrm N_i}$, $F_2^{\mathrm S_i}$, and 
cast these terms as a total derivative of a 3-form $\delta C_3$ that is linear
in the corresponding external gauge fields $A_1^\psi$, $A_1^\varphi$, $A_1^{\mathrm N_i}$,
$A_1^{\mathrm S_i}$. All the relevant charges of interest are then obtained
by integrating $\delta C_3$ on the 2-cycle wrapped by the M2-brane.

After these general preliminaries, we are in a position to outline the
results we obtain for the setups of interest in this work.
The 3-form $\delta C_3$ derived from $E_4$ takes the form
\beq
\delta C_3 \supset (\omega_{2,\beta})^{\rm g} \wedge A_1^\beta + (\Omega_{2,I})^{\rm g} \wedge A_1^I \ ,
\eeq
where the label $\beta$ runs over all independent 2-cycles in $M_6$,
while $I$ is a collective label for the isometry directions $\psi$, $\varphi$.
The 2-forms $(\omega_{2,\beta})^{\rm g}$ are obtained from the harmonic 2-forms on $M_6$
by means of the replacements  
$d\psi \rightarrow d\psi + A_1^\psi$, $d\varphi \rightarrow d\varphi + A_1^\varphi$.
The 2-forms $(\Omega_{2,I})^{\rm g}$ are derived in the process of constructing the
equivariant completion of the harmonic 4-forms on $M_6$. We refer the interested reader
to \cite{Bah:2021brs}, where the explicit expressions of 
$(\omega_{2,\beta})^{\rm g}$ and $(\Omega_{2,I})^{\rm g}$ can be found.

%

The   charges under the $\mathrm{U}(1)$ field strengths for flavor symmetries and isometries,  respectively $F_2^\beta$ and $F_2^I$, can then by computed from the integrals
\begin{equation}
(Q_\beta)^i= N \int_{\mathcal{C}_2^i} (\omega_{2,\beta})^\text{g} , \qquad (Q_I)^i=\int_{\mathcal{C}_2^i} (\Omega_{2,I})^{\rm g}.
\end{equation}
The results can be summarized in the following table,
up to an overall orientation choice,
\begingroup
\renewcommand{\arraystretch}{1.5}
\beq
\begin{array}{r | ccccc}
&  Q_{\psi}  &  Q_\varphi  &  Q_{\mathrm N_j}  &  Q_{\mathrm S_j}  & \text{mult.}  \\ \hline
 i=1, \dots, k-1   \;\;  & 
 \chi^{-1} N_{\mathrm N_i}   & 0  &
 -N \, (\mathsf A_{k-1})_{ij}  & 0 &     N_{\mathrm N_i} 
\\
 i=2k-1, \dots, k+1  \;\; & 
  \chi^{-1} N_{\mathrm S_{i}}  & 0 & 0 &  +N \, (\mathsf A_{k-1})_{ij}  & 
 N_{\mathrm S_{i}} 
\end{array}
\eeq
\endgroup
where we have written the flavor charges in terms of the
entries of the Cartan matrix of $\mathfrak{su}(k)$.
Notice that the charges
are given in a basis that corresponds to the external
gauge fields in the anomaly inflow computation.
Making use of \eqref{dictionary}, they are readily translated
to the basis used in the integration of the anomaly
polynomial of the 6d SCFT,
\begingroup
\renewcommand{\arraystretch}{1.5}
\beq \label{brane_charges}
\begin{array}{r | ccccc}
&  Q_{R'}  &  Q_t  &  Q_{\widetilde \beta_j}  &  Q_{\widetilde \gamma _j}  & \text{mult.}  \\ \hline
 i=1, \dots, k-1   \;\;  & 
0   & 0  &
 -N \, (\mathsf A_{k-1})_{ij}    & 0 &     N_{\mathrm N_i} 
\\
 i=2k-1, \dots, k+1  \;\; & 
0  & 0 & 0 &  -N \, (\mathsf A_{k-1})_{ij}  & 
 N_{\mathrm S_{i}} 
\end{array}
\eeq
\endgroup
In the above tables we have also reported the
multiplicity of the M2-brane operators,
which is derived as explained below.

\paragraph{Multiplicities of  M2-branes operators.}
The multiplicities, or degeneracies, of the states originating from an M2-brane 
probe wrapping a 2-cycle can be derived using a Landau-level argument
\cite{Gaiotto:2009gz}.
The probe M2-branes of interest in this work
wrap a resolution 2-cycle  $\mathcal C_2^i$ in $M_4$,
and sit at a point on $\Sigma_g$.
Moreover, a non-trivial $G_4$-flux is turned on along the 4-cycle that results
from combining the resolution 2-cycle $\mathcal C_2^i$ and the Riemann surface $\Sigma_g$.
As a result, the M2-brane behaves like a point particle on $\Sigma_g$,
in the presence of a non-zero magnetic field.
This quantum-mechanical system exhibits a well-known Landau degeneracy of states,
which is simply given by the total magnetic flux, measured in units of the minimal
magnetic flux that can be turned on. This quantized magnetic flux   is indeed identified with the
quantized $G_4$-flux through the 4-cycle obtained combining $\Sigma_g$ and $\mathcal C_2^i$. 
In conclusion, the expected degeneracies of the wrapped M2-brane operators of interest
are simply given by the values of the corresponding $G_4$-flux.

\paragraph{Comparison to flip fields.}

The anomaly polynomial $I_6^{\rm flip}$ is interpreted as a sum over
\emph{flip} fields.
The charges of the corresponding \emph{flipped} operators
are collected in \eqref{flipped_charges}, for a generic pair of labels
$a<b$, $a,b=1,\dots,k$.
These pairs 
correspond to all positive roots of $\mathfrak{su}(k)$. Those pairs $a<b$ with
$b = a+1$ correspond to the \emph{simple} roots of $\mathfrak{su}(k)$.
Based on standard intuition regarding 
M-theory on a $\mathbb Z_k$ singularity,
we expect the M2-branes wrapping the $(k-1)$ resolution 2-cycles at the north and south poles
to correspond to the simple roots of $\mathfrak{su}(k)_{\mathrm N}$, $\mathfrak{su}(k)_{\mathrm S}$.
(The full set of positive roots corresponds to BPS bound states of the M2-brane states
corresponding to the simple roots.)
Now, the identity
\beq
L_{a,a+1}^i = (\mathsf A_{k-1})_{ai} \ , \qquad a ,i = 1,\dots  , k-1 \ ,
\eeq
shows 
a match between the charges of the flipped operators in \eqref{flipped_charges},
and the charges of M2-branes wrapping resolution 2-cycles in \eqref{brane_charges}.

The multiplicities of the flip fields for generic $a<b$ are reported in \eqref{multiplicity}.
Let us specialize to $b = a+1$, and combine \eqref{multiplicity} with the relations \eqref{dictionary}
between the flavor fluxes and the resolution fluxes. We obtain
\beq
m^{(\beta)}_{a,a+1}  
= N_{\mathrm N_a} \ , \qquad
m^{(\gamma)}_{a,a+1}  
= N_{\mathrm S_a}  \ , \qquad a = 1,\dots,k-1 \ .
\eeq
We thus verify that, for pairs with $b=a+1$, corresponding to simple roots,
the multiplicities that enter $I_6^{\rm flip}$
coincide with the degeneracies given by the Landau-level argument
discussed above.

\paragraph{M-theory origin of the flipping mechanism.} 

Recall that the flip fields enter the 4d theory via the schematic 
superpotential coupling 
\beq \label{flip2}
W_{\rm flip} =\phi \,\cO \ .
\eeq
Based on the previous analysis, we identify the flipped operators
$\cO$ with the operators originating from M2-branes wrapping the
resolution 2-cycles in the internal space.
For generic $N$, the coupling \eqref{flip2} is irrelevant.
It is therefore important in the UV, where its effect is to project out
the operators $\cO$. This fits with the fact that the 6d parent SCFT admits
no blow-up modes for the $\mathbb C^2/\mathbb Z_k$ singularity \cite{Witten:1995zh,Ganor:1996pc,Hanany:1996ie,Brunner:1997gk,Blum:1997fw,Blum:1997mm,Intriligator:1997dh,Brunner:1997gf,Hanany:1997gh}.
In contrast, \eqref{flip2} is irrelevant in the deep IR, where the flip fields
$\phi$ become   free fields and decouple.
The operators $\cO$ are thus effectively reintroduced in the 4d theory.

 
\section{The case of genus one} \label{sec_genus_one}
In this section, we consider several explicit examples at genus one in order to gather evidence for a series of connected claims:
\begin{itemize}
\item Since $I_6^\text{v,t}$ drops out of \eqref{comparison_eq} when $\chi=0$, in this case the quantity $-I_6^\text{inflow}$ provides direct access to the anomaly polynomial of the corresponding 4d SCFT, 
\begin{equation}\label{SCFT_from_inflow_torus}
-I_6^\text{inflow}= I_6^\text{SCFT}.
\end{equation}
\item The quantity we have called $I_6^{\rm flip}$ in \eqref{comparison_eq} is indeed the anomaly polynomial of the flip fields that appear in the field-theoretic construction
on tori with fluxes.
\item The difference between $\int_{\Sigma_g} I_8^{\rm SCFT}$ and $I_6^{\rm flip}$ as given by our formula \eqref{dec_and_flip} is indeed equal to the anomaly
polynomial of the \emph{interacting} SCFT in four dimensions,
\beq \label{SCFT_from_difference}
\int_{\Sigma_g} I_8^{\rm SCFT} - I_6^{\rm flip} = I_6^{\rm SCFT} \ .
\eeq
\end{itemize}
Note that both the $E^3$ contribution to the inflow anomaly polynomial recorded in appendix \ref{app_E4cube} and the $E_4X_8$ contribution derived in appendix \ref{app_E4X8} are expressed in a form which only applies for higher-genus Riemann surfaces, i.e.~$\chi<0$. The genus-one result can be obtained either by first using \eqref{dictionary} and subsequently fixing $\chi=0$, or by repeating an analogous anomaly inflow computation as in \cite{Bah:2021brs} with cohomology class representatives chosen consistently with $\chi=0$ from the beginning (which we describe in appendix \ref{app_torus_inflow}). We have verified that both paths produce the same result, 
\begin{align}
	-I_6^\text{inflow}(\chi=0) & = - \frac{1}{2} \big[k N (N-1) c_1^2(R') - k N^2 c_1^2(t)\big] \sum_{i=1}^k \big[N_{\beta_i} c_1(\beta_i) + N_{\gamma_i} c_1(\gamma_i)\big]\nonumber\\
	& \phantom{=\ } + \frac{k N^2}{2} \sum_{i=1}^k \big[N_{\beta_i} c_1(t) \, c_1^2(\beta_i) - N_{\gamma_i} c_1(t) \, c_1^2(\gamma_i)\big]  \nn \\
		& \phantom{=\ } + \frac{N^2 (N-1)}{2} \sum_{i=1}^k c_1^2(\beta_i) \sum_{j=1}^k N_{\beta_j} c_1(\beta_j)\nonumber\\
	& \phantom{=\ } + \frac{N^2 (N-1)}{2} \sum_{i=1}^k c_1^2(\gamma_i) \sum_{j=1}^k N_{\gamma_j} c_1(\gamma_j)\nonumber\\
	& \phantom{=\ } + \frac{N^2}{2} \Bigg[\sum_{i=1}^k c_1^2(\beta_i) \sum_{j=1}^k N_{\gamma_j} c_1(\gamma_j) + \sum_i c_1^2(\gamma_i) \sum_{j=1}^k N_{\beta_j} c_1(\beta_j)\Bigg]\nonumber\\
	& \phantom{=\ } + \frac{k N^3}{6} \sum_{i=1}^k \big[N_{\beta_i} c_1^3(\beta_i) + N_{\gamma_i} c_1^3(\gamma_i)\big]\nonumber\\
	& \phantom{=\ } - \frac{k N}{24} \sum_{i=1}^k \big[N_{\beta_i} c_1(\beta_i) + N_{\gamma_i} c_1(\gamma_i)\big] \, p_1(T_4) \, \nn \\
	& \phantom{=\ } - \sum_{1\le a < b \le k}   m_{a,b}^{(\beta)} \, I_6^{\rm free}  \big(q_{R'} = 1; \, q_t = 0; \, q_{\widetilde \beta_i} = N \, L_{a,b}^i; \, 
q_{\widetilde \gamma_i} =  0 \big)
\nn \\
& \phantom{=\ } - \sum_{1\le a < b \le k}    m_{a,b}^{(\gamma)} \, I_6^{\rm free} \big( q_{R'} = 1; \, 
q_t = 0; \,
q_{\widetilde \beta_i} =  0;  \,
q_{\widetilde \gamma_i} =   N \, L_{a,b}^i  \big ) \, .\label{torus_expression}
\end{align}
We explore how the expression \eqref{torus_expression} can be used to verify the claims highlighted above in various examples.

\subsection[The $Y^{p,q}$ quiver theories from inflow]{\boldmath The $Y^{p,q}$ quiver theories from inflow}

Consider the case $k=2$ for genus one. 
Here we find that the equation \eqref{torus_expression} reproduces the anomaly polynomial of the $Y^{p,q}$ quiver gauge theories, which can be engineered on a stack of D3-branes at the tip of the Calabi-Yau cone over $Y^{p,q}$ \cite{Benvenuti2005}. 

The $Y^{p,q}$ are an infinite family of Sasaki-Einstein manifolds labeled by positive integers $p$ and $q$ with $0 \leq q \leq p$ \cite{Gauntlett:2004zh}. The holographic duals of the corresponding $AdS_5 \times Y^{p,q}$ solutions in Type IIB string theory were constructed in \cite{Benvenuti2005}, using an iterative procedure on the quiver for $Y^{p,p}$. The field content of this family of quiver gauge theories is summarized in table \ref{Ypq_fields}. The quiver associated with $Y^{p,q}$ has $2p$ gauge groups, represented diagrammatically by $2p$ nodes. All fields are in either a spin-$0$ or spin-$1/2$ representation of a global $\SU(2)$ symmetry. There are two additional global $\U(1)$'s, labeled here as $\text{U}(1)_B$ and $\text{U}(1)_F$. 

\begin{table}[t!]
	\centering
	\def\arraystretch{1.5}
	\begin{tabular}{|r || c|c|c|c|c|}
\hline
&    Degeneracy  & $\text{U}(1)_B$ & $\text{U}(1)_F$ & $\text{U}(1)_\varphi$ & $\text{U}(1)_{R_0}$ \\\hline \hline
$Y$ singlets & $p+q$ &$p-q$ & $-1$ &$0$ & $1$ \\ \hline
$Z$ singlets & $p-q$ &$p+q$ & $1$ &$0$ & $0$ \\ \hline
$U$ doublets & $p$ &$-p$ & $0$ & $\pm 1/2$ & $1/2$ \\ \hline
$V$ doublets & $q$ &$q$ & $1$ & $\pm 1/2$ & $1/2$ \\ \hline
	\end{tabular}
	\caption{Field content for the infinite family of $Y^{p,q}$ quiver gauge theories. See, e.g.~\cite{Benvenuti2005}.}
	\label{Ypq_fields}
\end{table}
The anomaly polynomial for general $Y^{p,q}$ quiver gauge theories can be computed directly from the field content and associated degeneracies, and is given by
\begin{align}
I_6^\text{SCFT}(Y_{pq}) &= \frac{N^2}{8}(p+q) \, c_1(R_0)^3
-\frac{1}{12}\left[4c_1(R_0)^2 -p \,\, p_1(TW_4) \right] c_1(R_0)
\nonumber\\
& \phantom{=\ } +\frac{N^2}{4}(2p-q)\, c_1(R_0)^2 c_1(F)+\frac{N^2}{4}(p^2-q^2)\, c_1(R_0)^2 c_1(B) 
\nonumber \\
& \phantom{=\ } -\frac{N^2}{8}(p+q)\, c_1(\varphi)^2c_1(R_0) +\frac{N^2q}{4}\, c_1(\varphi)^2c_1(F)-\frac{N^2}{4}(p^2-q^2)\, c_1(\varphi)^2c_1(B)
\nonumber\\
& \phantom{=\ } -\frac{N^2p}{2} \, \left[c_1(F)+p \, c_1(B)\right]^2 c_1(R_0)-\frac{N^2 p}{2}(p^2+pq-q^2) \, c_1(B)^2 c_1(R_0)
\nonumber
\\
& \phantom{=\ } +N^2 p^2 \left[c_1(F)+q \, c_1(B)\right]c_1(B) c_1(F) \, .
\end{align}
In virtue of a chain of dualities connecting the GMSW solution in 11d supergravity to the $AdS_5 \times Y^{p,q}$ solutions in Type IIB, the internal manifold $M_6$ in the corresponding inflow setup is defined by $k=2$ and $\chi =0$. The quantity $-I_6^\text{inflow}$ can be obtained for example from \eqref{torus_expression},
\begin{align}
-&I_6^\text{inflow}(k=2,\chi=0) =
\nn \\
& \,\,\, 2N^2 \left[ N_{\widetilde \beta_1} c_1(\widetilde \beta_1)+ N_{\widetilde \gamma_1} c_1(\widetilde \gamma_1) \right]\left(c_1(t)^2- c_1(R')^2\right)
\nonumber \\
&-4N^2\left[  N_{\widetilde \beta_1} c_1(\widetilde \beta_1)^2+ N_{\widetilde \gamma_1} c_1(\widetilde \gamma_1)^2 \right] c_1(R')
+2N^2\left[ N_{\widetilde \beta_1} c_1(\widetilde \beta_1)c_1(\widetilde \gamma_1)^2+ N_{\widetilde \gamma_1} c_1(\widetilde \gamma_1)c_1(\widetilde \beta_1)^2 \right] 
\nonumber \\
&-2N^2\left[ N_{\widetilde \beta_1}  c_1(\widetilde \beta_1)^3+ N_{\widetilde \gamma_1} c_1(\widetilde \gamma_1)^3 \right]-\frac{1}{12}(N_{\widetilde \beta_1}+N_{\widetilde \gamma_1})\left[4c_1(R')^2-p_1(TW_4)\right] c_1(R').
\end{align}
Under the field strength redefinitions,
\begin{equation}
	\begin{gathered}
		c_1(\widetilde \beta_1)=\frac{1}{2} \, c_1(F)-\frac{1}{2}(p-q)\, c_1(B)-\frac{1}{2}\, c_1(R_0) \, , \quad c_1(\widetilde \gamma_1)=-\frac{1}{2}\, c_1(F)-\frac{1}{2}(p+q)\, c_1(B) \, ,\nonumber\\
		c_1(R')=c_1(R_0) \, , \quad c_1(t)=-\frac{1}{k} \, c_1(\varphi) \, ,
	\end{gathered}
\end{equation}
and the identifications
\begin{equation}\label{Ypq_flux_identification}
N_{\widetilde \beta_1}=2(p+q), \qquad N_{\widetilde \gamma_1}=2(p-q)
\end{equation}
between the integers $p$, $q$ and the resolution flux quanta $N_{\widetilde \beta_1}$, $N_{\widetilde \gamma_1}$, we verify an exact match, $I_6^\text{SCFT}(Y_{pq})=-I_6^\text{inflow}$. This match supports our claim \eqref{SCFT_from_inflow_torus} that the topological and geometric data of $M_6$ fully characterize the anomaly polynomial of the corresponding (genus-one) 4d SCFT.


\subsection{More quiver theories and flip fields at genus one}
Next we revisit some explicit examples at genus one reported in \cite{Bah:2017gph} in order to provide further evidence for the interpretation of  $I_6^{\rm flip}$ in \eqref{comparison_eq} and the equality \eqref{SCFT_from_difference}. In these examples, we consider $N$ to be generic, but make the implicit assumption that $N$ is large enough to ensure that all the couplings between flip fields and baryons
are irrelevant. It would be interesting to consider in greater detail low values of $N$,
but we refrain from such analysis in this work.

\begin{figure}
\centering
\includegraphics[width = 15.2 cm]{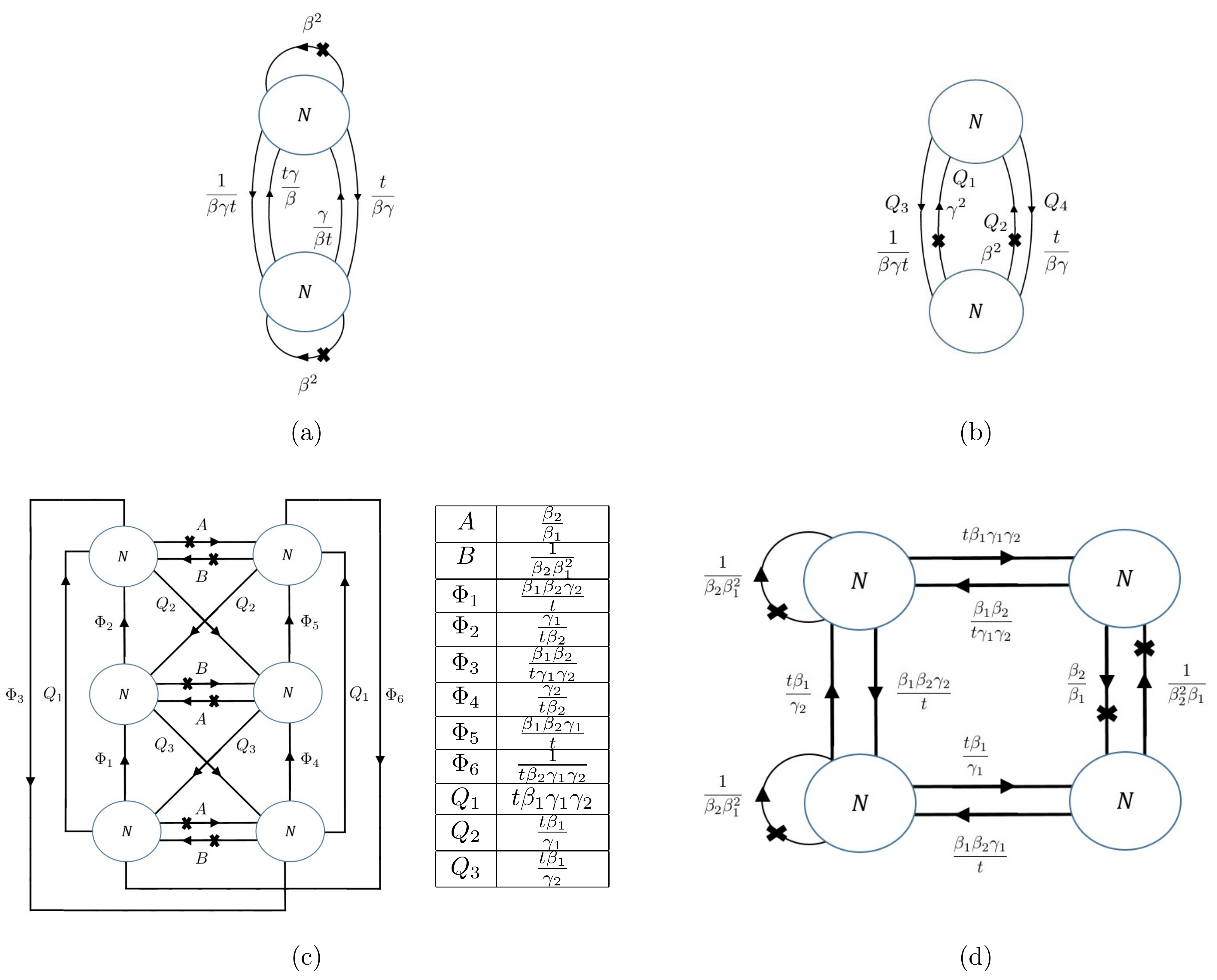}
\caption{
Explicit examples of quiver gauge theories of class $\mathcal S_k$
realized on a torus with fluxes, taken from \cite{Bah:2017gph}.
The gauge nodes are $\SU(N)$ groups. An arrow connecting two distinct nodes is a bifundamental
chiral multiplet. An arrow connecting a node to itself represents an adjoint-plus-singlet chiral multiplet. An ``X'' on an arrow signals that the baryon associated with that field
is flipped. The charges under the global, non-R-symmetries 
are given in terms of fugacities. The reference R-symmetry charges
for the various quivers are as follows. In quiver (a),
all fields (except the flip fields) have R-charge $2/3$.
In quiver (b), all fields (except the flip fields) have R-charge $1/2$.
In quiver (c), all fields (except the flip fields) have R-charge $2/3$.
In quiver (d), the bifundamentals have R-charge $1/2$ and the adjoint-plus-singlet's
have R-charge $1$.
For a description 
of the superpotential of these models, we refer the reader to \cite{Bah:2017gph}.
}
\label{fig_quivers}
\end{figure}

\paragraph{Example no.~1.} Consider the gauge theory with the quiver
depicted in figure 4 of \cite{Bah:2017gph}, reproduced for
convenience as quiver (a) in figure \ref{fig_quivers}. This theory corresponds to $k=2$. In our notation,
the flavor fluxes are
\beq \label{example1_fluxes}
N_{\widetilde \beta_1} = 1 \ , \qquad N_{\widetilde \gamma_1} = 0 \ .
\eeq
From the quiver  one extracts the  charges of all fields.
These charges are: the charge $\widehat q_{R}$ under a convenient reference R-symmetry
(given in the caption of figure \ref{fig_quivers});
the charges $\widehat q_t$, $\widehat q_\beta$, $\widehat q_\gamma$  extracted from the exponents
of the $t$, $\beta$, $\gamma$ fugacities in the quiver.
In order to compare the quiver data with our formulae, we need the map
between the charges $(\widehat q_R ; \widehat q_t ; \widehat q_\beta ; \widehat q_\gamma)$
of the quiver description, and the charges $(q_{R'} ; q_t ; q_{\widetilde \beta_1} ; q_{\widetilde \gamma_1} )$
used in this work,
\beq \label{example1_dictionary}
\widehat q_R = q_{R'} - \frac 13  \, q_{\widetilde \beta_1} \ , \qquad
\widehat q_t = q_t \ , \qquad
\widehat q_\beta = - q_{\widetilde \beta_1} \ , \qquad
\widehat q_{\gamma} = - q_{\widetilde \gamma_1} \ .
\eeq
The last two relations simply state how to relate our conventions for the flavor charges
to those of \cite{Bah:2017gph}. The first relation encodes how the reference R-symmetry
in the quiver compares to the reference R-symmetry $R'$ from six dimensions.

According to our general formula \eqref{dec_and_flip}, 
we have one species of  flip field  with  charges $(q_{R'} ; q_t ; q_{\widetilde \beta_1} ; q_{\widetilde \gamma_1}) = (1;0;2N;0)$ and multiplicity 2,
as computed from \eqref{multiplicity} using \eqref{example1_fluxes} (with labels $a = 1$, $b = 2$).
Making use of the dictionary \eqref{example1_dictionary},
the charges in the notation of the quiver are $(\widehat q_R ; \widehat q_t ; \widehat q_\beta ; \widehat q_\gamma) = (1 - \tfrac 23 N ; 0 ; -2N ; 0)$. These are indeed 
the correct charges for the (fermions in the chiral) fields that flip the baryons of the adjoints that carry an ``X'' in the quiver diagram.
 
Finally, using again  \eqref{example1_fluxes} and \eqref{example1_dictionary},
one can verify  \eqref{SCFT_from_difference}, where the anomaly of the SCFT is extracted from the quiver,
simply ignoring the flip fields. (The gauge singlets associated with the arrows that connect a node
to itself are kept, because they participate in relevant superpotential couplings.) Note that due to the restriction \eqref{example1_fluxes} on the flavor fluxes, there is a symmetry enhancement of the $\mathrm{U}(1)_{\widetilde \gamma_1}$ to an $\mathrm{SU}(2)$. This can be seen at the level of the anomaly polynomial in that $c_1(\widetilde \gamma_1)$  enters only quadratically, via the term
\begin{align}
I_6^\mathrm{SCFT} \supset 2N^2 c_1(\widetilde \beta_1)c_1(\widetilde \gamma_1)^2.
\end{align}
 
\paragraph{Example no.~2.} This example is the gauge theory with quiver
depicted in figure 5 of \cite{Bah:2017gph},
reproduced for
convenience as quiver (b) in figure \ref{fig_quivers}. This theory also corresponds to $k=2$. In our notation,
the flavor fluxes are
\beq \label{example2_fluxes}
N_{\widetilde \beta_1} = \frac  12 \ , \qquad N_{\widetilde \gamma_1} = \frac 12 \ .
\eeq
Once again, we need the dictionary between the charges 
$(\widehat q_R ; \widehat q_t ; \widehat q_\beta ; \widehat q_\gamma)$
of the quiver description, and the charges $(q_{R'} ; q_t ; q_{\widetilde \beta_1} ; q_{\widetilde \gamma_1} )$
used in this work,
\beq \label{example2_dictionary}
\widehat q_R = q_{R'} - \frac 14  \, q_{\widetilde \beta_1} - \frac 14  \, q_{\widetilde \beta_2} \ , \qquad
\widehat q_t = q_t \ , \qquad
\widehat q_\beta = - q_{\widetilde \beta_1} \ , \qquad
\widehat q_{\gamma} = - q_{\widetilde \gamma_1} \ .
\eeq
Our formula  \eqref{dec_and_flip}
predicts one species of flip fields  
with  charges $(q_{R'} ; q_t ; q_{\widetilde \beta_1} ; q_{\widetilde \gamma_1}) = (1;0;2N;0)$,
and one species with charges $(q_{R'} ; q_t ; q_{\widetilde \beta_1} ; q_{\widetilde \gamma_1}) = (1;0;0;2N)$.
According to \eqref{multiplicity} and \eqref{example2_fluxes}, both species have multiplicity 1.
In the notation of the quiver, the charges are
$(\widehat q_R ; \widehat q_t ; \widehat q_\beta ; \widehat q_\gamma) = (1- \tfrac 12 N ; 0 ; -2N ;0)$
and $(\widehat q_R ; \widehat q_t ; \widehat q_\beta ; \widehat q_\gamma) = (1- \tfrac 12 N ; 0 ; 0; -2N)$,
which indeed match with the charges of the   (fermions in the chiral) fields that flip the baryons
of the fields marked with an ``X'' in the quiver. We can verify \eqref{SCFT_from_difference} in this example as well, making use of \eqref{example2_fluxes} and \eqref{example2_dictionary}.

\paragraph{Example no.~3.} This example is the gauge theory with quiver
depicted in figure 6 of \cite{Bah:2017gph}, reproduced for
convenience as quiver (c) in figure \ref{fig_quivers}. This theory has $k=3$. In our notation,
the flavor fluxes are
\beq \label{example3_fluxes}
N_{\widetilde \beta_1 } = 2 \ , \qquad N_{\widetilde \beta_2} = 1 \ , \qquad
N_{\widetilde \gamma_1}  = 0 \ , \qquad
N_{\widetilde \gamma_2} = 0 \ .
\eeq
Let $(\widehat q_R; \widehat q_t ; \widehat q_{\beta_1} ; \widehat q_{\beta_2} ; \widehat q_{\gamma_1} ; \widehat q_{\gamma_2} )$ denote the charges extracted from the quiver (and its adjacent table). They are related to the charges $(q_{R'} ; q_t ; q_{\widetilde \beta_1} ; q_{\widetilde \beta_2} ; q_{\widetilde \gamma_1} ; q_{\widetilde \gamma_2} )$ used in this work via
\beq \label{example3_dictionary}
\widehat q_R = q_{R'} - \frac 49 \, q_{\widetilde \beta_1} - \frac  29 \, q_{\widetilde \beta_2}
+ \frac 19 \, q_t \ , \quad
\widehat q_t = q_t \ , \quad
\begin{array}{l}
\widehat q_{\beta_1} = q_{\widetilde \beta_1} + q_{\widetilde \beta_2} \\
\widehat q_{\beta_2} = q_{\widetilde \beta_2} 
\end{array} \ , \quad 
\begin{array}{l}
\widehat q_{\gamma_1} = q_{\widetilde \gamma_1} + q_{\widetilde \gamma_2} \\
\widehat q_{\gamma_2} = q_{\widetilde \gamma_2} 
\end{array} \ .
\eeq
As in the previous examples, using  \eqref{example3_fluxes}
and \eqref{example3_dictionary} we can verify that the charges and multiplicities
of flip fields given by our formulae match with the quiver data.
We can also verify \eqref{SCFT_from_difference}, where the SCFT anomaly is computed from the quiver,
simply ignoring all flip fields.

\paragraph{Example no.~4.} This example is the gauge theory with quiver
depicted in figure 7 of \cite{Bah:2017gph},
reproduced for
convenience as quiver (d) in figure \ref{fig_quivers}. This theory has $k=3$. In our notation,
the flavor fluxes are
\beq \label{example4_fluxes}
N_{\widetilde \beta_1 } = 1 \ , \qquad N_{\widetilde \beta_2} = 1 \ , \qquad
N_{\widetilde \gamma_1}  = 0 \ , \qquad
N_{\widetilde \gamma_2} = 0 \ .
\eeq
The dictionary between
$(\widehat q_R; \widehat q_t ; \widehat q_{\beta_1} ; \widehat q_{\beta_2} ; \widehat q_{\gamma_1} ; \widehat q_{\gamma_2} )$ and
$(q_{R'} ; q_t ; q_{\widetilde \beta_1} ; q_{\widetilde \beta_2} ; q_{\widetilde \gamma_1} ; q_{\widetilde \gamma_2} )$ in this example is
\beq \label{example4_dictionary}
\widehat q_R = q_{R'} - \frac 12 \, q_{\widetilde \beta_1} - \frac  12 \, q_{\widetilde \beta_2}
\ , \quad
\widehat q_t = q_t \ , \quad
\begin{array}{l}
\widehat q_{\beta_1} = q_{\widetilde \beta_1} + q_{\widetilde \beta_2} \\
\widehat q_{\beta_2} = q_{\widetilde \beta_2} 
\end{array} \ , \quad 
\begin{array}{l}
\widehat q_{\gamma_1} = q_{\widetilde \gamma_1} + q_{\widetilde \gamma_2} \\
\widehat q_{\gamma_2} = q_{\widetilde \gamma_2} 
\end{array} \ .
\eeq
Once again, we have a match of charges and multiplicities of flip fields, and \eqref{SCFT_from_difference} can be verified.

 
\section{Central charges} \label{sec_central_charges} 

We continue our study of the putative 4d $\mathcal{N}=1$ SCFTs of interest in this paper by computing their central charges using $a$-maximization \cite{Intriligator:2003jj}. In this section we will first describe the computational complexity of this $a$-maximization problem. We then study several important properties of the resulting central charges. Some exact results are presented for a special family of $G_4$-flux configurations with sufficiently few independent parameters that the maximization problem can be solved analytically. Finally, we treat the general case with a perturbative analysis in the regime where the ratios $N_{\mathrm{N}_i}/|\chi|N$ and $N_{\mathrm{S}_i}/|\chi|N$ are small. Most of the discussion in this section implicitly assumes that we are working with a $g \geq 2$ Riemann surface, unless otherwise specified.

\subsection{Computational setup}\label{amax_comput_setup}

The anomaly polynomial of a 4d $\mathcal{N}=1$ SCFT contains
\begin{equation}
	I_6^\mathrm{SCFT} \supset \frac{\mathrm{tr} R^3}{6} \bigg(\frac{F_2^R}{2\pi}\bigg)^{\!3} - \frac{\mathrm{tr} R}{24} \frac{F_2^R}{2\pi} \, p_1(TW_4) \, ,
\end{equation}
where $R$ is the generator of the superconformal R-symmetry and $F_2^R$ its background field strength, and $p_1(TW_4)$ is the first Pontryagin class of the 4d worldvolume $W_4$. Its central charges are \cite{Anselmi:1997am}
\begin{equation}
	a = \frac{3}{32} (3 \mathrm{tr} R^3 - \mathrm{tr} R) \, , \quad c = \frac{1}{32} (9 \mathrm{tr} R^3 - 5 \mathrm{tr} R) \, .\label{a_c_definitions}
\end{equation}
In the presence of additional flavor symmetries, one may also compute the associated flavor central charges \cite{Anselmi:1997ys},
\begin{equation}
	b_G \delta^{ab} = -6 \mathrm{tr} (R \, T_G^a \, T_G^b) \, ,
\end{equation}
where $T_G^a$ are generators of the flavor symmetry $G$, with the normalization $\mathrm{tr} (T_G^a \, T_G^b) = \delta^{ab}/2$ in the fundamental representation.

As described before, we can access $I_6^\mathrm{SCFT}$ through either of the equalities in \eqref{main_equation}, repeated here for convenience,
\begin{equation}
	I_6^\mathrm{SCFT} = -I_6^\mathrm{inflow} - I_6^\mathrm{v,t} = \int_{\Sigma_g} I_8^\mathrm{SCFT} - I_6^\mathrm{flip} \, .
\end{equation}
For concreteness, suppose we restrict our attention to the first equality above, we can define a trial R-symmetry as
\begin{equation}
	R_\mathrm{trial} = 2 T_\psi + q^\varphi T_\varphi + \sum_{i=1}^{k-1} (q^{\mathrm{N}_i} T_{\mathrm{N}_i} + q^{\mathrm{S}_i} T_{\mathrm{S}_i}) \, .\label{trial_R_symmetry}
\end{equation}
The various $T_G$ are the generators of the $\mathrm{U}(1)_G$ flavor symmetries, and the factor of $2$ in front of $T_\psi$ is inserted to ensure it has the appropriately normalized R-charge \cite{Gauntlett:2006ai}. This is equivalent to carrying out the following replacements in the anomaly polynomial(s),
\begin{equation}
	F_2^\psi \to 2 F_2^R \, , \quad F_2^\varphi \to F_2^\varphi + q^\varphi F_2^R \, , \quad F_2^{\mathrm{N}_i} \to F_2^{\mathrm{N}_i} + q^{\mathrm{N}_i} F_2^R \, , \quad F_2^{\mathrm{S}_i} \to F_2^{\mathrm{S}_i} + q^{\mathrm{S}_i} F_2^R \, ,\label{anomaly_polynomial_field_strength_replacements}
\end{equation}
from which we can construct a trial central charge,
\begin{equation}
	a_\mathrm{trial}(\{q^G\}) = \frac{27}{16} \, \mathcal{A}_{RRR} + \frac{9}{4} \, \mathcal{A}_R \, ,\label{a_trial_definition}
\end{equation}
where $\mathcal{A}_{RRR}$ and $\mathcal{A}_R$ are respectively the coefficients of $(F_2^R/2\pi)^3$ and $F_2^R/2\pi$ in the polynomial $I_6^\mathrm{SCFT}(\{q^G\})$. The central charge $a$ corresponds to the local maximum of $a_\mathrm{trial}(\{q^G\})$ with respect to the $2k-1$ real parameters $q^G$. Such a solution is hereafter denoted by $\{q^G_\mathrm{max}\}$. Up to an overall normalization, the flavor central charge $b_G$ for a given flavor symmetry $G$ can be extracted by plugging $\{q^G_\mathrm{max}\}$ into $I_6^\mathrm{SCFT}(\{q^G\})$, and then reading off the coefficient $\mathcal{A}_{RGG}$ in front of $(F_2^R/2\pi) (F_2^G/2\pi)^2$. We can also similarly find $c$.

Analytically searching for a local maximum of $a_\mathrm{trial}(\{q^G\})$ amounts to performing a two-step process: first solving the quadratic equations $\partial a_\mathrm{trial} / \partial q^G = 0$ simultaneously for all $G$, and then identifying a solution $\{q^G_\mathrm{max}\}$ that yields a negative-definite Hessian matrix $[\partial^2 a_\mathrm{trial} / \partial q^{G_i} \partial q^{G_j}]_{\{q^G\} = \{q^G_\mathrm{max}\}}$. As far as the former is concerned, Bezout's theorem asserts that $a_\mathrm{trial}(\{q^G\})$ has at most $2^{2k-1}$ (or infinitely many) critical points. Hence, the computational complexity of the problem scales roughly as $4^k$ at large $k$. This presents a formidable challenge even for state-of-the-art algebraic solvers, in which case we have to resort to numerical methods. In fact, the only scenario where we are able to analytically solve the $a$-maximization problem for generic combinations of $\chi$, $N$, $N_{\mathrm{N}_i}$, $N_{\mathrm{S}_i}$ for $i = 1,\dots,k-1$ is $k = 2$, in which we are able to fully reproduce the result of \cite{Bah:2019vmq}. As we will discuss later in this section, there exists a family of theories with certain special flux configurations $\{N_{\mathrm{N}_i},N_{\mathrm{S}_i}\}$ that significantly reduce the number of independent extremization parameters $q^G$, thus rendering the $a$-maximization problem analytically solvable for $k > 2$ as well.

\subsection{Salient properties of the central charge}

The central charge $a$ thus obtained for the SCFTs via $a$-maximization exhibits a number of crucial properties which we collect in this subsection.

\paragraph{Uniqueness.}

For a given choice of $k$, $\chi$, $N$, $N_{\mathrm{N}_i}$, $N_{\mathrm{S}_i}$, if the central charge $a$ exists then it is necessarily unique. The proof of this statement proceeds as follows. Given the replacements made in \eqref{anomaly_polynomial_field_strength_replacements}, the trial central charge $a_\mathrm{trial}(\{q^G\})$ defined in \eqref{a_trial_definition} is a cubic polynomial of the $2k-1$ variables $q^G$ where $G = \varphi, \mathrm{N}_i, \mathrm{S}_i$ for $i = 1,\dots,k-1$. Suppose a local maximum of $a_\mathrm{trial}$ exists at some point $\{q^G_\mathrm{max}\} \in \mathbb{R}^{2k-1}$. We can project the $(2k-1)$-dimensional vector $\vec{q} = (q^{G_1},\dots,q^{G_{2k-1}})$ onto an arbitrary line $\mathbb{R}_p \subset \mathbb{R}^{2k-1}$ passing through $\{q^G_\mathrm{max}\}$. Restricted to this line, the trial central charge becomes a univariate cubic polynomial $a_\mathrm{trial}(q_p) = a_\mathrm{trial}(\{q^G\})|_{\mathbb{R}_p}$ where $\vec{q}_p = \mathrm{proj}_{\mathbb{R}_p}\vec{q}$.

A univariate cubic polynomial admits at most one local maximum, so by construction $\{q^G_\mathrm{max}\}$ is the unique local maximum of $a_\mathrm{trial}(q_p)$ along any $\mathbb{R}_p$. Since $\mathbb{R}_p$ can be chosen to connect $\{q^G_\mathrm{max}\}$ to any other point in the parameter space, along which $\{q^G_\mathrm{max}\}$ is always the unique local maximum, we conclude that $a_\mathrm{trial}$ has at most one local maximum at $\{q^G\} = \{q^G_\mathrm{max}\}$.\footnote{To the best of our knowledge, this result concerning the number of local maxima/minima of a multivariate cubic polynomial was first explicitly proven by \cite{10.1023/A:1024778309049}.}

\paragraph{\boldmath Large-$N$ scaling relation.}

Recall that the central charges of the SCFT are encoded by the anomaly polynomial $I_6^\mathrm{SCFT} = -I_6^\mathrm{inflow} - I_6^\mathrm{v,t}$. The former can be further decomposed into an $\mathcal{O}\big(N^3,N_{\mathrm{N}_i,\mathrm{S}_i}^3\big)$ contribution $I_6^\text{inflow,$E_4^3$}$ and an $\mathcal{O}(N,N_{\mathrm{N}_i,\mathrm{S}_i})$ contribution $I_6^\text{inflow,$E_4 X_8$}$. Although $I_6^\mathrm{v,t}$ is of $\mathcal{O}(N^3)$ according to \eqref{dec_from6d}, we have checked that it always contribute only at $\mathcal{O}(N)$ to the central charge $a$. Hence in the large-$N,N_{\mathrm{N}_i,\mathrm{S}_i}$ limit (or loosely, the large-$N$ limit), $a$ can be effectively determined by performing $a$-maximization directly on $-I_6^\text{inflow,$E_4^3$}$.

We observe that the large-$N$ inflow anomaly polynomial $I_6^\text{inflow,$E_4^3$}$, constructed in \cite{Bah:2021brs} and reviewed in appendix \ref{app_E4cube}, satisfies an interesting identity. Suppose we have two distinct setups with generally different Euler characteristics, $\chi_A$, $\chi_B$, of the ($g \geq 2$) Riemann surfaces, and they are wrapped by different numbers, $N_A$, $N_B$, of M5-branes, then the corresponding large-$N$ inflow anomaly polynomials follow
\begin{align}
	& \phantom{=\ \ } I_6^\text{inflow,$E_4^3$}(k,\chi_A,N_A,\{N_{\mathrm{N}_i}\},\{N_{\mathrm{S}_i}\},F_2^\psi,F_2^\varphi,\{F_2^{\mathrm{N}_i}\},\{F_2^{\mathrm{S}_i}\})\nonumber\\
	& = \frac{\chi_A N_A^3}{\chi_B N_B^3} \, I_6^\text{inflow,$E_4^3$}\bigg(k,\chi_B,N_B,\bigg\{\frac{\chi_B N_B}{\chi_A N_A} \, N_{\mathrm{N}_i}\bigg\},\bigg\{\frac{\chi_B N_B}{\chi_A N_A} \, N_{\mathrm{S}_i}\bigg\},F_2^\psi,F_2^\varphi,\{F_2^{\mathrm{N}_i}\},\{F_2^{\mathrm{S}_i}\}\bigg) \, .\label{I6_scaling_relation}
\end{align}
Consequently, we have an analogous scaling relation for the trial central charge,
\begin{align}
	& \phantom{=\ \ } a_\mathrm{trial}(k,\chi_A,N_A,\{N_{\mathrm{N}_i}\},\{N_{\mathrm{S}_i}\},q^\varphi,\{q^{\mathrm{N}_i}\},\{q^{\mathrm{S}_i}\})\nonumber\\
	& = \frac{\chi_A N_A^3}{\chi_B N_B^3} \, a_\mathrm{trial}\bigg(k,\chi_B,N_B,\bigg\{\frac{\chi_B N_B}{\chi_A N_A} \, N_{\mathrm{N}_i}\bigg\},\bigg\{\frac{\chi_B N_B}{\chi_A N_A} \, N_{\mathrm{S}_i}\bigg\},q^\varphi,\{q^{\mathrm{N}_i}\},\{q^{\mathrm{S}_i}\}\bigg) \, .
\end{align}
This motivates the definitions of ``reduced flux quanta,''
\begin{equation}
	n_{\mathrm{N}_i} = \frac{N_{\mathrm{N}_i}}{|\chi| N} \, , \qquad n_{\mathrm{S}_i} = \frac{N_{\mathrm{S}_i}}{|\chi| N} \, ,
\end{equation}
such that the central charge, which is the (unique) local maximum of $a_\mathrm{trial}$, obeys
\begin{equation}
	\frac{1}{\chi_A N_A^3} \, a(k,\chi_A,N_A,\{n_{\mathrm{N}_i}\},\{n_{\mathrm{S}_i}\}) = \frac{1}{\chi_B N_B^3} \, a(k,\chi_B,N_B,\{n_{\mathrm{N}_i}\},\{n_{\mathrm{S}_i}\})\label{a_scaling_relation}
\end{equation}
in the large-$N$ limit. It follows from \eqref{I6_scaling_relation} that the same scaling relation applies to the other central charges $c$ and $b_G$ as well.

\paragraph{Existence and flux positivity.}

As alluded to earlier, a central charge does not necessarily exist for an arbitrary combination of $k$, $\chi$, $N$, $N_{\mathrm{N}_i}$, $N_{\mathrm{S}_i}$. This is possible because while Bezout's theorem states that there can be at most $2^{2k-1}$ critical points of $a_\mathrm{trial}(\{q^G\})$, they may be all saddle points (possibly with a local minimum swapped in). Moreover, given a choice of orientation for the internal space $M_6$, we define the charge (or more precisely the number of M5-branes) $N$ to be positive under this orientation. Similarly, we can define the rest of the flux quanta using the same choice of orientation for the relevant cycles.\footnote{Alternatively, flipping the orientation for $M_6$ amounts to flipping the signs of all the flux quanta, including $N$.} For a supersymmetric theory, we expect all of these flux quanta to have the same sign as $N$. It is therefore important for us to examine the range of parameters within which our construction admits a central charge.

\begin{figure}[t!]
	\begin{subfigure}{0.24\textwidth}
		\centering
		\includegraphics[width=\textwidth]{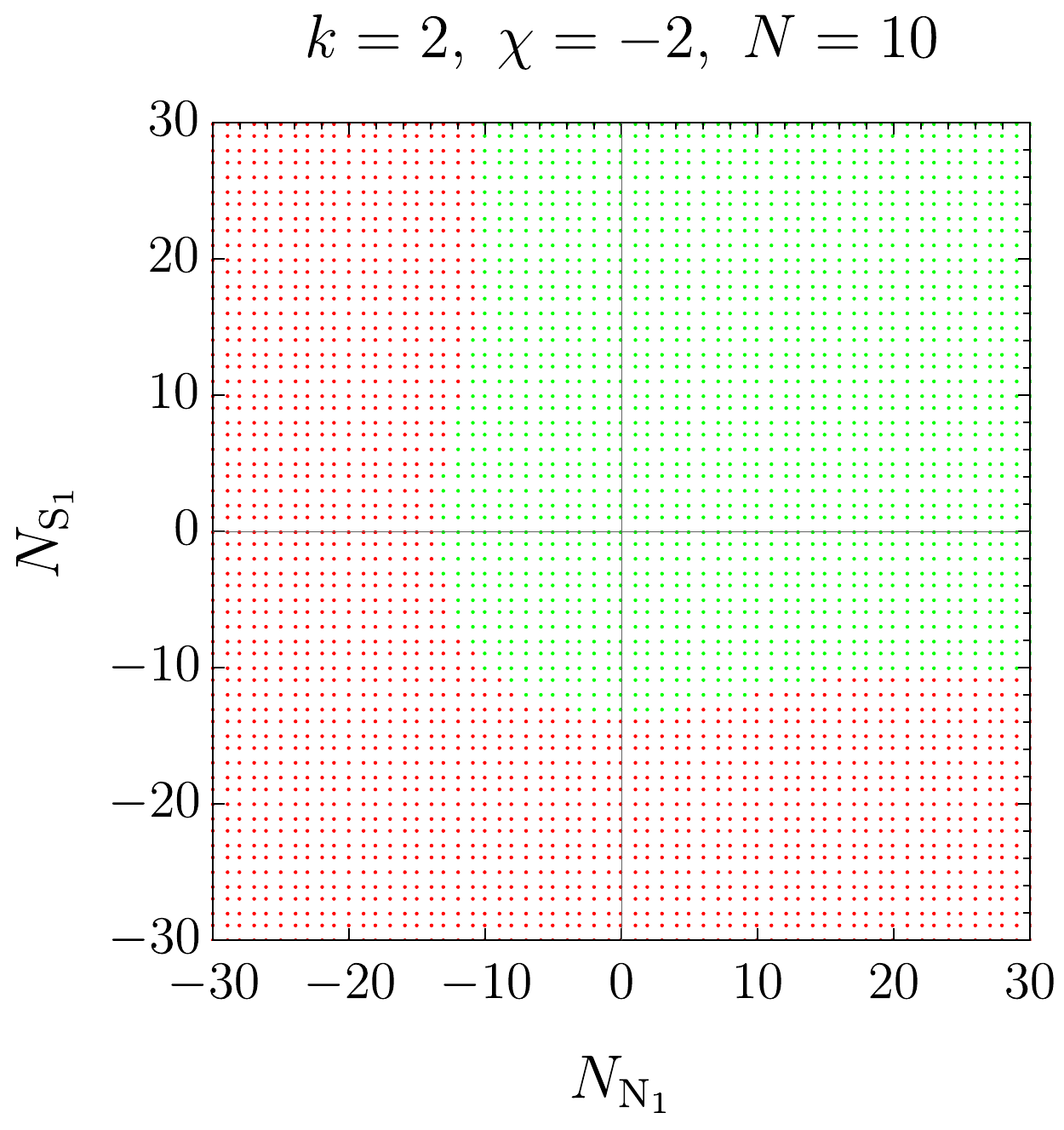}
	\end{subfigure}
	\hfill
	\begin{subfigure}{0.24\textwidth}
		\centering
		\includegraphics[width=\textwidth]{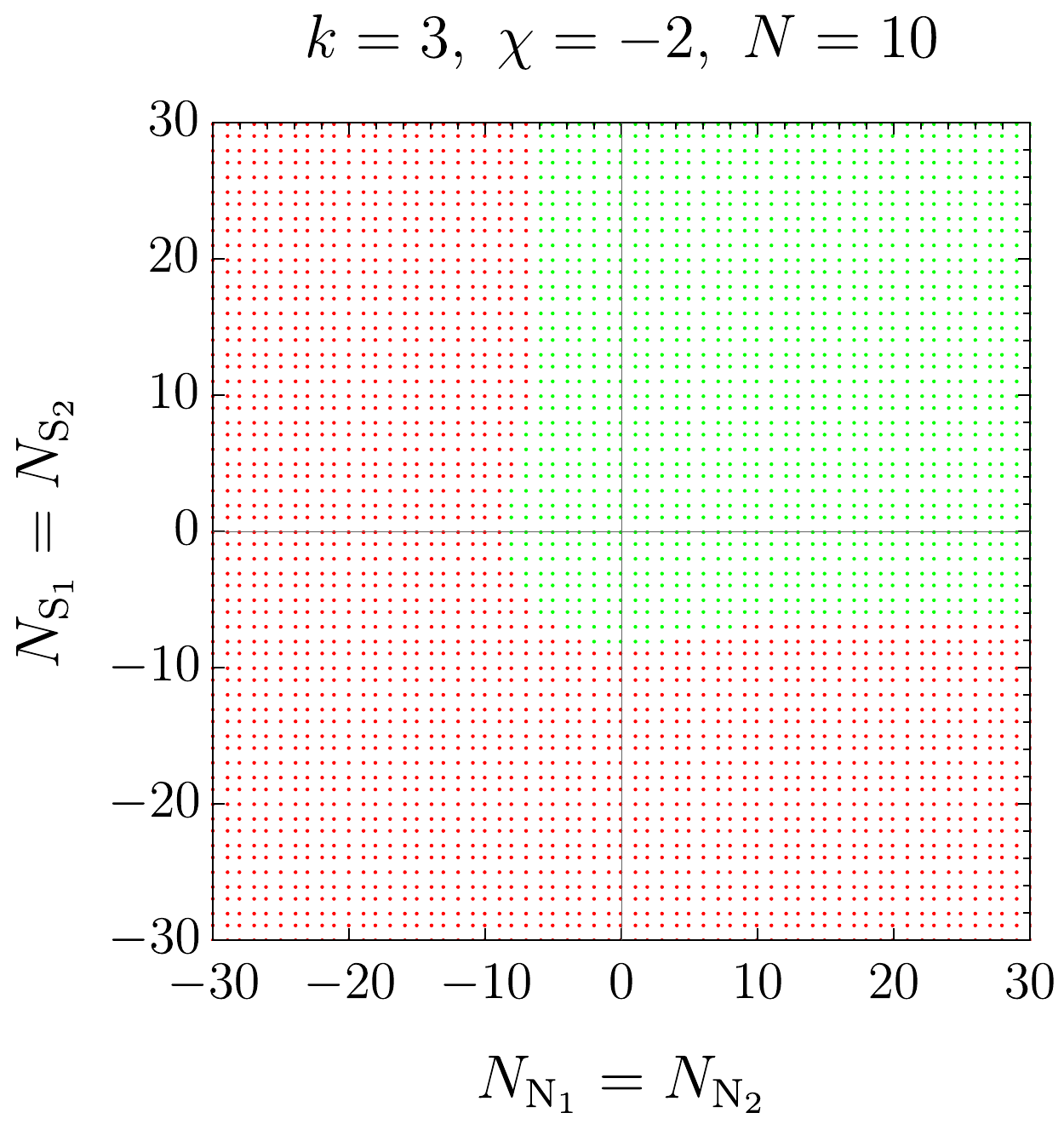}
	\end{subfigure}
	\hfill
	\begin{subfigure}{0.24\textwidth}
		\centering
		\includegraphics[width=\textwidth]{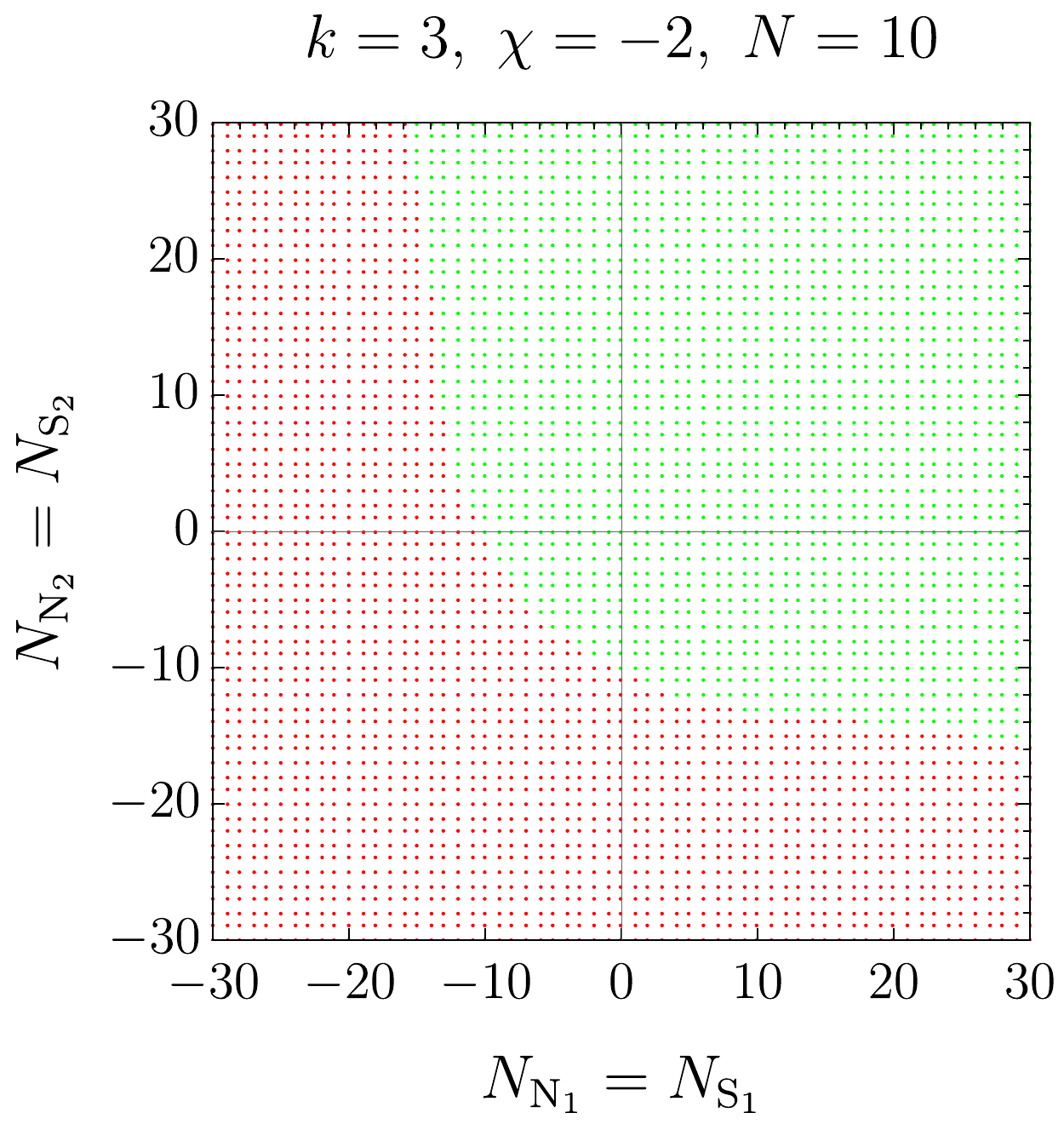}
	\end{subfigure}
	\hfill
	\begin{subfigure}{0.24\textwidth}
		\centering
		\includegraphics[width=\textwidth]{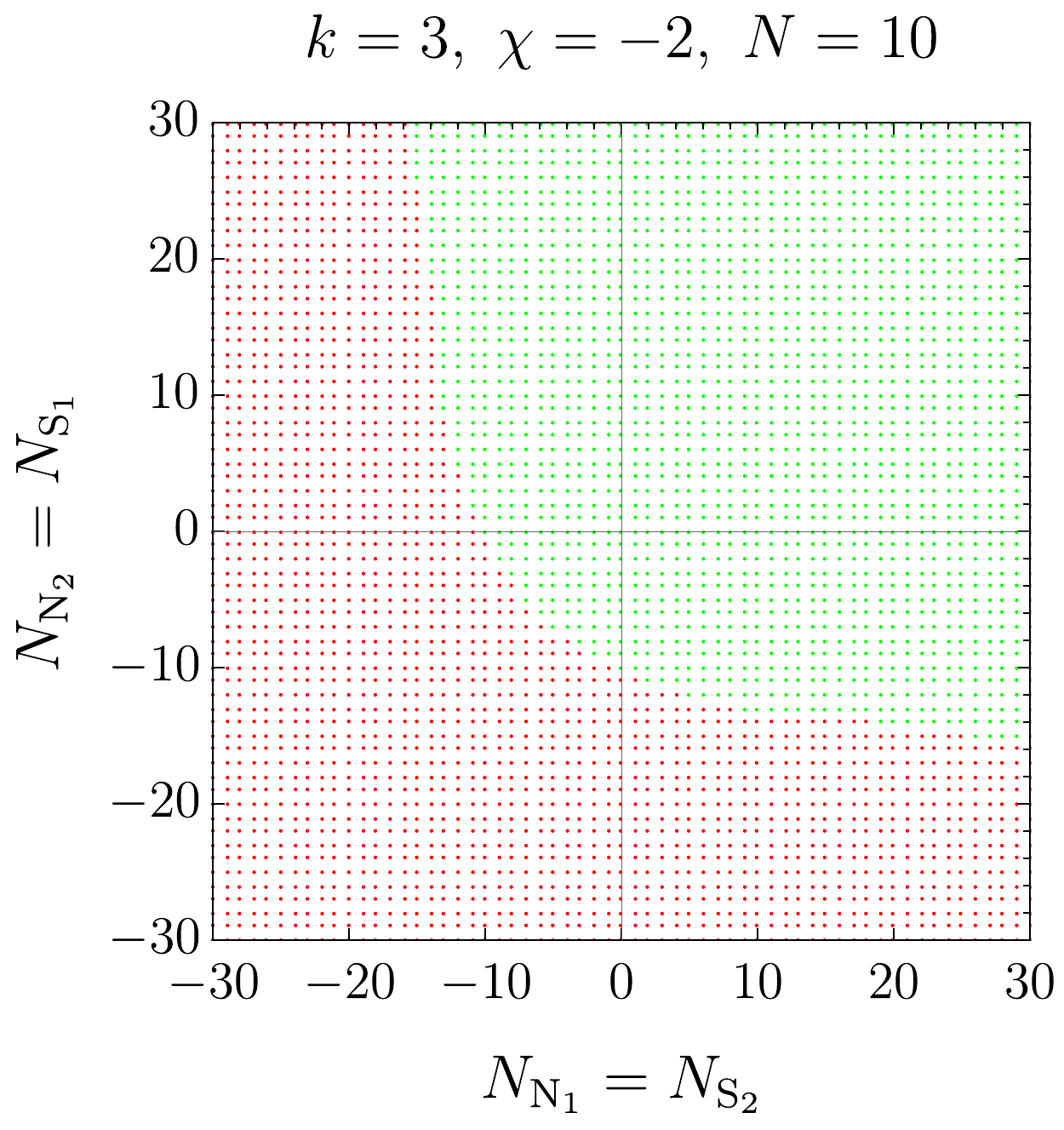}
	\end{subfigure}
	\par\bigskip
	\begin{subfigure}{0.24\textwidth}
		\centering
		\includegraphics[width=\textwidth]{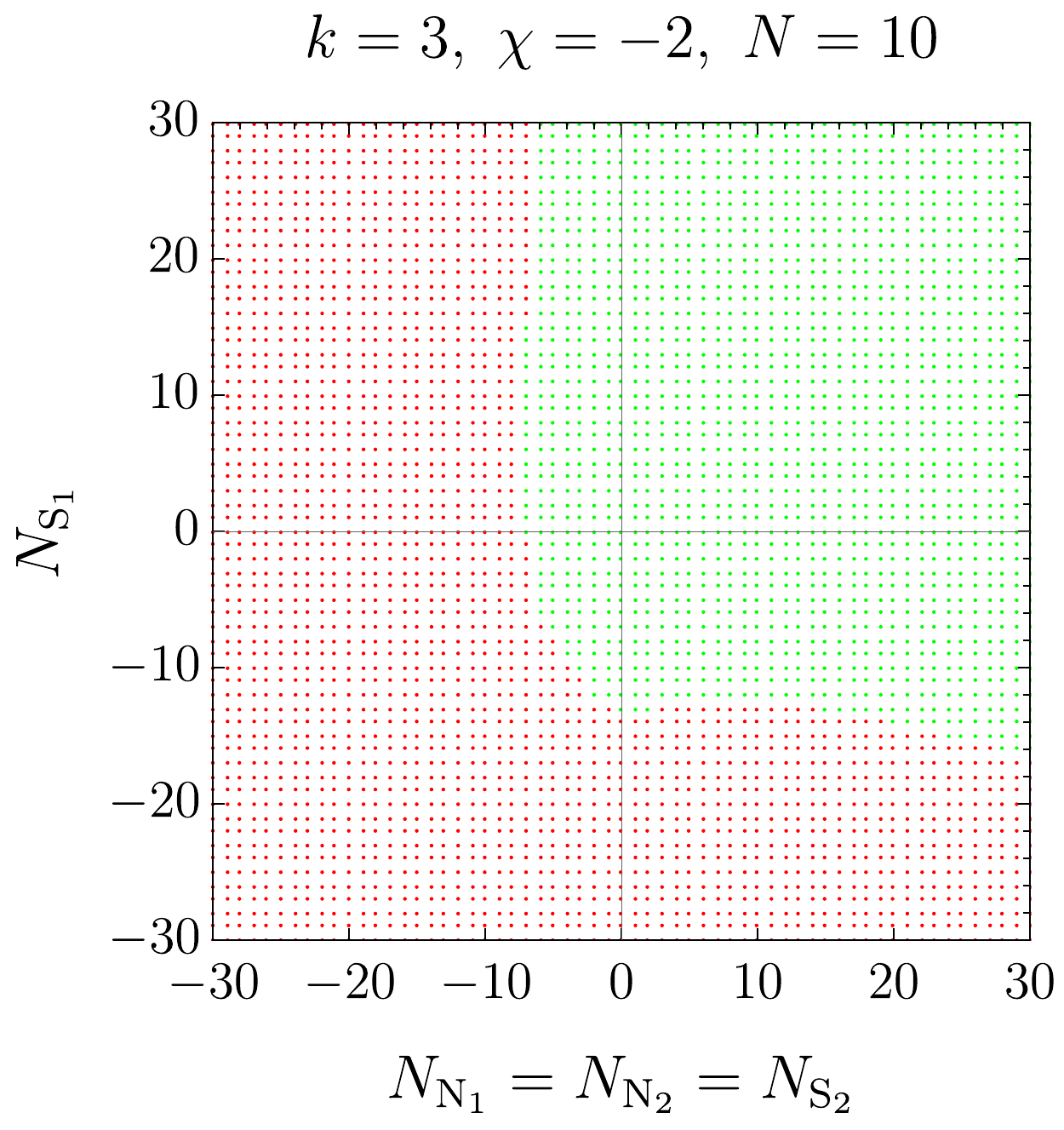}
	\end{subfigure}
	\hfill
	\begin{subfigure}{0.24\textwidth}
		\centering
		\includegraphics[width=\textwidth]{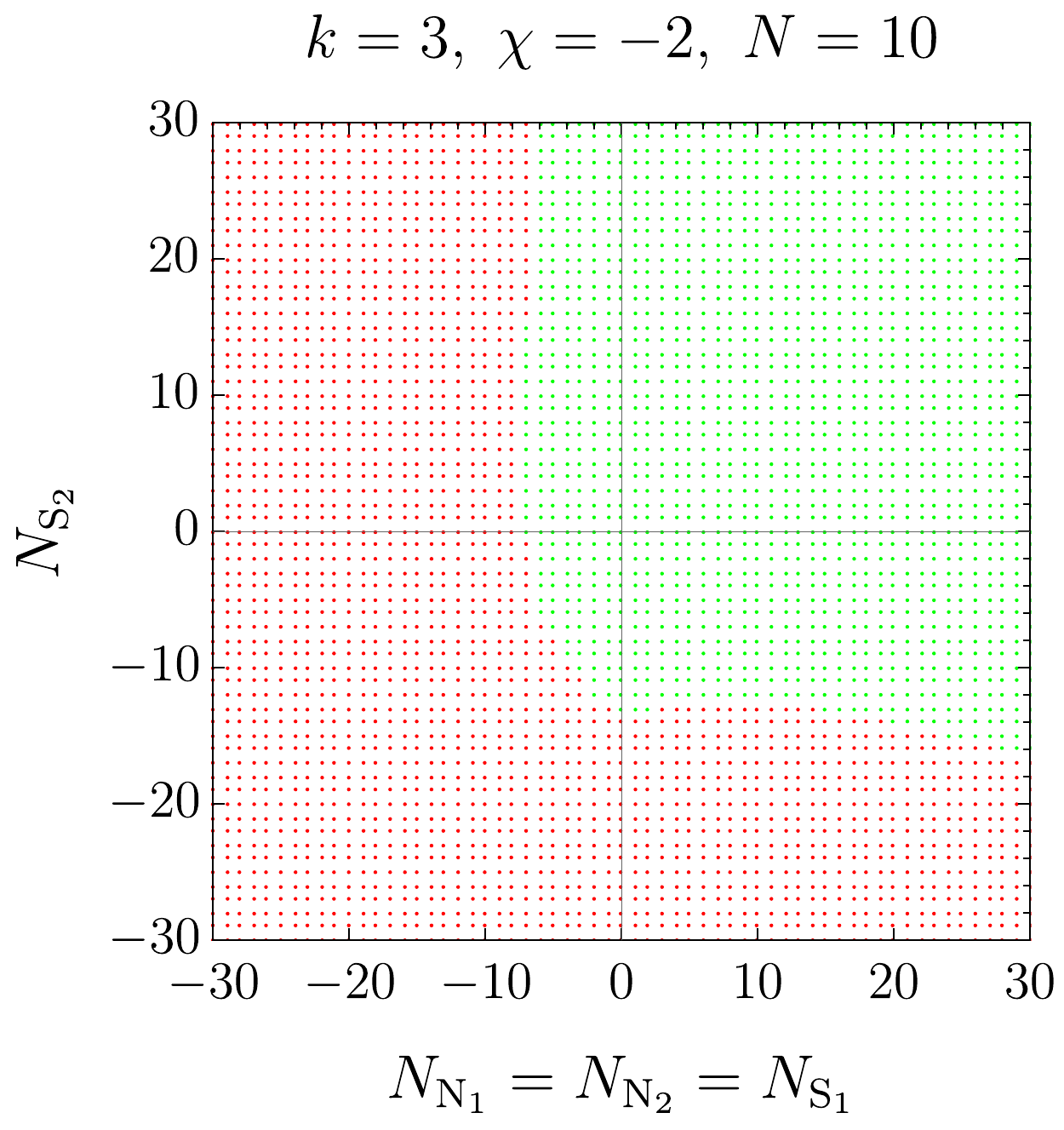}
	\end{subfigure}
	\hfill
	\begin{subfigure}{0.24\textwidth}
		\centering
		\includegraphics[width=\textwidth]{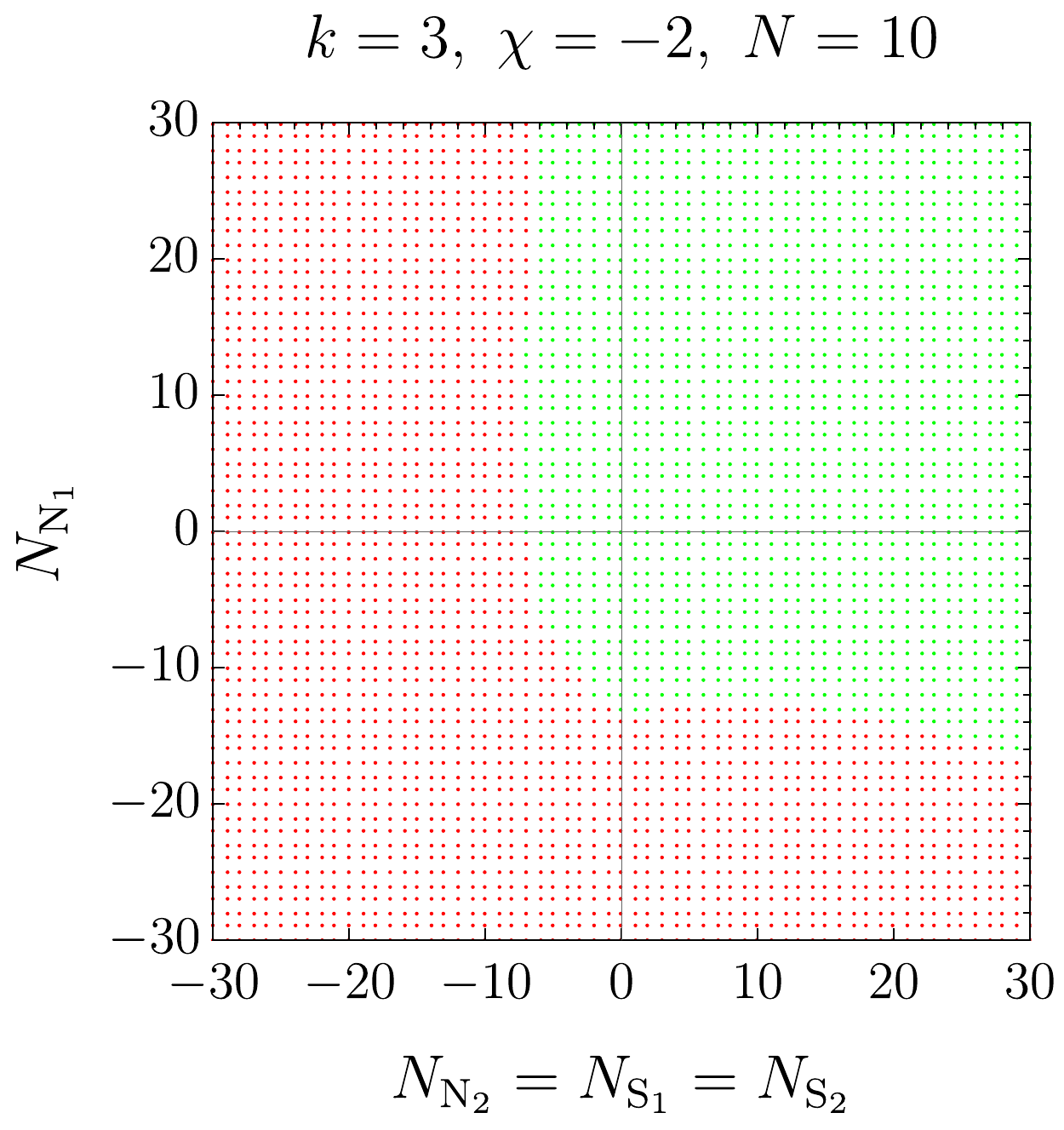}
	\end{subfigure}
	\hfill
	\begin{subfigure}{0.24\textwidth}
		\centering
		\includegraphics[width=\textwidth]{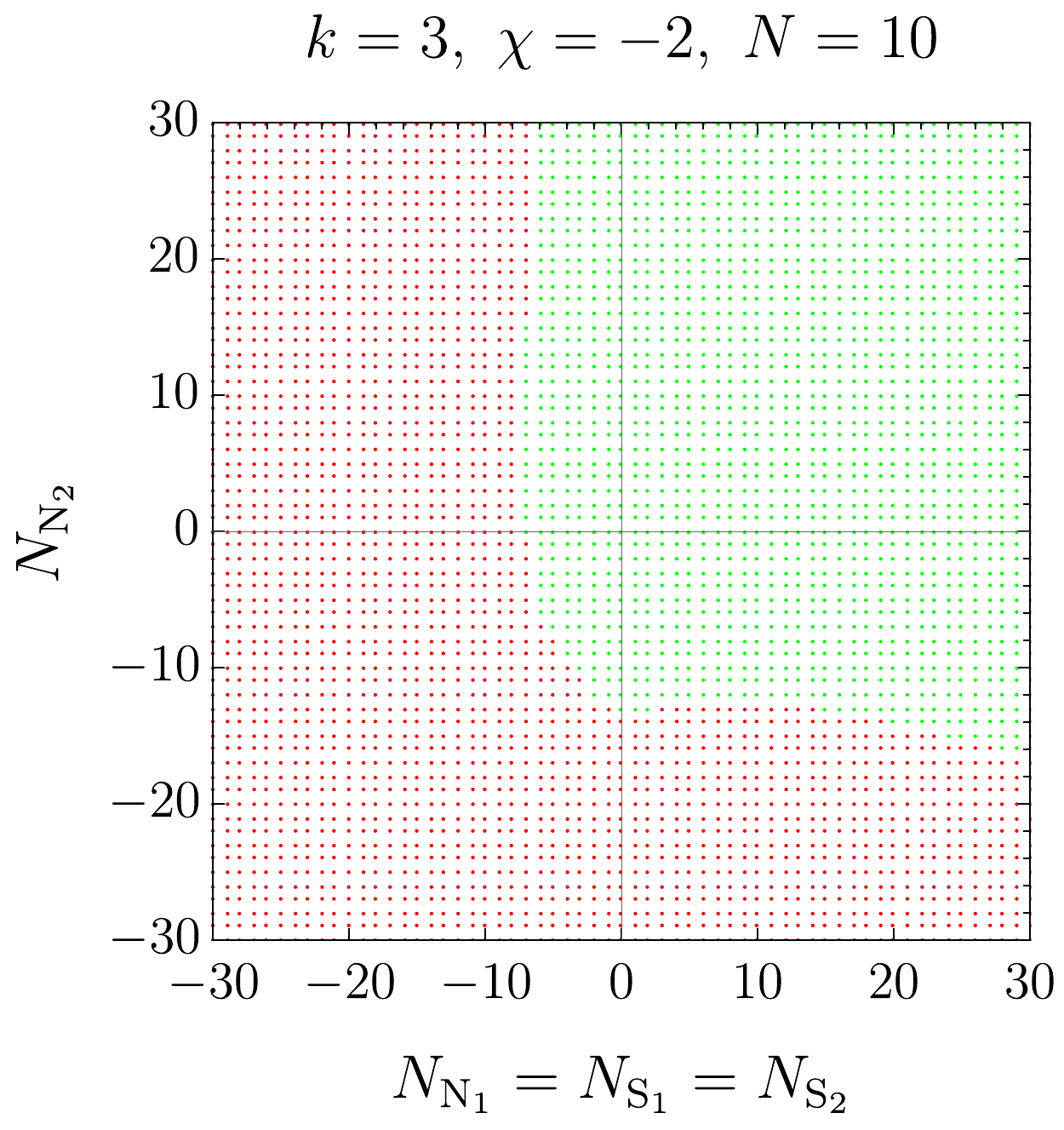}
	\end{subfigure}
	\caption{2d exclusion plots visualizing the existence/nonexistence of the central charge $a$ for $k=2$ and $k=3$. The choices of $k$, $\chi$, $N$ are labeled on top of each plot, whereas the specific configuration of the (integral) flux quanta, $N_{\mathrm{N}_i}$, $N_{\mathrm{S}_i}$, can be read off from the coordinates of a given dot. A dot is green if $a$ exists for the corresponding combination of flux quanta; it is red if $a$ does not exist. Note that the upper four plots are symmetric with respect to the exchange of axes as a consequence of the $D_2$ symmetry of $M_4$.}
	\label{a_exclusion_plots}
\end{figure}

In figure \ref{a_exclusion_plots} we illustrate the range of existence of the central charge $a$ for a variety of specific configurations of $N_{\mathrm{N}_i}$ and $N_{\mathrm{S}_i}$, given fixed $k$, $\chi$, $N$. For $k=2$ there is only one trivial pair of axes, i.e.~$N_{\mathrm{N}_1}$ and $N_{\mathrm{S}_1}$, but for $k=3$ there are four independent flux quanta, so we show here several representative 2d cross-sections in the (discrete) space of flux quanta. In all cases considered we observe a clear-cut boundary separating the regions of the flux lattice with or without a central charge. We also note that $a$ always exists when the flux quanta are strictly nonnegative; the red ``exclusion region'' always lies in the three quadrants with at least one negative flux quantum. In fact, we find that the same qualitative feature applies to general $k$: setups characterized by strictly nonnegative flux configurations have (unique) central charges.

Furthermore, we can deduce from the large-$N$ scaling relation \eqref{a_scaling_relation} that as we decrease $|\chi|$ or $N$, the boundary of the exclusion region retreats towards the two positive axes but never crosses them. This is because the ratio $\chi_B N_B / \chi_A N_A$ cannot change sign as long as we keep $g \geq 2$. In other words, the central charge is guaranteed to exist everywhere in the first quadrant (where all flux quanta are positive) once the signs of $\chi$ and $N$ are fixed, thus conforming to our expectation that all fluxes in SCFTs should be of uniform sign.

For the case of $k=2$ studied in \cite{Bah:2019vmq}, we can see that the relative signs of fluxes are fixed directly in the holographic dual, namely, the GMSW solution \cite{Gauntlett:2004zh}. If explicit holographic duals are found for $k>2$ in the future, we anticipate the same uniform-flux-sign condition to hold.

\paragraph{Dihedral symmetry.}

It was noted in \cite{Bah:2021brs} that the inflow anomaly polynomial is invariant (up to a sign) under the two parity transformations of $M_4$. Unsurprisingly, this nice property carries over to the central charge. Let us first consider the action of a north-south involution, the trial central charge $a_\mathrm{trial}(\{N_{\mathrm{N}_i}\},\{N_{\mathrm{S}_i}\},q^\varphi,\{q^{\mathrm{N}_i}\},\{q^{\mathrm{S}_i}\})$ is invariant under
\begin{equation}
	N_{\mathrm{N}_i} \leftrightarrow N_{\mathrm{S}_i} \, , \quad q^\varphi \to q^\varphi \, , \quad q^{\mathrm{N}_i} \leftrightarrow -q^{\mathrm{S}_i}\label{north_south_reflection_exchanges}
\end{equation}
for $i = 1,\dots,k-1$. Similarly, if we consider the action of an east-west involution, $a_\mathrm{trial}$ is invariant under
\begin{equation}
	N_{\mathrm{N}_i} \leftrightarrow N_{\mathrm{N}_{k-i}} \, , \quad N_{\mathrm{S}_i} \leftrightarrow N_{\mathrm{S}_{k-i}} \, , \quad q^\varphi \to -q^\varphi \, , \quad  q^{\mathrm{N}_i} \leftrightarrow q^{\mathrm{N}_{k-i}} \, , \quad  q^{\mathrm{S}_i} \leftrightarrow q^{\mathrm{S}_{k-i}} \, .\label{east_west_reflection_exchanges}
\end{equation}
Omitting the explicit functional dependence on $k$, $\chi$, and $N$, we see that the central charge exhibits a dihedral symmetry, 
\begin{equation}
	a(\{N_{\mathrm{N}_i}\},\{N_{\mathrm{S}_i}\}) = a(\{N_{\mathrm{S}_i}\},\{N_{\mathrm{N}_i}\}) = a(\{N_{\mathrm{N}_{k-i}}\},\{N_{\mathrm{S}_{k-i}}\}) = a(\{N_{\mathrm{S}_{k-i}}\},\{N_{\mathrm{N}_{k-i}}\}) \, ,\label{a_parity_invariance}
\end{equation}
where the arguments should be understood to be ordered.

When fluxes are configured symmetrically, we can further use this symmetry of $a_\mathrm{trial}$ to place powerful constraints on the values of the parameters $q^G_\mathrm{max}$ that determine the exact R-symmetry.
For example, consider a flux configuration with $N_{\mathrm{N}_i} = N_{\mathrm{S}_i}$ for all $i$, \eqref{north_south_reflection_exchanges} implies that local maxima of $a_\mathrm{trial}$ should exist in pairs distributed symmetrically across the fixed locus of this involution in the $(2k-1)$-dimensional parameter space, i.e.~if a local maximum is located at some point $(q^\varphi_\mathrm{max},\{q^{\mathrm{N}_i}_\mathrm{max}\},\{q^{\mathrm{S}_i}_\mathrm{max}\})$, then there must also be another local maximum at $(q^\varphi_\mathrm{max},\{-q^{\mathrm{S}_i}_\mathrm{max}\},\{-q^{\mathrm{N}_i}_\mathrm{max}\})$. However, since $a_\mathrm{trial}$ has at most one local maximum, all (pairs of) local maxima must coincide and lie on the fixed locus $q^{\mathrm{N}_i}=-q^{\mathrm{S}_i}$ of the involution \eqref{north_south_reflection_exchanges}. Applying the same reasoning to \eqref{east_west_reflection_exchanges}, we conclude that for any choice of $k$, $\chi$, $N$,
\begin{enumerate}
	\item $N_{\mathrm{N}_i} = N_{\mathrm{S}_i} \Rightarrow q^{\mathrm{N}_i}_\mathrm{max} = -q^{\mathrm{S}_i}_\mathrm{max} \, ;$
	\item $N_{\mathrm{N}_i} = N_{\mathrm{N}_{k-i}} \, , \, N_{\mathrm{S}_i} = N_{\mathrm{S}_{k-i}} \Rightarrow q^\varphi_\mathrm{max} = 0 \, , \, q^{\mathrm{N}_i}_\mathrm{max} = q^{\mathrm{N}_{k-i}}_\mathrm{max} \, , \, q^{\mathrm{S}_i}_\mathrm{max} = q^{\mathrm{S}_{k-i}}_\mathrm{max} \, ;$
	\item $N_{\mathrm{N}_i} = N_{\mathrm{N}_{k-i}} = N_{\mathrm{S}_i} = N_{\mathrm{S}_{k-i}} \Rightarrow q^\varphi_\mathrm{max} = 0 \, , \, q^{\mathrm{N}_i}_\mathrm{max} = q^{\mathrm{N}_{k-i}}_\mathrm{max} = -q^{\mathrm{S}_i}_\mathrm{max} = -q^{\mathrm{S}_{k-i}}_\mathrm{max} \, .$
\end{enumerate}
Note that $k=2$ setups are described by just two resolution flux quanta $N_{\mathrm{N}_1}$ and $N_{\mathrm{S}_1}$, so they automatically fall under the second family. The invariance of $a_\mathrm{trial}$ under \eqref{east_west_reflection_exchanges} then fixes $q^\varphi_\mathrm{max} = 0$. Indeed, this is consistent with the fact that for $k=2$ the $\mathrm{U}(1)_\varphi$ flavor symmetry is enhanced to $\mathrm{SU}(2)_\varphi$, whose non-abelian nature prohibits it from mixing with the R-symmetry, as argued in \cite{Intriligator:2003jj}. For general $k$, the last family of flux configurations listed above is of the most interest because the central charge can be much more efficiently determined through a modified $a$-maximization problem. Instead of \eqref{trial_R_symmetry} the trial R-symmetry for such flux configurations can be expressed as
\begin{equation}
	R_\mathrm{trial} = 2 T_\psi + \sum_{i=1}^{\lceil (k-1)/2 \rceil} q^{\mathrm{N}_i} (T_{\mathrm{N}_i} + T_{\mathrm{N}_{k-i}} - T_{\mathrm{S}_i} - T_{\mathrm{S}_{k-i}}) \, .
\end{equation}
In this way, the dimension of the parameter space is reduced from $2k-1$ to $\lceil (k-1)/2 \rceil$, i.e. there are roughly a factor of four fewer degrees of freedom. We refer to such setups as bisymmetric flux configurations. In the next subsection we will study a special case of bisymmetric flux configurations in which all flux quanta $N_{\mathrm{N}_i}$ and $N_{\mathrm{S}_i}$ are equal.

\subsection{Exact results for uniform flux configurations}

The $a$-maximization problem is much more tractable for uniform flux configurations, that is, $N_{\mathrm{N}_i} = N_{\mathrm{S}_i} \coloneqq N_\mathrm{N}$ for $i=1,2,\dots,k-1$, than for arbitrary configurations. It is indeed exactly solvable for sufficiently small $k$. As described earlier, finding the central charges for $k=2$ and $k=3$ with uniform flux configurations is effectively a 1d maximization problem, and in both cases the various central charges can be written as reasonably compact closed-form expressions. For $k=2$, we get
\begin{align}
	a & = \frac{\big[\chi^2 (9 N^2 - 8) - 18 N_\mathrm{N} (\chi N - 2 N_\mathrm{N})\big]^{3/2} - 9 \chi^2 N_\mathrm{N} (9 N^2 - 6) + 54 N_\mathrm{N}^2 (3 \chi N - 4 N_\mathrm{N})}{48 \chi^2} \, ,\label{k=2_a_expression}\\
	c & = \frac{\big[\chi^2 (9 N^2 - 8) - 18 N_\mathrm{N} (\chi N - 2 N_\mathrm{N})\big]^{3/2} - 9 \chi^2 N_\mathrm{N} (9 N^2 - 6) + 54 N_\mathrm{N}^2 (3 \chi N - 4 N_\mathrm{N})}{48 \chi^2}\nonumber\\
	& \phantom{=\ } + \frac{\sqrt{\chi^2 (9 N^2 - 8) - 18 N_\mathrm{N} (\chi N - 2 N_\mathrm{N})}}{48} \, ,\label{k=2_c_expression}
\end{align}
\begin{align}
	\mathcal{A}_{R\varphi\varphi} & = \frac{\chi^2 N (N^2 - 1) + N N_\mathrm{N} \big[3 (\chi N - 2 N_\mathrm{N}) + \sqrt{\chi^2 (9 N^2 - 8) - 18 N_\mathrm{N} (\chi N - 2 N_\mathrm{N})}\big]}{12 \chi} \, ,\label{k=2_b_phi_expression}\\
	\mathcal{A}_{Rii} & = \frac{N (\chi N - N_\mathrm{N}) \sqrt{\chi^2 (9 N^2 - 8) - 18 N_\mathrm{N} (\chi N - 2 N_\mathrm{N})} - 3 N N_\mathrm{N} (\chi N - 2 N_\mathrm{N})}{3 \chi} \, ,\label{k=2_b_i_expression}
\end{align}
whereas for $k=3$, we have
\begin{align}
	a & = \frac{9 \chi^2 (N - 1)^2 (27 N^2 + 27 N - 14) - 6 \chi N_\mathrm{N} (135 N^3 - 253 N + 118) + 12 N_\mathrm{N}^2 (81 N^2 - 80)}{64 \big[10 N_\mathrm{N} - 3 \chi (N - 1)\big]} \, ,\label{k=3_a_expression}\\
	c & = \frac{3 \chi^2 (N - 1)^2 (81 N^2 + 81 N - 28) - 2 \chi N_\mathrm{N} (405 N^3 - 641 N + 236) + N_\mathrm{N}^2 (972 N^2 - 640)}{64 \big[10 N_\mathrm{N} - 3 \chi (N - 1)\big]} \, ,\label{k=3_c_expression}
\end{align}
\vspace{-5ex}
\begin{align}
	\mathcal{A}_{R\varphi\varphi} & = \frac{3 \chi^2 N (N - 1) (N^2 - 1) - 10 \chi N N_\mathrm{N} (N^2 - 1) + 24 N^2 N_\mathrm{N}^2}{120 N_\mathrm{N} - 36 \chi (N - 1)} \, ,\label{k=3_b_phi_expression}\\
	\mathcal{A}_{Rii} & = \frac{9 \chi^2 N^2 (N - 1)^2 - 51 \chi N^2 N_\mathrm{N} (N - 1) + 64 N^2 N_\mathrm{N}^2}{20 N_\mathrm{N} - 6 \chi (N - 1)} \, .\label{k=3_b_i_expression}
\end{align}
The divergence of the expressions above when $N = 1$ and $N_\mathrm{N} = 0$ shall not worry us as long as we are working in the large-$N$ limit. Moreover, it can be easily checked that $a$ and $c$ are the same at leading order. Interestingly, we observe that $\mathcal{A}_{R \, \mathrm{N}_1 \mathrm{N}_1}=\mathcal{A}_{R \, \mathrm{S}_1 \mathrm{S}_1}$ for $k=2$, and $\mathcal{A}_{R \, \mathrm{N}_1 \mathrm{N}_1}=\mathcal{A}_{R \, \mathrm{N}_2 \mathrm{N}_2}=\mathcal{A}_{R \, \mathrm{S}_1 \mathrm{S}_1}=\mathcal{A}_{R \, \mathrm{S}_2 \mathrm{S}_2}$ for $k=3$. This pattern has a natural generalization for higher $k$, as we will soon see.

\begin{figure}[t!]
	\begin{subfigure}{0.5\textwidth}
		\centering
		\includegraphics[width=\textwidth]{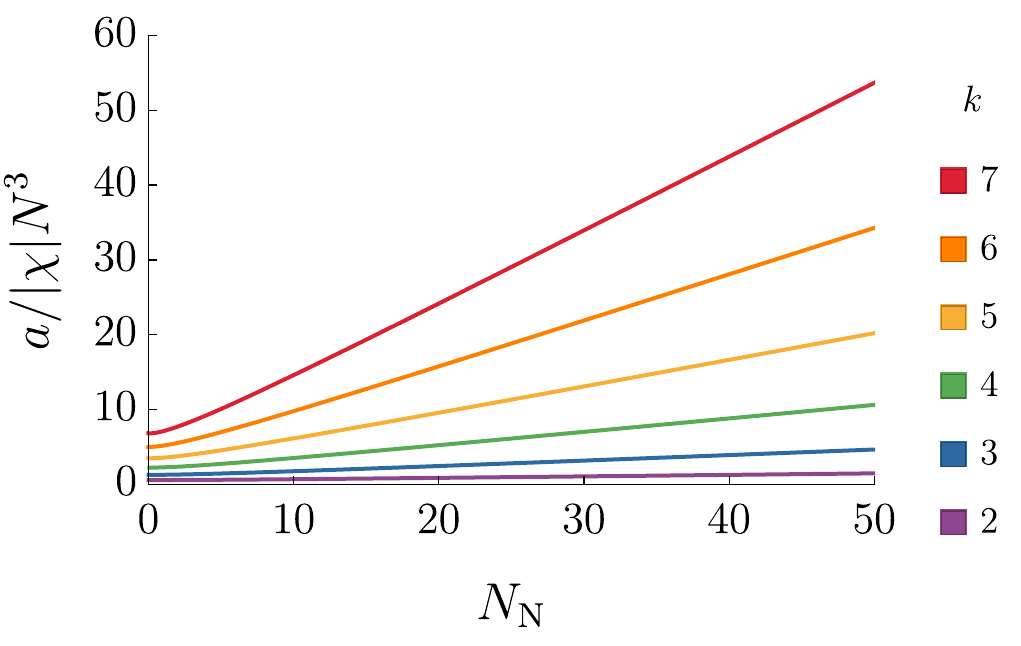}
	\end{subfigure}
	\hfill
	\begin{subfigure}{0.5\textwidth}
		\centering
		\includegraphics[width=\textwidth]{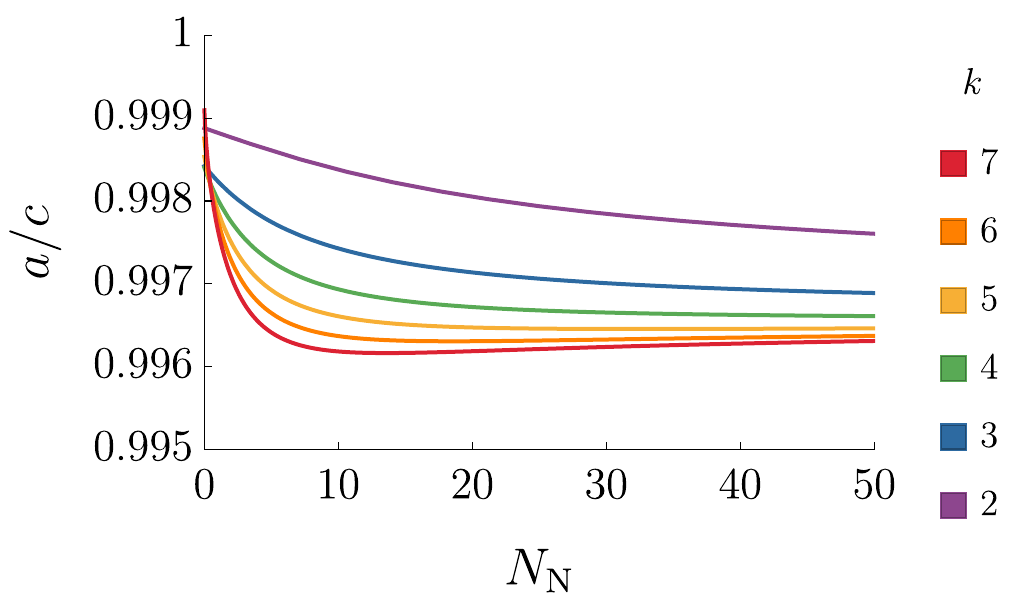}
	\end{subfigure}
	\par\bigskip
	\begin{subfigure}{0.5\textwidth}
		\centering
		\includegraphics[width=\textwidth]{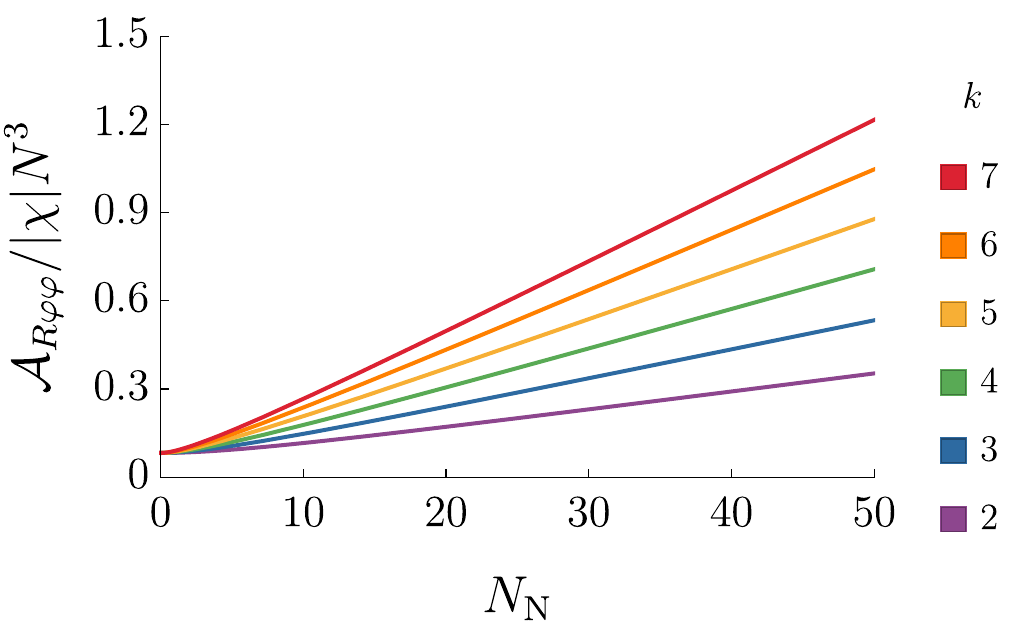}
	\end{subfigure}
	\hfill
	\begin{subfigure}{0.5\textwidth}
		\centering
		\includegraphics[width=\textwidth]{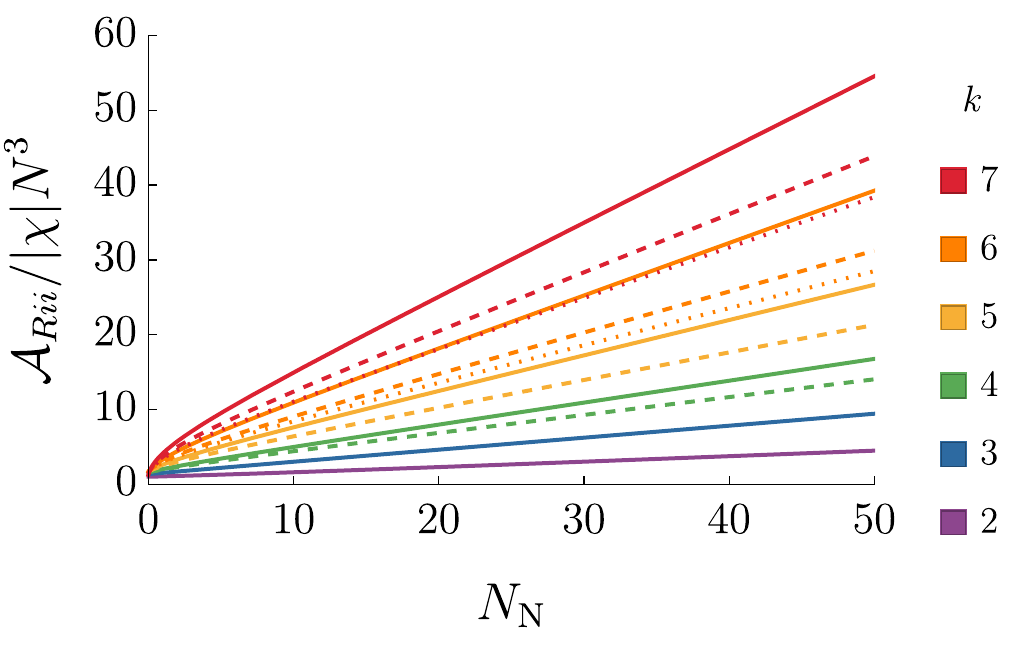}
	\end{subfigure}
	\caption{Plots of $a$, $a/c$, $\mathcal{A}_{R\varphi\varphi}$, $\mathcal{A}_{Rii}$ for $k=2$ to $k=7$ with identical flux quanta $N_{\mathrm{N}_i} = N_{\mathrm{S}_i} \coloneqq N_\mathrm{N}$ for all $i = 1,\dots,k-1$. $N_\mathrm{N}$ is treated as a continuous real parameter for visualization purposes. All of the plots are evaluated with $\chi = -2$ and $N = 10$. For a given $k$, there are $\lceil (k-1)/2 \rceil$ independent $\mathcal{A}_{Rii}$ anomaly coefficients, which are proportional to the flavor central charges $b_i$ associated with the resolution cycles of $M_4$. We use solid lines for $i=1$, dashed lines for $i=2$, and dotted lines for $i=3$ where applicable.}
	\label{central_charge_plots}
\end{figure}

While being fully analytic, the central charges we find for $k \geq 4$ cannot be reduced to similarly compact forms. Nevertheless, figure \ref{central_charge_plots} illustrates the functional dependence of $a$, $a/c$, $\mathcal{A}_{R\varphi\varphi}$, $\mathcal{A}_{Rii}$ on $N_\mathrm{N}$ for a range of $k$. Note that the scaling relation \eqref{I6_scaling_relation} implies up to $\mathcal{O}(N^3,N_\mathrm{N}^3)$, changing $\chi$ and $N$ amounts to rescaling the axes of the plots of $a$, $\mathcal{A}_{R\varphi\varphi}$, $\mathcal{A}_{Rii}$ without altering their qualitative behavior. It can be seen that all the central charges are monotonic in $k$ and $N_\mathrm{N}$. We also note that the ratio $a/c$ is well within the Hofman-Maldacena bounds \cite{Hofman:2008ar} on $\mathcal{N} = 1$ SCFTs,
\begin{equation}
	\frac{1}{2} \leq \frac{a}{c} \leq \frac{3}{2} \, .
\end{equation}

Let us briefly comment on the flavor central charge $b_i \propto \mathcal{A}_{Rii}$. In general, because of the $D_2$ symmetry of $M_4$, there are $\lceil (k-1)/2 \rceil$ independent $\mathcal{A}_{Rii}$ when all the resolution flux quanta are equal, i.e.
\begin{equation}
	\mathcal{A}_{R \, \mathrm{N}_i \mathrm{N}_i} = \mathcal{A}_{R \, \mathrm{N}_{k-i} \mathrm{N}_{k-i}} = \mathcal{A}_{R \, \mathrm{S}_i \mathrm{S}_i} = \mathcal{A}_{R \, \mathrm{S}_{k-i} \mathrm{S}_{k-i}}\label{flavor_charge_symmetry}
\end{equation}
for $i = 1,2,\dots,\lceil (k-1)/2 \rceil$, hence the notation $\mathcal{A}_{Rii} \coloneqq \mathcal{A}_{R \, \mathrm{N}_i \mathrm{N}_i} = \mathcal{A}_{R \, \mathrm{S}_i \mathrm{S}_i}$. Specifically, there is one independent $\mathcal{A}_{Rii}$ for $k=2,3$, two for $k=4,5$, and three for $k=6,7$. It is evident from the separation between lines of like color in figure \ref{central_charge_plots} that
\begin{equation}
	\mathcal{A}_{Rii} \geq \mathcal{A}_{R(i+1)(i+1)} \geq \cdots \geq \mathcal{A}_{R\lceil (k-1)/2 \rceil\lceil (k-1)/2 \rceil} \, .
\end{equation}
The inequalities are simultaneously saturated when $N_\mathrm{N} = 0$.

\subsection{Perturbative analysis}

Even though it is exceptionally challenging to analytically determine the central charge through $a$-maximization for arbitrary combinations of $k$, $\chi$, $N$, $N_{\mathrm{N}_i}$, $N_{\mathrm{S}_i}$, we can use perturbation theory to solve the equations $\partial a_\mathrm{trial} / \partial q^G = 0$ order by order in the regime where
\begin{equation}
	\frac{N_{\mathrm{N}_i}^3}{\chi^2} \, , \frac{N_{\mathrm{S}_i}^3}{\chi^2} \gg |\chi| N \gg N_{\mathrm{N}_i}, N_{\mathrm{S}_i} \gg 1 \, .\label{4d_theory_large-N_limit}
\end{equation}
The first and the last inequalities are required to ensure that the $\mathcal{O}(N)$ contributions from $I_6^\text{inflow,$E_4 X_8$}$ and $I_6^\mathrm{v,t}$ are negligible.\footnote{Appropriate powers of $\chi$ are inserted in the inequalities here based on the facts that $N_{\mathrm{N}_i}/|\chi|N, N_{\mathrm{S}_i}/|\chi|N \sim 1$ are the ``characteristic scales'' and that $I_6^\text{inflow,$E_4^3$}$ scales as $\chi N^3$.} We find that the perturbative expansions of the central charges $a=c$ and the anomaly coefficient $\mathcal{A}_{R\varphi\varphi}$ can be written as
\begin{align}
	a = c & = -\frac{9 k^2 \chi N^3}{64} - \frac{27 N}{64\chi} \sum_{i,j=1}^{k-1} i (k-j) (N_{\mathrm{N}_i} N_{\mathrm{N}_j} + N_{\mathrm{S}_i} N_{\mathrm{S}_j}) + \mathcal{O}\big(N_{\mathrm{N}_i,\mathrm{S}_i}^3\big) \, ,\label{4d_central_charge}\\
	\mathcal{A}_{R\varphi\varphi} & = -\frac{\chi N^3}{12} - \frac{N}{2 k^2 \chi} \sum_{i,j=1}^{k-1} i (k-j) (N_{\mathrm{N}_i} N_{\mathrm{N}_j} + N_{\mathrm{S}_i} N_{\mathrm{S}_j}) + \mathcal{O}\big(N_{\mathrm{N}_i,\mathrm{S}_i}^3\big) \, .\label{4d_phi_flavor_charge}
\end{align}
For uniform flux configurations, these perturbative expansions have been checked to be consistent with the previously shown exact expressions \eqref{k=2_a_expression}, \eqref{k=2_b_phi_expression}, \eqref{k=3_a_expression}, \eqref{k=3_b_phi_expression} derived for $k=2$ and $k=3$.

The symmetry \eqref{flavor_charge_symmetry} between flavor central charges $b_{\mathrm{N}_i,\mathrm{S}_i} \propto \mathcal{A}_{R(\mathrm{N}_i,\mathrm{S}_i)(\mathrm{N}_i,\mathrm{S}_i)}$ no longer holds for nonuniform flux configurations. We list below the perturbative expansions of various flavor central charges for $k=2$ and $k=3$, so the reader can compare them to their uniform-flux analogs \eqref{k=2_b_i_expression} and \eqref{k=3_b_i_expression}. For $k=2$, we obtain
\begin{align}\label{tHooft_coeffs_k2}
	\mathcal{A}_{R \, \mathrm{N}_1 \mathrm{N}_1} & = -\chi N^3 + N^2 N_{\mathrm{N}_1} - \frac{N N_{\mathrm{S}_1}^2}{2\chi} + \mathcal{O}\big(N_{\mathrm{N}_i,\mathrm{S}_i}^3\big) \, ,\\
	\mathcal{A}_{R \, \mathrm{S}_1 \mathrm{S}_1} & = -\chi N^3 + N^2 N_{\mathrm{S}_1} - \frac{N N_{\mathrm{N}_1}^2}{2\chi} + \mathcal{O}\big(N_{\mathrm{N}_i,\mathrm{S}_i}^3\big) \, ,
\end{align}
whereas for $k=3$, we obtain
\begin{align}\label{tHooft_coeffs_k3}
	\mathcal{A}_{R \, \mathrm{N}_1 \mathrm{N}_1} & = -\frac{3 \chi N^3}{2} + \frac{N^2 (5 N_{\mathrm{N}_1} + 2 N_{\mathrm{N}_2})}{4}\nonumber\\
	& \phantom{=\ } - \frac{N \big[5 N_{\mathrm{N}_1}^2 + 26 N_{\mathrm{N}_1} N_{\mathrm{N}_2} + 2 N_{\mathrm{N}_2}^2 + 16 (N_{\mathrm{S}_1}^2 + N_{\mathrm{S}_1} N_{\mathrm{S}_2} + N_{\mathrm{S}_2}^2)\big]}{24\chi} + \mathcal{O}\big(N_{\mathrm{N}_i,\mathrm{S}_i}^3\big) \, ,\\
	\mathcal{A}_{R \, \mathrm{N}_2 \mathrm{N}_2} & = -\frac{3 \chi N^3}{2} + \frac{N^2 (2 N_{\mathrm{N}_1} + 5 N_{\mathrm{N}_2})}{4}\nonumber\\
	& \phantom{=\ } - \frac{N \big[2 N_{\mathrm{N}_1}^2 + 26 N_{\mathrm{N}_1} N_{\mathrm{N}_2} + 5 N_{\mathrm{N}_2}^2 + 16 (N_{\mathrm{S}_1}^2 + N_{\mathrm{S}_1} N_{\mathrm{S}_2} + N_{\mathrm{S}_2}^2)\big]}{24\chi} + \mathcal{O}\big(N_{\mathrm{N}_i,\mathrm{S}_i}^3\big) \, ,\\
	\mathcal{A}_{R \, \mathrm{S}_1 \mathrm{S}_1} & = -\frac{3 \chi N^3}{2} + \frac{N^2 (5 N_{\mathrm{S}_1} + 2 N_{\mathrm{S}_2})}{4}\nonumber\\
	& \phantom{=\ } - \frac{N \big[5 N_{\mathrm{S}_1}^2 + 26 N_{\mathrm{S}_1} N_{\mathrm{S}_2} + 2 N_{\mathrm{S}_2}^2 + 16 (N_{\mathrm{N}_1}^2 + N_{\mathrm{N}_1} N_{\mathrm{N}_2} + N_{\mathrm{N}_2}^2)\big]}{24\chi} + \mathcal{O}\big(N_{\mathrm{N}_i,\mathrm{S}_i}^3\big) \, ,\\
	\mathcal{A}_{R \, \mathrm{S}_2 \mathrm{S}_2} & = -\frac{3 \chi N^3}{2} + \frac{N^2 (2 N_{\mathrm{S}_1} + 5 N_{\mathrm{S}_2})}{4}\nonumber\\
	& \phantom{=\ } - \frac{N \big[2 N_{\mathrm{S}_1}^2 + 26 N_{\mathrm{S}_1} N_{\mathrm{S}_2} + 5 N_{\mathrm{S}_2}^2 + 16 (N_{\mathrm{N}_1}^2 + N_{\mathrm{N}_1} N_{\mathrm{N}_2} + N_{\mathrm{N}_2}^2)\big]}{24\chi} + \mathcal{O}\big(N_{\mathrm{N}_i,\mathrm{S}_i}^3\big) \, .
\end{align}

\subsection{Genus-one cases}

The expressions reported earlier in this section are not applicable to the cases where the Riemann surface is a torus, although we shall remark that our previous arguments leading to the uniqueness theorem of central charges continue to hold here. To determine the central charges in such cases, we can perform $a$-maximization on the expression \eqref{torus_expression} obtained from re-writing the inflow anomaly polynomial using \eqref{dictionary}. Note that $I_6^\mathrm{v,t}$ vanishes in this limit, so that the anomalies of the interacting 4d SCFT can be read off directly, as stated in \eqref{SCFT_from_inflow_torus}.

Let us first focus on the case of $k=2$. We find that the the various central charges admit qualitatively different expressions depending on whether the two independent flux quanta, $N_{\mathrm{N}_1}$ and $N_{\mathrm{S}_1}$, are the same or different. Specifically, if $N_{\mathrm{N}_1} \neq N_{\mathrm{S}_1}$, we recover the central charges of the $Y^{p,q}$ theories obtained by \cite{Benvenuti2005} for $q > 0$, using the identifications \eqref{Ypq_flux_identification}. On the other hand, if $N_{\mathrm{N}_1} = N_{\mathrm{S}_1} \coloneqq N_\mathrm{N}$, or equivalently, $q = 0$, we instead have
\begin{equation}
	\begin{gathered}
		a = \frac{3 (9 N^2 - 8) N_\mathrm{N}}{64} \, , \qquad c = \frac{(27 N^2 - 16) N_\mathrm{N}}{64} \, ,\\
		\mathcal{A}_{R\varphi\varphi} = \frac{N^2 N_\mathrm{N}}{8} \, , \qquad \mathcal{A}_{R \, \mathrm{N}_1 \mathrm{N}_1} = \mathcal{A}_{R \, \mathrm{S}_1 \mathrm{S}_1} = \frac{3 N^2 N_\mathrm{N}}{2} \, .
	\end{gathered}
\end{equation}

Similarly to the higher-genus cases, it is technically challenging to analytically solve the $a$-maximization problem corresponding to generic flux configurations for $k > 2$, except when all the resolution flux quanta are equal. For these uniform flux configurations, the central charges admit rather simple functional forms as follows,
\begin{equation}
	a = \mathsf{a}_1(k) N^2 N_\mathrm{N} - \mathsf{a}_2(k) N_\mathrm{N} \, , \quad c = \mathsf{c}_1(k) N^2 N_\mathrm{N} - \mathsf{c}_2(k) N_\mathrm{N} \, , \quad \mathcal{A}_{R\varphi\varphi} = \mathsf{b}(k) N^2 N_\mathrm{N} \, .
\end{equation}
We record in table \ref{genus_one_central_charge_coefficients_table} the values of these coefficients for $k=2$ to $k=7$. Note that $\mathsf{a}_1(k) = \mathsf{c}_1(k)$ as expected.

\begin{table}[t!]
	\centering
	\begin{tabular}{|c||cccccc|} 
		\hline
		$k$ & $2$ & $3$ & $4$ & $5$ & $6$ & $7$\\
		\hline
		\hline
		$\mathsf{a}_1$ & $27/64$ & $243/160$ & $81/22$ & $14175/1952$ & $120285/9536$ & $17199/856$\\
		$\mathsf{a}_2$ & $3/8$ & $3/2$ & $15/4$ & $15/2$ & $105/8$ & $21$\\
		$\mathsf{c}_1$ & $27/64$ & $243/160$ & $81/22$ & $14175/1952$ & $120285/9536$ & $17199/856$\\
		$\mathsf{c}_2$ & $1/4$ & $1$ & $5/2$ & $5$ & $35/4$ & $14$\\
		$\mathsf{b}$ & $1/8$ & $1/5$ & $3/11$ & $21/61$ & $495/1192$ & $52/107$\\
		\hline
	\end{tabular}
	\caption{Central charge coefficients (with genus one) for various $k$.}
	\label{genus_one_central_charge_coefficients_table}
\end{table}

 
\section{Conclusion and outlook} \label{outlook}

In this work, we have identified a geometric origin of flip fields in 4d $\mathcal N = 1$
SCFTs of  class $\mathcal S_k$, by  adopting   an 11d perspective on these models.
The comparison between anomaly inflow from M-theory, and integration of the 6d anomaly 
polynomial, led us to the  relation \eqref{main_equation}, which is central to our analysis.
The charges and multiplicities in $I_6^{\rm flip}$ are then interpreted in terms of
M2-brane operators, associated to blow-up modes of the $\mathbb C^2/\mathbb Z_k$ singularity. 
We thus get a physical picture of the role of flip fields:
they are necessary to interpolate between six dimensions,
where such blow-up modes are not present, and four dimensions,
where they are part of the SCFT.

The results of this paper suggest  several  
directions for future research. Firstly, it would be interesting to
find explicit $AdS_5$ solutions in 11d supergravity,
which generalize the GMSW solutions from $k=2$ to higher values of $k$.
This work shows that the topology and flux configuration of $M_6$
give rise, via inflow, to the anomaly polynomial of a
4d SCFT of class $\mathcal S_k$
with   fluxes for the $\SU(k)_{b}$, $\SU(k)_{c}$ flavor symmetries.
This observation is a strong hint that $AdS_5$ solutions
should exist, whose internal space has the topology and $G_4$-flux quanta
of $M_6$.

The special case in which the Riemann surface is a torus also deserves
further investigation. For $k=2$, we have obtained a precise match between the M-theory
inflow anomaly polynomial, and the anomaly polynomial of 
the SCFT
realized by $N$ D3-branes at the tip of the cone 
over the   Sasaki-Einstein space $Y^{p,q}$ (with $p$, $q$ determined
by the flavor fluxes in the M-theory construction). It is natural to study   generalizations to higher values of $k$,
for instance exploring possible connections to other families of explicit Sasaki-Einstein metrics,
such as \cite{Cvetic:2005ft,Cvetic:2005vk}.

Another natural direction for further study is to consider 4d theories
of class $\mathcal S_\Gamma$, i.e.~theories obtained from reduction 
of the 6d (1,0)
SCFT realized by $N$ M5-branes  probing the   singularity $\mathbb C^2/\Gamma$,
with $\Gamma$ an ADE subgroup of $\SU(2)$.
Based on our results, we conjecture that the pattern of charges of
flip fields for these models should be given in terms of the roots and Cartan
matrix of $\mathfrak g_\Gamma$, the ADE Lie algebra associated to $\Gamma$.
It would be interesting to perform explicit checks of this conjecture, for instance
 against the Lagrangian models
 of \cite{Kim:2018lfo}.













\acknowledgments

We are grateful to Patrick Jefferson,
Zohar Komargodski,
Shlomo Razamat, Evyatar Sabag, Alessandro Tomasiello, 
Thomas Waddleton, and 
Gabi Zafrir
 for interesting
conversations and correspondence. 
The work of IB, EL, and PW is supported in part by NSF grant PHY-2112699. The work of IB is also supported in part by the Simons Collaboration on Global Categorical Symmetries.
FB is supported by STFC Consolidated Grant ST/T000864/1.
FB is supported by the European Union’s Horizon 2020 Framework: ERC Consolidator Grant 682608.


\appendix
\addtocontents{toc}{\protect\setcounter{tocdepth}{1}}



\section{\boldmath Review of the $E_4^3$ contribution to anomaly inflow} \label{app_E4cube}
In this appendix, we record the $E_4^3$ contribution to $-I_6^\text{inflow}$, derived in \cite{Bah:2021brs} for $\chi<0$,
\begin{align}
    & -\!I_6^\text{inflow,$E_4^3$} =\nonumber\\
    & - \frac{2}{3\chi^2} \sum_{i=1}^{2k} N_i \left[ \ell_i^2 N_i^2+3U_i U_{i+1}\right] \bigg(\frac{F_2^\psi}{2\pi}\bigg)^{\!3}\nonumber\\
    & + \frac{3}{2\chi} \sum_{i=1}^{2k}\left[ \frac{2\ell_i}{3\chi}N_i^3+N_i(U_i Y_{i+1}^\psi+U_{i+1} Y_i^\psi)+\frac{N}{k}(U_iU_{i+1}+\ell_i^2 N_i^2)(\delta_i^{2k}-\delta_i^{k})\right] \bigg(\frac{F_2^\psi}{2\pi}\bigg)^{\!2} \frac{F_2^\varphi}{2\pi}\nonumber\\
    & - \sum_{i=1}^{2k} \bigg[\frac{N_i^3}{3\chi^2}+N_i Y_i^\psi Y_{i+1}^\psi-\frac{N N_i}{2k \chi}(\ell_{i+1}-\ell_i)(U_i+U_{i+1})\nonumber\\
    & \phantom{- \sum_{i=1}^{2k} \bigg[ } + \frac{N}{k \chi}(\delta_i^{2k}-\delta_i^{k})\left(N_i U_i+ \chi U_i Y_i^\psi +\chi U_{i+1} Y_{i+1}^\psi\right) \bigg] \frac{F_2^\psi}{2\pi} \bigg(\frac{F_2^\varphi}{2\pi}\bigg)^{\!2}\nonumber\\
    & + \frac{3N}{2\chi} \sum_{i=1}^{2k} N_i(U_i+U_{i+1})w_\alpha^i \, \bigg(\frac{F_2^\psi}{2\pi}\bigg)^{\!2} \frac{F_2^\alpha}{2\pi} -\frac{\chi N^2}{2k} \sum_{i=1}^{2k} Y_{i+1}^\psi (\ell_{i+1}-\ell_i)(w_\alpha^{i+1}-w_\alpha^i) \bigg(\frac{F_2^\varphi}{2\pi}\bigg)^{\!2} \frac{F_2^\alpha}{2\pi}\nonumber\\
    & - 2N \sum_{i=1}^{2k} \left[ N_i\left(\frac{N_i}{2\chi}+Y_{i+1}^\psi\right) w_\alpha^i -\frac{N}{k \chi}(\delta_i^{2k}-\delta_i^k) \delta_\alpha^\mathrm{C} U_i \right] \frac{F_2^\psi}{2\pi} \frac{F_2^\varphi}{2\pi} \frac{F_2^\alpha}{2\pi}\nonumber\\
    & - N^2 \sum_{i=1}^{2k} \left[N_i w_\alpha^i w_\beta^i - (\ell_{i+1}-\ell_i)(w_\alpha^{i+1}-w_\alpha^i)(w_\beta^{i+1}-w_\beta^i)U_{i+1}\right] \frac{F_2^\psi}{2\pi}  \frac{F_2^\alpha}{2\pi} \frac{F_2^\beta}{2\pi}\nonumber\\
    & - \frac{\chi N^2}{2} \sum_{i=1}^{2k} (\ell_{i+1}-\ell_i) (w_\alpha^{i+1}-w_\alpha^i)(w_\beta^{i+1}-w_\beta^i) Y_{i+1}^\psi \, \frac{F_2^\varphi}{2\pi} \frac{F_2^\alpha}{2\pi} \frac{F_2^\beta}{2\pi}\nonumber\\
    & - \frac{\chi N^3}{6}\sum_{i=1}^{2k} (w_\alpha^{i+1}-w_\alpha^i)(w_\beta^{i+1}-w_\beta^i)\left[ \ell_{i+1} w_\gamma^i-\ell_i w_\gamma^{i+1}+(\ell_{i+1}-\ell_i)(w_\gamma^{i+1}+w_\gamma^i)\right] \frac{F_2^\alpha}{2\pi} \frac{F_2^\beta}{2\pi} \frac{F_2^\gamma}{2\pi} \, ,\label{E4cubed}
\end{align}
where $U_i \equiv N_\alpha U_0^\alpha(t_i)$, $Y_i^\psi \equiv N_\alpha Y_{0,\psi}^\alpha (t_i)$ are values of functions parameterizing basis forms in cohomology at the positions of the monopoles, while $\ell_i$, $w_\alpha^i$ are constants on the intervals composing $\partial M_2$. Explicitly, in the basis introduced in \eqref{natural_four_cycles}, we have
\begin{equation}
    \begin{gathered}
        U_i = U_1-\sum_{j=1}^{i-1} \ell_j N_j \, , \quad Y^\psi_{i=1, \cdots, k}=Y^\psi_1-\sum_{j=1}^{i-1} \frac{N_j}{\chi} \, ,  \quad  Y^\psi_{i=k+1, \cdots, 2k}=Y^\psi_1+\frac{N}{2}-\sum_{j=1}^{i-1} \frac{N_j}{\chi} \, ,\\
        w_{i=1, \cdots,k}^\alpha= -\delta_{\mathrm{N}_i}^\alpha -\frac{\delta_C^\alpha}{\chi} \, , \quad w_{i=k+1, \cdots,2k-1}^\alpha= -\delta_{\mathrm{S}_{2k-i}}^\alpha -\frac{\delta_C^\alpha}{\chi} \, , \quad w_{i=k}^\alpha=w_{i=2k}^\alpha=-\frac{\delta_C^\alpha}{\chi} \, ,\\
        \ell_{i=1, \cdots, k}=i-\frac{k}{2} \, , \qquad \ell_{i=k+1, \cdots, 2k}=\frac{3k}{2}-i \, .
    \end{gathered}
\end{equation}
To study perturbative anomalies for continuous symmetries, topological mass terms from the 11d effective action must be integrated out. As discussed in \cite{Bah:2021brs}, this can be accomplished by imposing the condition 
\begin{equation}
\sum_{\alpha} N_\alpha F_2^\alpha =0
\end{equation}
on the original set of $2k-1$ $\text{U}(1)$ field strengths associated with the non-trivial cohomology classes of $M_6$.


\section{\boldmath Computation of the $E_4 X_8$ contribution to anomaly inflow} \label{app_E4X8}

Recall that the internal space $M_6$ is a fibration of $M_4$ over the Riemann surface
$\Sigma_g$, where $M_4$ is the resolved orbifold $S^4/\mathbb Z_k$.
In order to evaluate the $E_4 X_8$ contribution to the inflow anomaly polynomial,
it is convenient to regard $M_4$ as consisting of three regions:
the region near the north pole, the region near the south pole, and the
central region. In this appendix we present a computation of the contribution from each region, using the same parameterization of $E_4$ as employed in \cite{Bah:2021brs}. Note that this parameterization assumes $\chi<0$. After the field redefinitions \eqref{dictionary}, however, the final result can be extended to the case $\chi=0$.

\subsection{Contribution from the resolved orbifold singularities} 

The contributions to $X_8$ originating from the polar regions
can be evaluated recalling that $M_4$ is an $S^1_\varphi$ fibration
over a 3d space parametrized by the $\psi$ circle and the 2d space $M_2$.
The $S^1_\varphi$ fibration has monopole sources located along the boundary of $M_2$,
grouped into a collection of $k$ monopoles with charge $+1$ in the region near the north pole,
and a collection of $k$ monopoles with charge $-1$ in the region near the south pole.
In the vicinity of each monopole, $M_4$ is locally approximated by  single-center  ALF Taub-NUT metric.
This 4d space has self-dual curvature, and correspondingly it has only one independent
Chern root. If we denote the Chern roots as $\lambda_1$, $\lambda_2$,
we have
\beq \label{TN_roots}
\lambda_1 + \lambda_2 = 0 \ , \qquad \lambda_1 - \lambda_2 = 2 \, \lambda_1 \coloneqq 2 \, \lambda \ .
\eeq
The independent Chern root $\lambda$
can be identified with the first Chern class of the $S^1_\varphi$ bundle,
which is effectively localized at the center of the Taub-NUT space.
The relations \eqref{TN_roots} also apply
if we consider a
multicenter Taub-NUT metric to model the union of the northern and southern regions.
In this case, $\lambda$ is supported at the locations of the various monopoles.

It is   useful to recall that, for a single-center ALF Taub-NUT space ${\rm TN}_n$ with
monopole charge $n$, we have \cite{Gibbons:1979gd}
\beq \label{ALF}
p_1({\rm TN}_n) = \lambda_1^2 + \lambda_2^2 = 2\, \lambda^2 \ , \qquad
\int_{{\rm TN}_n} p_1({\rm TN}_n)  = 2 \, n  \ .
\eeq
If we consider a multi-center Taub-NUT,
we can write
\beq \label{lambda_sum}
2 \, \lambda^2 = \sum_{i=1}^{2k} \, p_1({\rm TN}_{n_i}) \ ,
\eeq
where the charge $n_i$ is $+1$ for $1 \le i \le k$ (northern region)
and $-1$ for $k+1 \le i \le 2k$ (sourthern region).
We observe that the 4-form $\lambda^2$ has legs along $M_4$ only,
and is supported at the locations of the monopoles.

Upon twisting $M_4$ over the Riemann surface, and activating the external gauge fields,
the Chern roots $\lambda_1$, $\lambda_2$ get shifted: 
the sum $\lambda_1 + \lambda_2$, which is associated with the angle $\psi$
in the base of the Taub-NUT $S^1_\varphi$ fibration,
is shifted by the total connection for the angle $\psi$,
consisting of a contribution along the Riemann surface
(implementing the topological twist), and a contribution along the external spacetime,
\beq
\lambda_1 + \lambda_2 \rightarrow  \lambda_1 + \lambda_2 - \chi \, V_2^\Sigma + \frac{F_2^\psi}{2\pi} \ .
\eeq
The difference $\lambda_1 - \lambda_2$, which is instead associated with the $S^1_\varphi$
fiber, is shifted by the external gauge field for the $\varphi$ isometry,
\beq
\lambda_1 - \lambda_2 \rightarrow \lambda_1 - \lambda_2+ \frac{F_2^\varphi}{2\pi} \ .
\eeq
Recalling \eqref{TN_roots}, the Chern roots of (the polar regions of) $M_4$ after 
twisting and gauging take the form
\beq \label{final_roots}
\lambda_{1,2} =  \pm \lambda - \frac \chi 2 \, V_2^\Sigma + \frac 12 \, \frac{F_2^\psi}{2\pi}
\pm \frac 12 \, \frac{F_2^\varphi}{2\pi} \ .
\eeq

After these preliminaries, we can proceed with the computation of $X_8$
that captures the northern and southern caps of $M_4$.
To compute the Pontryagin classes of the total 11d spacetime,
we can apply the splitting principle, with reference to the schematic decomposition
\beq
TM_{11} \rightarrow TW_4 \oplus T\Sigma_g \oplus TM_4 \  ,
\eeq
using \eqref{final_roots} for the Chern roots of the last summand.
We obtain
\beq
p_1(TM_{11}) = p_1(TW_4) + \lambda_1^2 + \lambda_2^2 \ , \qquad
p_2(TM_{11}) = p_1(TW_4) \, \big( \lambda_1^2 + \lambda_2^2 \big) + \lambda_1^2 \, \lambda_2^2   \ ,
\eeq
and hence (neglecting terms with more than six legs in the external spacetime)
\beq \label{totalX8}
X_8 = \frac{1}{192} \, \bigg[ p_1(TM_{11})^2 - 4 \, p_2(TM_{11}) \bigg]
= \frac{1}{192} \, \big(  \lambda_1^2 - \lambda_2^2 \big)^2
- \frac{1}{96} \, p_1(TW_4) \, \big( \lambda_1^2 + \lambda_2^2 \big) \ .
\eeq

As noted above, $\lambda$ is supported at the locations of the monopoles.
Upon expanding \eqref{totalX8}, we encounter terms without $\lambda$, terms linear in $\lambda$,
and terms with $\lambda^2$. Higher powers of $\lambda$ vanish, because they have too
many legs along $M_4$.
Next, we observe that the terms that are linear in $\lambda$ do not contribute to $E_4\, X_8$.
This is due to the fact that, in our construction of $E_4$,
we have imposed that $E_4$ be regular as we approach the locations of the monopoles.
As a result, $E_4$ localized at a monopole cannot provide the additional $M_4$ legs
that would be necessary (together with $\lambda$) to saturate the integral in the $M_4$ directions.
Furthermore, we can drop all terms in $X_8$ that have purely external legs.
Taking these considerations into account, we see that
the relevant terms in $X_8$ are given by
\begin{align}
X_8 = \frac{\chi}{96} \, V_2^\Sigma \, \frac{F_2^\psi}{2\pi} \, \bigg[
p_1(TW_4)
- \bigg(  \frac{F_2^\varphi}{2\pi} \bigg)^2 \bigg]
+ \frac{1}{48}\,  \lambda^2 \,  \bigg[
\bigg(  \frac{F_2^\psi}{2\pi} \bigg)^2 - p_1(TW_4)
 \bigg]
-  \frac{\chi}{24} \, \lambda^2 \, V_2^\Sigma \, \frac{F_2^\psi}{2\pi}  \ .
\end{align}
We may now consider the parametrization of $E_4$ given by equation (4.8) of \cite{Bah:2021brs}, and
reported here for convenience,
\begin{align}
E_4 & = N_\alpha \, (\Omega_4^\alpha)^{\rm g}
+ N_\alpha \, (\Omega_{2,I}^\alpha)^{\rm g} \, \frac{F_2^I}{2\pi}
+ N_\alpha \, \Omega_{0,IJ}^\alpha \,  \frac{F_2^I}{2\pi} \,  \frac{F_2^J}{2\pi}
\nn \\
& \phantom{=\ } + N \, \frac{F_2^\alpha}{2\pi} \, (  \omega_{2,\alpha} )^{\rm g}
+ N \,  \frac{F_2^\alpha}{2\pi} \, \omega_{0,\alpha I} \,   \frac{F_2^I}{2\pi} + N \,\frac{ \gamma_4 }{2\pi} \ .
\end{align}
For the explicit parameterization and properties of the various forms appearing in this expansion we refer the reader to appendix B of \cite{Bah:2021brs}.
Upon  
expanding $E_4 \, X_8$  and collecting the terms that can saturate
the $M_6$ integration, we arrive at
\begin{align} \label{relevant_E4X8}
E_4 X_8 \supset & -\frac{\chi}{96} \, N_\alpha \,  \Omega_4^\alpha \,V_2^\Sigma \bigg[
\bigg(  \frac{F_2^\varphi}{2\pi}  \bigg)^2 -p_1(TW_4)
\bigg] \frac{ F_2^\psi }{2\pi} 
\nonumber \\
&+ \frac{1}{48} \, \lambda^2 \,  \bigg[
N_\alpha \, (\Omega_{2,I}^\alpha)^{\rm g} \, \frac{F_2^I}{2\pi}  
+  N \, \frac{F_2^\alpha}{2\pi} \, (  \omega_{2,\alpha} )^{\rm g}
\bigg ] \bigg[
 \bigg( \frac{ F^\psi_2 }{2\pi} \bigg)^2 - p_1(TW_4)
 \bigg] 
\nonumber \\
& -\frac{\chi}{24} \, \lambda^2 \, V_2^\Sigma \, \bigg(
 N \,  \frac{F_2^\alpha}{2\pi} \, \omega_{0,\alpha I} \,   \frac{F_2^I}{2\pi} 
+ N_\alpha \, \Omega_{0,IJ}^\alpha \,  \frac{F_2^I}{2\pi} \,  \frac{F_2^J}{2\pi}
\bigg) \, \frac{F_2^\psi}{2\pi} \ .
\end{align}
The terms with $\gamma_4$ are omitted, as
it can be easily verified that they drop from the computation,
given our prescription to compute integrals of $\lambda^2$ described below.

For the first line of \eqref{relevant_E4X8} we just need the integral
\begin{equation}
N_\alpha \int_{M_6} \Omega_4^\alpha \, V_2^\Sigma= N_\alpha \int_{M_2} dT_1^\alpha=N_\alpha \, a_
{\mathrm C}^\alpha  = N \ .
\end{equation}
The terms with $\lambda^2$ are handled recalling \eqref{ALF} and \eqref{lambda_sum},
which imply the prescription
$\lambda^2 \, \cZ \rightarrow  \sum_{i=1}^{2k} \, n_i \, \cZ(t_i)$.
Here $\cZ$ stands for an arbitrary quantity on $M_4$,
$\cZ(t_i)$ denotes $\cZ$ evaluated at the $i$-th monopole, 
and $n_i=+1$ for $i=1,\cdots,k$ and $n_i=-1$ for $i=k+1,\cdots , 2k$
are the monopole charges.
We also need to recall that
\begin{equation}
  \begin{gathered}
    N_\alpha \, \Omega^\alpha_{2,\psi} = 2\, N_\alpha \, U^\alpha_0 \, V_2^\Sigma + \dots \, , \qquad N_\alpha \, \Omega^\alpha_{2,\varphi} = - \chi \, N_\alpha \, Y^\alpha_{0,\psi} \, V_2^\Sigma + \dots \, ,\\
    \omega_{2,\alpha} = - \chi \, W^\psi_{0,\alpha}  \, V_2^\Sigma  + \dots \, , \quad \Omega^\alpha_{0,\psi \psi} = - \frac 1 \chi \, U^\alpha_0 \ , \quad 2 \, \Omega^\alpha_{0, \psi \varphi} = Y^\alpha_{0,\psi} \, , \quad \Omega^\alpha_{0, \varphi \varphi} = Y^\alpha_{0,\varphi} \, ,
  \end{gathered}
\end{equation}
where the omitted terms are not relevant for the $M_6$ integration. We also need
\beq
\omega_{0,\alpha I} = W_{0,\alpha I} \ ,
\eeq
where the $W$'s are the 0-forms that enter the parametrization of the harmonic
2-forms $\omega_{2,\alpha}$.

We find it convenient to introduce the shorthand notation
\begin{align}
  \begin{gathered}
    U_i \coloneqq N_\alpha\, U^\alpha_0 (t_i) \, , \qquad Y_i^\psi \coloneqq N_\alpha \, Y^\alpha_{0,\psi} (t_i) \, , \qquad Y_i^\varphi \coloneqq N_\alpha \, Y^\alpha_{0,\varphi} (t_i) \, ,\\
    (W_\alpha^\psi)_i \coloneqq W_{0,\alpha \psi}(t_i) \, , \qquad (W_\alpha^\varphi)_i \coloneqq W_{0,\alpha \varphi}(t_i) \, .
  \end{gathered}
\end{align}
Collecting the various contributions to the integral of \eqref{relevant_E4X8}, we obtain
\begin{align}
&\int_{M_6} E_4 X_8 \supset -\frac{\chi N}{96} \, \bigg[
\bigg( \frac{F_2^\varphi}{2\pi} \bigg)^2 -p_1(TW_4)
\bigg] \, \frac{ F_2^\psi }{2\pi} 
  \\
&+ \frac{1}{48} \, \bigg[ 
\bigg( \frac{ F_2^\psi }{2\pi} \bigg)^2  -  p_1(TW_4)
\bigg] \sum_{i=1}^k \left[ 2 \, \frac{F_2^\psi}{2\pi}   \left(U_i-U_{k+i} \right)
- \chi \, \frac{F_2^\varphi}{2\pi} \big(Y_i^\psi-Y_{k+i}^\psi \big)\right]
\nonumber \\
&-\frac{\chi N }{48} \,\bigg[ 
\bigg(   \frac{F_2^\psi}{2\pi}  \bigg)^2  -  p_1(TW_4)
\bigg] \,
\frac{ F_2^\alpha }{2\pi} \, \sum_{i=1}^k  \left[(W_\alpha^\psi)_i-(W_\alpha^\psi )_{k+i}\right]
\nonumber \\
& -\frac{\chi}{24} \, \frac{F_2^\psi}{2 \pi} \, \sum_{i=1}^k \left[ -\frac{1}{\chi} \,
\bigg(  \frac{ F_2^\psi }{2\pi}  \bigg)^2 \left(U_i-U_{k+i} \right)   
+ \frac{ F_2^\psi }{2\pi}  \, \frac{ F_2^\varphi}{2\pi }  \big(Y_i^\psi-Y_{k+i}^\psi \big)
+\bigg(  \frac{F_2^\varphi}{2\pi} \bigg )^2 \left(Y_i^\varphi-Y_{k+i}^\varphi \right)\right]
\nonumber \\
& -\frac{\chi N}{24} \, \bigg ( \frac{F _2^\psi }{2\pi}  \bigg)^2 \, 
\frac{ F_2^\alpha }{2\pi} \,  \sum_{i=1}^k \left[(W_{\alpha}^\psi)_i-(W_{\alpha}^\psi)_{k+i} \right] 
-\frac{\chi N}{24} \, \frac{F_2^\psi }{2\pi} \,  \frac{F_2^\varphi}{2\pi} \,
\frac{ F_2^\alpha }{2\pi} \sum_{i=1}^k \left[(W_{\alpha}^{\varphi})_i-(W_{\alpha}^\varphi)_{k+i} \right]  \ . \nn 
\end{align}
To evaluate the above sums, we need to recall
some relations regarding the quantities  $U$, $Y$, $W$
for our choice of basis of 4- and 2-cycles in the parameterization for $E_4$,
\begin{align}
(Y_\varphi^\alpha)_{i=1, \cdots, k}=\frac{a_\mathrm C^\alpha}{2k}  \ , & \qquad  (Y_\varphi^\alpha)_{i=k+1, \cdots, 2k}=-\frac{a_\mathrm C^\alpha}{2k} \ , 
\nonumber \\
(Y_\psi^\alpha)_{i=1, \cdots, k}=(Y_\psi^\alpha)_1-\sum_{j=1}^{i-1} \frac{a_j^\alpha}{\chi} \ , & \qquad  (Y_\psi^\alpha)_{i=k+1, \cdots, 2k}=(Y_\psi^\alpha)_1+\frac{a_\mathrm{C}^\alpha}{2}-\sum_{j=1}^{i-1} \frac{a_j^\alpha}{\chi} \ , 
\nonumber \\
U_i^\alpha = U_1^\alpha-\sum_{j=1}^{i-1} \ell_j a_j^\alpha \ , \qquad (W_\alpha^\varphi)_i&=\frac{ w_\alpha^i -w_\alpha^{i-1}}{\ell_i - \ell_{i-1}} \ , \qquad
(W_\alpha^\psi)_i=\frac{\ell_i  w_\alpha^{i-1}- \ell_{i-1} w_\alpha^i}{\ell_i  - \ell_{i-1}} \ .
\end{align}
Note also that $\ell_i- \ell_{i-1}=-(\ell_{k+i}- \ell_{k+i-1})=+1$ for $i=1, \cdots, k$. The quantities
$w_\alpha^i$ are in turn given by  
\begin{equation}
w_{i=1, \cdots,k}^\alpha= -\delta_{\mathrm{N}_i}^\alpha -\frac{\delta_\mathrm C^\alpha}{\chi} \ , \quad w_{i=k+1, \cdots,2k-1}^\alpha= -\delta_{\mathrm{S}_{2k-i}}^\alpha -\frac{\delta_\mathrm C^\alpha}{\chi} \ , \quad w_{i=k}^\alpha=w_{i=2k}^\alpha=-\frac{\delta_\mathrm C^\alpha}{\chi}  \ . 
\end{equation}
The above relations imply in particular
\begin{equation}
  \begin{gathered}
    \sum_{i=1}^k (Y_i^\varphi-Y_{k+i}^\varphi) = N \, , \quad \sum_{i=1}^k (Y_i^\psi-Y_{k+i}^\psi) =0 \, , \sum_{i=1}^k (U_i-U_{k+i}) = \sum_{i=1}^k \sum_{j=i}^{k+i-1} \ell_j N_j \, ,\\
    \sum_{i=1}^k \left[(W_{\alpha, \varphi})_i-(W_{\alpha, \varphi})_{k+i} \right] = 0 \, , \quad
    \sum_{i=1}^k \left[(W_{\alpha}^\psi)_i-(W_{\alpha}^\psi)_{k+i} \right]= \sum_{i=1}^k 2\left(\delta_{\mathrm{S}_{k-i}}^\alpha-\delta_{\mathrm{N}_i}^\alpha\right) \, .
  \end{gathered}
\end{equation}
Furthermore, we need the following identity,
\begin{align}
  \sum_{i=1}^k \sum_{j=i}^{i+k-1} \ell_j N_j & = \frac{k^2}{4} \, \chi N-2 \sum_{i,j=1}^{k-1} (\mathsf{A}_{k-1}^{-1})_{ij}(N_{\mathrm{N}_i}+N_{\mathrm{S}_{k-j}}) \nn \\
  & = \frac{k^2}{4} \, \chi N-2 \sum_{i,j=1}^{k-1} (N_{\tilde{\beta}_i}+N_{\tilde{\gamma}_i}) \, ,
\end{align}
where $\mathsf A_{k-1}$ is the Cartan matrix of $\mathfrak{su}(k)$ and 
  in the second step we have used the dictionary \eqref{dictionary}
between the flux quanta in the M-theory setup,
and the flavor fluxes of the class $\cS_k$ reduction.

It follows that the contribution from $E_4 X_8$ to $- I_6^\text{inflow,$E_4 X_8$}$ capturing the polar caps of the resolved orbifold $M_4$ can be written as
\begin{align} \label{polar_caps}
(- I_6^\text{inflow,$E_4 X_8$})^\text{polar caps} & = -\frac{\chi N}{96}  \, \bigg[\bigg(\frac{F_2^\varphi}{2\pi}\bigg)^2 - p_1(TW_4)\bigg] \, \frac{F_2^\psi}{2\pi} -\frac{\chi N}{24} \,  \bigg(  \frac{ F_2^\varphi }{2\pi} \bigg)^2 \, \frac{F_2^\psi}{2\pi}\nonumber\\
& \phantom{=\ } + \frac{1}{24} \,\bigg[2 \, \bigg(\frac{F_2^\psi}{2\pi}\bigg)^2 - p_1(TW_4)\bigg] \, \frac{F_2^\psi}{2\pi} \, \bigg[\frac{k^2 \chi N}{4} - 2 \sum_{i,j=1}^{k-1} (N_{\tilde{\beta}_i}+N_{\tilde{\gamma}_i})\bigg]\nonumber\\
& \phantom{=\ } + \frac{\chi N}{24} \, \bigg[3 \, \bigg(\frac{F_2^\psi}{2\pi}\bigg)^2 -  p_1(TW_4)\bigg] \sum_{i=1}^{k-1} \bigg(\frac{F_2^{\mathrm{N}_i}}{2\pi} - \frac{F_2^{\mathrm{S}_i}}{2\pi}\bigg) \, . 
\end{align}

\subsection{Other contributions and final result}

The remaining contribution to consider is associated with the central region in $M_4$,
between the two polar caps. We may extract this contribution as follows,
\beq \label{central}
(- I_6^\text{inflow,$E_4X_8$})^\text{central region} 
= (- I_6^\text{inflow,$E_4X_8$})^\text{$S^4$}  - (- I_6^\text{inflow,$E_4X_8$})^\text{polar caps}_{k=1}  \ .
\eeq
The quantity $(- I_6^\text{inflow,$E_4X_8$})^\text{polar caps}_{k=1}$ is simply
\eqref{polar_caps} evaluated for $k=1$, with the convention that the sums over $i$
with range $1$ to $k-1$ be simply dropped.
The quantity $(- I_6^\text{inflow,$E_4X_8$})^\text{$S^4$}$ is the $E_4 X_8$
contribution to anomaly inflow for an (unorbifolded) $S^4$ fibered over the Riemann surface.
The rationale behind \eqref{central} is that the central region can be obtained
starting from $S^4$ and removing the polar caps without orbifold, which
are captured as the special case $k=1$ of the computation of the previous subsection.

The quantity $(- I_6^\text{inflow,$E_4X_8$})^\text{$S^4$}$ 
corresponds to a special case of the 
 Bah-Beem-Bobev-Wecht (BBBW)
   setup \cite{Bah:2012dg},
which is in general parameterized by two integer twist parameters
$q_1$, $q_2$, subject to the constraint $q_1 + q_2 = - \chi/2$.
In this work, we are turning off the flavor twist parameter $\zeta$
(governing a twist of the $\varphi$ isometry along the Riemann surface),
which corresponds to the case $q_1 = q_2 = - \chi/2$.
Both the $E_4^3$ and the $E_4 X_8$ contributions to 
anomaly inflow were studied for general $q_1$, $q_2$ in \cite{Bah:2019rgq},
with the results
\begin{align} \label{BBBW}
(- I_6^\text{inflow, $E_4^3$})^\text{$S^4$} & = \frac 23 \, N^3 \,
(q_1 \, n_1 \, n_2^2 + q_2 \, n_2 \, n_1^2)
 \ ,   \\
(- I_6^\text{inflow, $E_4X_8$})^\text{$S^4$} & = 
- \frac{N}{24} \, (q_1 \, n_1 + q_2 \, n_2) \, p_1(TW_4)
+ \frac N 6 \, (q_1 \, n_1^3 + q_2 \, n_2^3)\nn\\
& \phantom{=\ } - \frac N6 \, (q_1 \, n_1 \, n_2^2 + q_2 \, n_2 \, n_1^2) \ . \nn
\end{align}
The quantities $n_1$, $n_2$ are the first Chern classes of the external
gauge fields associated with the $U(1)^2$ isometry of the $S^4$ fibration over $\Sigma_g$
that is visible 
for generic $q_1$, $q_2$. The relation to the background fields $F_2^\psi$, $F_2^\varphi$ 
considered in this work is
\beq \label{BBBW_dictionary}
n_1 = \frac 14 \, \frac{F_2^\psi}{2\pi} +\frac 12 \, \frac{F_2^\varphi}{2\pi}  \ , \qquad 
n_2 = \frac 14 \, \frac{F_2^\psi}{2\pi} - \frac 12 \, \frac{F_2^\varphi}{2\pi} \ .
\eeq

We are now in a position to write down the final answer for the $E_4 X_8$
contribution to anomaly inflow. This quantity is the sum of \eqref{polar_caps}
and \eqref{central}, with $(- I_6^\text{inflow,$E_4X_8$})^\text{$S^4$}$ extracted from
\eqref{BBBW} with the identifications \eqref{BBBW_dictionary} and $q_1 = q_2 = - \chi/2$.
In conclusion,
we arrive at
\begin{align}
-I_6^\text{inflow, $E_4 X_8$}
& = - \frac{N \, \chi}{24} \, \frac{ F_2^\psi}{2\pi} \, \bigg (  \frac{F_2^\varphi}{2\pi} \bigg)^2
+ \frac{N \, \chi}{24} \, \bigg[ 3 \, \bigg( \frac{F_2^\psi}{2\pi} \bigg)^2 - p_1(TW_4) \bigg] \, 
\sum_{i=1}^{k-1} \bigg( \frac{ F_2^{\mathrm N_i} }{2\pi} -  \frac{ F_2^{\mathrm S_i}  }{2\pi}\bigg)
\nn \\
& \phantom{=\ } + \frac{F_2^\psi}{2\pi} \, p_1(TW_4) \, \bigg[ 
- \frac{N \, \chi}{96} \, (k^2-2) + \frac{1}{12} \, 
\sum_{i=1}^{k-1} \Big( N_{\widetilde \beta_i} + N_{\widetilde \gamma_i}  \Big)
\bigg]
\nn \\
& \phantom{=\ } + \bigg( \frac{F_2^\psi}{2\pi} \bigg)^3 \, \bigg[
 \frac{N \, \chi}{48} \, (k^2 - 1)  - \frac 16 \, \sum_{i = 1}^{k-1}  \Big( N_{\widetilde \beta_i} + N_{\widetilde \gamma_i}  \Big)
\bigg] \ .
\end{align}

\subsection{Comparison with \cite{Bah:2019vmq}}

The full inflow anomaly polynomial $- I_6^{\rm inflow}$ for general $k$,
including the $E_4 X_8$ contribution evaluated above, 
agrees with the results of \cite{Bah:2019vmq} for $k=2$.
To verify this, one needs to perform a redefinition of flux parameters   and external
gauge fields. The quantities in \cite{Bah:2019vmq} are related to those in this paper
and \cite{Bah:2021brs} as in the following table.
\begingroup
\renewcommand{\arraystretch}{2.}
\begin{center}
\begin{tabular}{cccc}
notation of \cite{Bah:2019vmq} & & notation of this work and \cite{Bah:2021brs}  \\ \hline
$N_\mathrm N$ & $=$ & $- N_{\mathrm S_1}$    \\
$N_\mathrm S$ & $=$ & $+ N_{\mathrm N_1}$    \\
$c_1^\mathrm N$  & $=$ & $\displaystyle    
- \frac{F_2^{\mathrm S_1}}{2\pi} + \frac{ (N \, \chi - N_{\mathrm S_1} ) ( N \, \chi - N_{\mathrm N_1} - N_{\mathrm S_1} ) }{ 2 \, N \, \chi \, (2 \, N \, \chi - N_{\mathrm N_1} - N_{\mathrm S_1}) } \, \frac{F_2^\psi}{2\pi}
$   \\
$c_1^\mathrm S$  & $=$ & $\displaystyle
+ \frac{ F_2^{\mathrm N_1}}{2\pi} + \frac{ (N \, \chi - N_{\mathrm N_1} ) ( N \, \chi - N_{\mathrm N_1} - N_{\mathrm S_1} ) }{ 2 \, N \, \chi \, (2 \, N \, \chi - N_{\mathrm N_1} - N_{\mathrm S_1}) } \, \frac{F_2^\psi }{2\pi}
$   \\
$c_1^\psi$  & $=$ & $\displaystyle \frac 12 \,  \frac{F^\psi}{2\pi}$   \\
$p_1^\varphi$  & $=$ & $ \bigg(\displaystyle \frac{F_2^\varphi}{2\pi} \bigg)^2$  
\end{tabular}
\end{center}
\endgroup

 
\section{Anomaly inflow computation for the torus case}\label{app_torus_inflow}

We derive in this appendix the $E_4^3$ contribution to the inflow anomaly polynomial, $I_6^\text{inflow,$E_4^3$}$, for the case of genus one, following the general philosophy adopted by \cite{Bah:2021brs} for the construction of the higher-genus inflow anomaly polynomial. We are going to describe only the essential elements which distinguish this computation from the original one, and we refer the reader to \cite{Bah:2021brs} for a detailed discussion of the formalism.

The Euler characteristic of the torus is $\chi = 2 - 2g = 0$, so it makes the topological twist of the $\mathrm{U}(1)_\psi$ bundle trivial while preserving $\mathcal{N} = 1$ supersymmetry \cite{Bah:2015fwa,Bah:2017wxp}. Specifically, the global angular forms associated with the isometries reduce respectively to
\begin{equation}
	(D\psi)^\mathrm{g} = d\psi + A_1^\psi \, , \qquad (D\varphi)^\mathrm{g} = d\varphi - L \, (d\psi + A_1^\psi) + A_1^\varphi \, .
\end{equation}
Cohomology class representatives of $H^2(M_6)$ and $H^4(M_6)$ can be written as
\begin{align}
	\omega_{2,\alpha} & = (dW_{0,\alpha}^\psi + L \, dW_{0,\alpha}^\varphi) \wedge \frac{d\psi}{2\pi} + dW_{0,\alpha}^\varphi \wedge \frac{D\varphi}{2\pi} + W_{0,\alpha}^\Sigma V_2^\Sigma \, ,\\
	\Omega_4^\alpha & = dT_1^\alpha \wedge \frac{d\psi}{2\pi} \wedge \frac{D\varphi}{2\pi} + (dU_{0,\psi}^\alpha + L \, dU_{0,\varphi}^\alpha) \wedge \frac{d\psi}{2\pi} \wedge V_2^\Sigma + dU_{0,\varphi}^\alpha \wedge \frac{D\varphi}{2\pi} \wedge V_2^\Sigma \, ,
\end{align}
where $W_{0,\alpha}^\Sigma$ is a constant such that $\omega_{2,\alpha}$ is closed. Note especially that the expression for $\Omega_4^\alpha$ is not the same as the na\"{i}ve $\chi = 0$ limit of (B.17) in \cite{Bah:2021brs}, otherwise the flux quanta $N_i = \int_{\mathcal{C}_{4,i}} N_\alpha \Omega_4^\alpha$ for $i = 1,2,\dots,2k$ are ill-defined. This is also necessary so as to recover the sum rules,
\begin{equation}
	\sum_{i=1}^{2k} a_i^\alpha = 0 \, , \qquad \sum_{i=1}^{2k} \ell_i a_i^\alpha = 0 \, ,\label{torus_sum_rules}
\end{equation}
that are consistent with their higher-genus cousins. As a reminder, we choose to parameterize a given basis of (co)homology classes using the expansion coefficients,
\begin{equation}
	\mathcal{C}_{4,\mathrm{C}} = a_\mathrm{C}^\alpha \, \mathcal{C}_{4,\alpha} \, , \quad \mathcal{C}_{4,i} = a_i^\alpha \, \mathcal{C}_{4,\alpha} \, , \quad \mathcal{C}_2^{\Sigma,i} = b^{\Sigma,i}_\alpha \, \mathcal{C}_2^\alpha \, , \quad \mathcal{C}_2^i = b^i_\alpha \, \mathcal{C}_2^\alpha \, ,
\end{equation}
where the various cycles are introduced in section \ref{sec_review_flux}. These coefficients can be expressed in terms of various auxiliary differential forms defined earlier,
\begin{equation}
	a_\mathrm{C}^\alpha = \int_{\partial M_2} T_1^\alpha \, , \quad a_i^\alpha = U_{0,\varphi}^\alpha(t_{i+1}) - U_{0,\varphi}^\alpha(t_i) \, , \quad b^{\Sigma,i}_\alpha = W_{0,\alpha}^\Sigma \, , \quad b^i_\alpha = W_{0,\alpha}^\varphi(t_{i+1}) - W_{0,\alpha}^\varphi(t_i) \, .
\end{equation}

Recall that the four-form flux $E_4$ (restricting only to continuous zero-form symmetries) can be expanded as follows,
\begin{equation}
	E_4 = N_\alpha \big(\Omega_4^\alpha\big)^\mathrm{g} + N_\alpha \big(\Omega_{2,I}^\alpha\big)^\mathrm{g} \, \frac{F_2^I}{2\pi} + N (\omega_{2,\alpha})^\mathrm{g} \, \frac{F_2^\alpha}{2\pi} + N_\alpha \Omega_{0,IJ}^\alpha \, \frac{F_2^I}{2\pi} \frac{F_2^J}{2\pi} + N \omega_{0,\alpha I} \, \frac{F_2^I}{2\pi} \frac{F_2^\alpha}{2\pi} \, .
\end{equation}
One can show that the following choice of forms,
\begin{equation}
	\begin{gathered}
		\Omega_{2,\psi}^\alpha = (dX_0^\alpha - L T_1^\alpha) \wedge \frac{d\psi}{2\pi} - T_1^\alpha \wedge \frac{D\varphi}{2\pi} + U_{0,\psi}^\alpha V_2^\Sigma \, ,\\
		\Omega_{2,\varphi}^\alpha = (dU_{0,\varphi}^\alpha + L \, dY_0^\alpha + T_1^\alpha) \wedge \frac{d\psi}{2\pi} + dY_0^\alpha \wedge \frac{D\varphi}{2\pi} + U_{0,\varphi}^\alpha V_2^\Sigma \, ,\\
		\Omega_{0,\psi\psi}^\alpha = X_0^\alpha \, , \qquad \Omega_{0,\psi\varphi}^\alpha = \frac{1}{2} \, U_{0,\varphi}^\alpha \, , \qquad \Omega_{0,\varphi\varphi}^\alpha = Y_0^\alpha \, ,
	\end{gathered}
\end{equation}
are compatible with the closure and regularity constraints on $E_4$. Applying the convention
\begin{equation}
	N_\alpha N_\beta \mathcal{J}_I^{\alpha\beta} = 0\label{decoupling_convention}
\end{equation}
for each $I \in \{\psi,\varphi\}$, we can uniquely fix the reference values,
\begin{align}
	\tilde{U}_{0,\psi}(t_1) & = \frac{1}{N} \sum_{i=2}^{2k} \Big[N_i + \ell_i \big(\tilde{Y}_{0,\varphi}(t_{i+1}) - \tilde{Y}_{0,\varphi}(t_i)\big)\Big] \sum_{j=1}^{i-1} (\ell_i - \ell_j) N_j \, ,\\
	\tilde{U}_{0,\varphi}(t_1) & = \frac{1}{N} \bigg[\sum_{i=1}^{2k} \ell_i N_i \tilde{Y}_{0,\varphi}(t_{i+1}) + \sum_{i=2}^{2k} \ell_i \big(\tilde{Y}_{0,\varphi}(t_{i+1}) - \tilde{Y}_{0,\varphi}(t_i)\big) \sum_{j=1}^{i-1} N_j\bigg] \, ,
\end{align}
where we used the shorthand notation $N_i = N_\alpha a_i^\alpha$, $\tilde{U}_{0,\psi} = N_\alpha U_{0,\psi}^\alpha$, $\tilde{U}_{0,\varphi} = N_\alpha U_{0,\varphi}^\alpha$, $\tilde{Y}_0 = N_\alpha Y_0^\alpha$.

The resulting inflow anomaly polynomial can be written as
\begin{align}
	& I_6^\text{inflow,$E_4^3$} =\nonumber\\
	& \sum_{i=1}^{2k} \tilde{u}_i \bigg[\big(\tilde{U}_{0,\varphi} + \ell_i \tilde{Y}_0\big) \bigg[X_0 + \frac{1}{2} \, \ell_i \big(\tilde{U}_{0,\varphi} + \ell_i \tilde{Y}_0\big)\bigg]\bigg]^{t_{i+1}}_{t_i} \bigg(\frac{F_2^\psi}{2\pi}\bigg)^{\!3}\nonumber\\
	& + \sum_{i=1}^{2k} \bigg[\frac{1}{2} \, \ell_i \tilde{U}_{0,\varphi}^3 + \tilde{u}_i \bigg(\tilde{X}_0 + \ell_i \tilde{U}_{0,\varphi} + \frac{1}{2} \, \ell_i^2 \tilde{Y}_0\bigg) \tilde{Y}_0\bigg]^{t_{i+1}}_{t_i} \bigg(\frac{F_2^\psi}{2\pi}\bigg)^{\!2} \frac{F_2^\varphi}{2\pi}\nonumber\\
	& + \sum_{i=1}^{2k} \ell_i \bigg[\bigg(\tilde{U}_{0,\varphi}^2 + \frac{1}{2} \, \tilde{u}_i \tilde{Y}_0\bigg) \tilde{Y}_0\bigg]^{t_{i+1}}_{t_i} \frac{F_2^\psi}{2\pi} \bigg(\frac{F_2^\varphi}{2\pi}\bigg)^{\!2}\nonumber\\
	& + \sum_{i=1}^{2k} \frac{1}{2} \bigg[\tilde{U}_{0,\varphi} \tilde{Y}_0 \big(\tilde{U}_{0,\varphi} + \ell_i \tilde{Y}_0\big)\bigg]^{t_{i+1}}_{t_i} \bigg(\frac{F_2^\varphi}{2\pi}\bigg)^{\!3}\nonumber\\
	& + \sum_{i=1}^{2k} N \bigg[W_{0,\alpha}^\psi \tilde{U}_{0,\varphi} \tilde{X}_0 + \tilde{u}_i \Big[W_{0,\alpha}^\varphi \tilde{X}_0 + w_{\alpha,i} \big(\tilde{U}_{0,\varphi} + \ell_i \tilde{Y}_0\big)\Big]\nonumber\\
	& \phantom{+ \sum_{i=1}^{2k} N \bigg[\ } + b^\Sigma_\alpha \ell_i \bigg[\tilde{X}_0 \tilde{Y}_0 + \frac{1}{2} \big(\tilde{U}_{0,\varphi} + \ell_i \tilde{Y}_0\big)^2\bigg]\bigg]^{t_{i+1}}_{t_i} \bigg(\frac{F_2^\psi}{2\pi}\bigg)^{\!2} \frac{F_2^\alpha}{2\pi}\nonumber\\
	& + \sum_{i=1}^{2k} N \bigg[\frac{1}{2} \big(w_{\alpha,i} + \ell_i W_{0,\alpha}^\varphi\big) \tilde{U}_{0,\varphi}^2 + \ell_i w_{\alpha,i} \tilde{u}_i Y_0 + \frac{1}{2} \, b^\Sigma_\alpha \big(\tilde{U}_{0,\varphi} + \ell_i \tilde{Y}_0\big)^2\bigg]^{t_{i+1}}_{t_i} \frac{F_2^\psi}{2\pi} \frac{F_2^\varphi}{2\pi} \frac{F_2^\alpha}{2\pi}\nonumber\\
	& + \sum_{i=1}^{2k} N \bigg[\ell_i W_{0,\alpha}^\varphi \tilde{U}_{0,\varphi} \tilde{Y}_0 + b^\Sigma_\alpha \bigg(\tilde{U}_{0,\varphi} + \frac{1}{2} \, \ell_i \tilde{Y}_0\bigg) \tilde{Y}_0\bigg]^{t_{i+1}}_{t_i} \bigg(\frac{F_2^\varphi}{2\pi}\bigg)^{\!2} \frac{F_2^\alpha}{2\pi}\nonumber\\
	& + \sum_{i=1}^{2k} N^2 \bigg[\tilde{u}_i \bigg(W_{0,\alpha}^\psi + \frac{1}{2} \, \ell_i W_{0,\alpha}^\varphi\bigg) W_{0,\beta}^\varphi + b^\Sigma_\alpha w_{\beta,i} \big(\tilde{U}_{0,\varphi} + \ell_i \tilde{Y}_0\big)\bigg]^{t_{i+1}}_{t_i} \frac{F_2^\psi}{2\pi} \frac{F_2^\alpha}{2\pi} \frac{F_2^\beta}{2\pi}\nonumber\\
	& + \sum_{i=1}^{2k} N^2 \ell_i \bigg[\frac{1}{2} \, W_{0,\alpha}^\varphi W_{0,\beta}^\varphi \tilde{U}_{0,\varphi} + b^\Sigma_\alpha W_{0,\beta}^\varphi \tilde{Y}_0\bigg]^{t_{i+1}}_{t_i} \frac{F_2^\varphi}{2\pi} \frac{F_2^\alpha}{2\pi} \frac{F_2^\beta}{2\pi}\nonumber\\
	& + \sum_{i=1}^{2k} \frac{N^3}{2} \, b^\Sigma_\alpha \ell_i \bigg[W_{0,\beta}^\varphi W_{0,\gamma}^\varphi\bigg]^{t_{i+1}}_{t_i} \frac{F_2^\alpha}{2\pi} \frac{F_2^\beta}{2\pi} \frac{F_2^\gamma}{2\pi} \, ,\label{torus_inflow_anomaly_polynomial}
\end{align}
subject to the condition that $N_\alpha F_2^\alpha = 0$. The values of the various auxiliary functions evaluated at any $t_i$ are given by
\begin{align}
	\ell_i & = \begin{cases} \displaystyle i - \frac{k}{2} & \mathrm{if} \ 1 \leq i \leq k \, ,\\[2ex] \displaystyle \frac{3k}{2} - i & \mathrm{if} \ k + 1 \leq i \leq 2k \, ,\end{cases}\\
	\tilde{u}_i & = \tilde{U}_{0,\psi}(t_1) + \ell_i \tilde{U}_{0,\varphi}(t_1) +  \sum_{j=1}^{i-1} (\ell_i - \ell_j) N_j \, ,\\
	\tilde{U}_{0,\psi}(t_i) & = \tilde{U}_{0,\psi}(t_1) -  \sum_{j=1}^{i-1} \ell_j N_j \, ,\\
	\tilde{U}_{0,\varphi}(t_i) & = \tilde{U}_{0,\varphi}(t_1) +  \sum_{j=1}^{i-1} N_j \, ,\\
	\tilde{Y_0}(t_i) & = \begin{cases} \displaystyle \frac{N}{2k} & \, \mathrm{if} \ 1 \leq i \leq k \, ,\\[2ex] \displaystyle -\frac{N}{2k} & \, \mathrm{if} \ k + 1 \leq i \leq 2k \, ,\end{cases}\\
	\tilde{X_0}(t_i) & = - \sum_{j=1}^{i-1} \Big[\ell_j N_j + \ell_j^2 \big(\tilde{Y_0}(t_{j+1}) - \tilde{Y_0}(t_j)\big)\Big] \, ,\\
	W_{0,\alpha}^\psi(t_i) & = \frac{\ell_i w_{\alpha,i-1} - \ell_{i-1} w_{\alpha,i}}{\ell_i - \ell_{i-1}} \, ,\\
	W_{0,\alpha}^\varphi(t_i) & = \frac{w_{\alpha,i} - w_{\alpha,i-1}}{\ell_i - \ell_{i-1}} \, ,
\end{align}
and in a Poincar\'{e}-dual basis of (co)homology classes as introduced in section \ref{sec_review_flux}, we have
\begin{equation}
	w_{\alpha,i=1,\dots,k-1} = -\delta_{\alpha,\mathrm{N}_i} \, , \quad w_{\alpha,i=k+1,\dots,2k-1} = -\delta_{\alpha,\mathrm{S}_{2k-i}} \, , \quad w_{\alpha,k} = w_{\alpha,2k} = 0 \, , \quad b^\Sigma_\alpha = \delta_{\alpha,\mathrm{C}} \, .
\end{equation}

A prescription \eqref{dictionary} is provided in the main text to obtain an inflow anomaly polynomial that has a well-defined $\chi = 0$ limit. It turns out that we can exactly reproduce this anomaly polynomial by taking \eqref{torus_inflow_anomaly_polynomial} and carrying out the replacements,
\begin{equation}
	\begin{gathered}
		\frac{F_2^\psi}{2\pi} = 2 c_1(R') \, , \qquad \frac{F_2^\varphi}{2\pi} = -k c_1(t) \, ,\\
		\frac{F_2^{\mathrm{N}_i}}{2\pi} = c_1(\widetilde{\beta}_i) - 2 N_{\widetilde{\beta}_i} c_1(R') \, , \qquad \frac{F_2^{\mathrm{S}_{k-i}}}{2\pi} = -c_1(\widetilde{\gamma}_i) + 2 N_{\widetilde{\gamma}_i} c_1(R') \, ,\\
		N_{\mathrm{N}_i} = \sum_{j=1}^{k-1} (\mathsf{A}_{k-1})_{ij} \, N_{\widetilde{\beta}_j} \, , \qquad N_{\mathrm{S}_{k-i}} = \sum_{j=1}^{k-1} (\mathsf{A}_{k-1})_{ij} \, N_{\widetilde{\gamma}_j} \, .
	\end{gathered}
\end{equation}
To conclude, this independent derivation for genus one from first principles confirms the validity of using \eqref{dictionary} to acquire an inflow anomaly polynomial for arbitrary $g \geq 1$, including the case $\chi = 0$.









\section{Change of basis between flavor and resolution flux quanta}
\label{app_change_of_basis}

From the perspective of an 11d flux background probed by an M5-brane stack, the flux quanta $N_\alpha$ appearing in the expansion of $E_4$ represent the amount of flux threading the 4-cycles $\mathcal{C}_{4,\alpha}$. Therefore, it is natural to express these flux quanta with respect to a (co)homology basis determined by an intuitive choice of 4-cycles, namely,
\begin{equation}
	\mathcal{C}_{4,\alpha=\mathrm{N}_1, \dots, \mathrm{N}_{k-1}} = \mathcal{C}_{4}^{i=1,\dots,k-1}
	\, , \quad \mathcal{C}_{4,\alpha=C} = \mathcal{C}_{4}^{\mathrm{C}} \, , \quad \mathcal{C}_{4,\alpha=\mathrm{S}_1,\dots, \mathrm{S}_{k-1}}= \mathcal{C}_4^{i=2k-1,\dots,k+1} \, ,
\end{equation}
as in \eqref{natural_four_cycles}. One can then proceed to define the ``natural'' basis 2-cycles $\mathcal{C}_2^\alpha$ to be those that are Poincar\'{e}-dual to these basis 4-cycles. For simplicity, the basis of (co)homology classes described above will hereafter be referred to as the ``resolution flux basis.'' On the other hand, as explained in section \ref{basis_interp}, the starting point for the basis  implicitly used in the compactification of the 6d $\mathcal{N}=(1,0)$ theories is instead an intuitive set of basis 2-cycles,
\begin{equation}
	\big(\mathcal{C}_{2}^{\alpha=\mathrm{N}_1, \dots, \mathrm{N}_{k-1}}\big)' = \mathcal{C}_{2}^{i=1,\dots,k-1}
	\, , \quad \big(\mathcal{C}_{2}^{\alpha=C}\big)' = \mathcal{C}_{2}^{\mathrm{C}} \, , \quad \big(\mathcal{C}_2^{\alpha=\mathrm{S}_1,\dots, \mathrm{S}_{k-1}}\big)'= \mathcal{C}_2^{i=2k-1,\dots,k+1} \, ,
\end{equation}
where $\mathcal{C}_2^{\mathrm{C}}$ is defined to be the Poincar\'{e} dual of $\mathcal{C}_{4,\mathrm{C}} = M_4$. The rest of the basis 4-cycles $(\mathcal{C}_{4,\alpha})'$ can be similarly defined to be the Poincar\'{e} duals of $(\mathcal{C}_2^\alpha)'$. We will hereafter refer to this basis of (co)homology classes as the ``flavor flux basis.''

In this appendix we find the transformation matrix that relates the resolution flux basis and the flavor flux basis. Using the relations following from Poincar\'{e} duality and regularity of the basis two-forms in appendix C of \cite{Bah:2021brs}, we can express the basis 2-cycles $\mathcal{C}_2^\alpha$ in the resolution flux basis in terms of familiar 2-cycles. Let us start with the case of $k=2$. One finds that
\begin{equation}
	\begin{gathered}
		\mathcal{C}_2^{\Sigma,1} = \mathcal{C}_2^{\Sigma,2} = \chi \, \mathcal{C}_2^{\mathrm{N}_1} + \mathcal{C}_2^\mathrm{C} \, , \qquad \mathcal{C}_2^{\Sigma,3} = \mathcal{C}_2^{\Sigma,4} = \chi \, \mathcal{C}_2^{\mathrm{S}_1} + \mathcal{C}_2^C \, ,\\
		\mathcal{C}_2^{i=1} = 2 \, \mathcal{C}_2^{\mathrm{N}_1} \, , \qquad \mathcal{C}_2^{i=2} = \mathcal{C}_2^{i=4} = -\mathcal{C}_2^{\mathrm{N}_1} + \mathcal{C}_2^{\mathrm{S}_1} \, , \qquad \mathcal{C}_2^{i=3} = -2 \, \mathcal{C}_2^{\mathrm{S}_1} \, ,
	\end{gathered}
\end{equation}
which can be inverted to give
\begin{equation}
	\begin{aligned}
		\mathcal{C}_2^{\mathrm{N}_1} & = \frac{1}{2} \, \mathcal{C}_2^{i=1} \, ,\\
		\mathcal{C}_2^\mathrm{C} & = \mathcal{C}_2^{\Sigma,1} - \frac{\chi}{2} \, \mathcal{C}_2^{i=1} = \mathcal{C}_2^{\Sigma,3} + \frac{\chi}{2} \, \mathcal{C}_2^{i=3} \, ,\\
		\mathcal{C}_2^{\mathrm{S}_1} & = -\frac{1}{2} \, \mathcal{C}_2^{i=3} \, .
	\end{aligned}
\end{equation}
Meanwhile, the basis 2-cycles in the flavor basis are
\begin{equation}
	\begin{aligned}
		\big(\mathcal{C}_2^{\mathrm{N}_1}\big)' & = \mathcal{C}_2^{i=1} = 2 \, \mathcal{C}_2^{\mathrm{N}_1} \, ,\\
		\big(\mathcal{C}_2^\mathrm{C}\big)' & = \mathcal{C}_2^\mathrm{C} \, ,\\
		\big(\mathcal{C}_2^{\mathrm{S}_1}\big)' & = \mathcal{C}_2^{i=3} = -2 \, \mathcal{C}_2^{\mathrm{S}_1} \, ,
	\end{aligned}
\end{equation}
so we can express the basis transformation compactly as $(\mathcal{C}_2^{\alpha})' = (\mathcal{R}_2^{-1})^\alpha_\beta \, \mathcal{C}_2^\beta$ where
\begin{equation}
	\mathcal{R}_2^{-1} = \begin{pmatrix} 2 & 0 & 0 \\ 0 & 1 & 0 \\ 0 & 0 & -2 \end{pmatrix} \, .
\end{equation}
The orthonormal pairing between cycles and cohomology representatives is preserved if $(\omega_{2,\beta})' = (\mathcal{R}_2)^\alpha_\beta \, \omega_{2,\alpha}$, such that
\begin{equation}
	\int_{(\mathcal{C}_2^\alpha)'} (\omega_{2,\beta})' = \int_{(\mathcal{R}_2^{-1})^\alpha_\gamma \, \mathcal{C}_2^\gamma} (\mathcal{R}_2)^\delta_\beta \, \omega_{2,\delta} = (\mathcal{R}_2^{-1})^\alpha_\gamma \, (\mathcal{R}_2)^\delta_\beta \, \delta^\gamma_\delta = \delta^\alpha_\beta \, .
\end{equation}
Demanding Poincar\'{e} duality between the 2-cycles $(\mathcal{C}_2^{\alpha})'$ and some set of 4-cycles $(\mathcal{C}_{4,\alpha})'$ amounts to requiring
\begin{equation}
	\delta^\alpha_\beta = \int_{M_6} (\Omega_4^\alpha)' \wedge (\omega_{2,\beta})' = \int_{M_6} (\mathcal{R}_4)^\alpha_\gamma \, \Omega_4^\gamma \wedge (\mathcal{R}_2)_\beta^\delta \, \omega_{2,\delta} = (\mathcal{R}_4 \mathcal{R}_2)^\alpha_\beta \, ,
\end{equation}
which means that the 4-cycles that are Poincar\'{e}-dual to the basis 2-cycles in the flavor basis can be expressed as $(\mathcal{C}_{4,\alpha})' = (\mathcal{R}_4^{-1})^\alpha_\beta \, \mathcal{C}_{4,\beta}$ with
\begin{equation}
	\mathcal{R}_4^{-1} = \mathcal{R}_2 = \begin{pmatrix} \displaystyle \frac{1}{2} & 0 & 0 \\[1.5ex] 0 & 1 & 0 \\[1.5ex] 0 & 0 & \displaystyle -\frac{1}{2} \end{pmatrix} \, .
\end{equation}
As discussed in appendix D of \cite{Bah:2021brs}, the flux quanta associated with the new and old bases of 4-cycles are related by $N_\alpha' = (\mathcal{R}_4^{-1})_\alpha^\beta \, N_\beta$ to make the inflow anomaly polynomial invariant, and hence we get\footnote{Note that the requirement of having no topological mass term associated with $\gamma_4$ in the inflow anomaly polynomial is preserved, i.e.~$N_\alpha' (F_2^\alpha)' = (\mathcal{R}_4^{-1})_\alpha^\gamma N_\gamma \, (\mathcal{R}_2^{-1})^\alpha_\delta F_2^\delta = (\mathcal{R}_2 \mathcal{R}_2^{-1})^\gamma_\delta N_\gamma F_2^\delta = N_\alpha F_2^\alpha = 0$, assuming that we only restrict to the zero-form symmetries and also follow the convention that $N_\alpha N_\beta \mathcal{J}_I^{\alpha\beta} = 0$.}
\begin{equation}
N_{\mathrm{N}_1}' = \frac{1}{2} \, N_{\mathrm{N}_1} \, , \qquad N_\mathrm{C}' = N_\mathrm{C} = N \, , \qquad N_{\mathrm{S}_1}' = -\frac{1}{2} \, N_{\mathrm{S}_1} \, .
\end{equation}
As noted previously, in order to preserve positivity of the flux quanta an additional overall sign flip must be included in the basis transformation on southern quantities.

We now move on to the case of $k=3$. By repeating the same exercise as before, we derive that in the resolution flux basis,
\begin{equation}
	\begin{aligned}
		\mathcal{C}_2^{\mathrm{N}_1} & = \frac{1}{3} \big(2 \, \mathcal{C}_2^{i=1} + \mathcal{C}_2^{i=2}\big) \, ,\\
		\mathcal{C}_2^{\mathrm{N}_2} & = \frac{1}{3} \big(\mathcal{C}_2^{i=1} + 2 \, \mathcal{C}_2^{i=2}\big) \, ,\\
		\mathcal{C}_2^\mathrm{C} & = \mathcal{C}_2^{\Sigma,1} - \frac{\chi}{2} \big(2 \, \mathcal{C}_2^{i=1} + \mathcal{C}_2^{i=2}\big) \, ,\\
		\mathcal{C}_2^{\mathrm{S}_2} & = -\frac{1}{3} \big(2 \, \mathcal{C}_2^{i=4} + \mathcal{C}_2^{i=5}\big) \, ,\\
		\mathcal{C}_2^{\mathrm{S}_1} & = -\frac{1}{3} \big(\mathcal{C}_2^{i=4} + 2 \, \mathcal{C}_2^{i=5}\big) \, .
	\end{aligned}
\end{equation}
On the other hand, the basis 2-cycles in the flavor flux basis are
\begin{equation}
	\begin{aligned}
		\big(\mathcal{C}_2^{\mathrm{N}_1}\big)' & = \mathcal{C}_2^{i=1} = 2 \, \mathcal{C}_2^{\mathrm{N}_1} - \mathcal{C}_2^{\mathrm{N}_1} \, ,\\
		\big(\mathcal{C}_2^{\mathrm{N}_2}\big)' & = \mathcal{C}_2^{i=2} = -\mathcal{C}_2^{\mathrm{N}_1} + 2 \, \mathcal{C}_2^{\mathrm{N}_2} \, ,\\
		\big(\mathcal{C}_2^\mathrm{C}\big)' & = \mathcal{C}_2^\mathrm{C} \, ,\\
		\big(\mathcal{C}_2^{\mathrm{S}_2}\big)' & = \mathcal{C}_2^{i=4} = -2 \, \mathcal{C}_2^{\mathrm{S}_2} + \mathcal{C}_2^{\mathrm{S}_1} \, ,\\
		\big(\mathcal{C}_2^{\mathrm{S}_1}\big)' & = \mathcal{C}_2^{i=5} = \mathcal{C}_2^{\mathrm{S}_2} - 2 \, \mathcal{C}_2^{\mathrm{S}_1} \, ,
	\end{aligned}
\end{equation}
so the corresponding basis transformation matrix is
\begin{equation}
	\mathcal{R}_2^{-1} = \begin{pmatrix} 2 & -1 & 0 & 0 & 0 \\ -1 & 2 & 0 & 0 & 0 \\ 0 & 0 & 1 & 0 & 0 \\ 0 & 0 & 0 & -2 & 1 \\ 0 & 0 & 0 & 1 & -2 \end{pmatrix} \, .
\end{equation}
The flux quanta in the two bases are then related by
\begin{equation}
	\begin{gathered}
		N_{\mathrm{N}_1}' = \frac{1}{3} (2 N_{\mathrm{N}_1} + N_{\mathrm{N}_2}) \, , \qquad N_{\mathrm{N}_2}' = \frac{1}{3} (N_{\mathrm{N}_1} + 2 N_{\mathrm{N}_2}) \, , \qquad N_\mathrm{C}' = N \, ,\\
		N_{\mathrm{S}_2}' = -\frac{1}{3} (2 N_{\mathrm{S}_2} + N_{\mathrm{S}_1}) \, , \qquad N_{\mathrm{S}_1}' = -\frac{1}{3} (N_{\mathrm{S}_2} + 2 N_{\mathrm{S}_1}) \, .
	\end{gathered}
\end{equation}

Following the analogous exercise for $k \geq 4$,
one can inductively find that the basis transformation matrix $\mathcal{R}_2^{-1}$ is given by
\begin{equation}
	\mathcal{R}_2^{-1} = \begin{pmatrix} \mathsf{A}_{k-1} & 0 & 0 \\ 0 & 1 & 0 \\ 0 & 0 & -\mathsf{A}_{k-1} \end{pmatrix} \, ,
\end{equation}
where $\mathsf{A}_{k-1}$ is the Cartan matrix of $\mathfrak{su}(k)$. Note that this formula applies also to $k=2$ and $k=3$.

To summarize, the flux quanta $N_{\mathrm{N}_i}, N_{\mathrm{S}_i}$ in the resolution flux basis can be expressed in terms of the flux quanta $N_{\mathrm{N}_i}', N_{\mathrm{S}_i}'$ in the flavor flux basis as
\begin{equation}
	N_{\mathrm{N}_i} = \sum_{i=1}^{k-1} (\mathsf{A}_{k-1})_{ij} \, N_{\mathrm{N}_j}' \, , \qquad N_{\mathrm{S}_i} = -\sum_{i=1}^{k-1} (\mathsf{A}_{k-1})_{ij}\,  N_{\mathrm{S}_j}' \, ,
\end{equation}
 Alternately, we have
\begin{equation}
	N_{\mathrm{N}_i}' = \sum_{i=1}^{k-1} (\mathsf{A}_{k-1}^{-1})_{ij} \, N_{\mathrm{N}_j} \, , \qquad N_{\mathrm{S}_i}' = -\sum_{i=1}^{k-1} (\mathsf{A}_{k-1}^{-1})_{ij} \, N_{\mathrm{S}_j} \, ,
\end{equation}
where the inverse $\mathsf{A}_{k-1}$ given by \cite{Wei_2017}
\begin{equation}
	(\mathsf{A}_{k-1}^{-1})_i^j = \mathrm{min}\{i,j\} - \frac{ij}{k} \, .
\end{equation}
We can thus interpret the flux quanta identifications in \eqref{dictionary} as a change of basis between a natural 4-cycle and a natural 2-cycle basis, along with a reversal of indexing and sign flip in the south to preserve positivity of the fluxes, i.e.
\begin{equation}
	N_{\widetilde \beta_i}=N_{\mathrm{N}_i}' \, , \qquad N_{\widetilde \gamma_i}=-N_{\mathrm{S}_{k-i}}' \, .
\end{equation}






\bibliographystyle{./JHEP}
\bibliography{./references}


\end{document}